\documentclass[twocolumn]{aastex62}

\newcommand{\Msun}{$M_{\odot}$}
\newcommand{\Mbh}{$M_{\rm BH}$}

\newcommand{\Lsun}{$L_\odot$}

\renewcommand{\deg}{\ensuremath{^{\circ}}}
\newcommand{\ml}{\emph{M/L}}
\newcommand{\mleff}{\emph{M/L$_{\rm eff}$}}
\newcommand{\hst}{\emph{HST}}
\newcommand{\kms}{km~s$^{-1}$}
\newcommand{\se}{$\arcsec$}
\newcommand{\mm}{$\arcmin$}

\newcommand{\go}{G$_{\rm outer}$}

\newcommand{\Mppc}{\Msun/${\rm pc^2}$}

\usepackage{multirow}
\usepackage{footnote}

\graphicspath{{./}{figures/}}

\received{September 17, 2018}
\revised{December 18, 2018}
\accepted{January 14, 2019}
\submitjournal{ApJ}

\shorttitle{Improved the Dynamical Masses of the Central BHs in Nearby Low-mass ETGs}
\shortauthors{Nguyen et al.}

\begin{document}

\title{\Large{\bf Improved Dynamical Constraints on the Masses of the Central Black Holes in Nearby Low-mass Early-type Galactic Nuclei And the First Black Hole Determination for NGC 205}}

\correspondingauthor{Dieu D. Nguyen, Anil C. Seth}
\email{dieu.nguyen@utah.edu, aseth@astro.utah.edu}

\author[0000-0002-5678-1008]{Dieu D. Nguyen}
\affil{Department of Physics and Astronomy, University of Utah, 115 South 1400 East, Salt Lake City, UT 84112, USA}
\affiliation{National Astronomical Observatory of Japan (NAOJ), National Institute of Natural Sciences (NINS), 2-21-1 Osawa, Mitaka, Tokyo 181-8588, Japan}

\author{Anil C. Seth}
\affiliation{Department of Physics and Astronomy, University of Utah, 115 South 1400 East, Salt Lake City, UT 84112, USA}

\author[0000-0002-6922-2598]{Nadine Neumayer}
\affiliation{Max Planck Institut f\"ur Astronomie (MPIA), K\"onigstuhl 17, D-69121 Heidelberg, Germany}

\author{Satoru Iguchi}
\affiliation{National Astronomical Observatory of Japan (NAOJ), National Institute of Natural Sciences (NINS), 2-21-1 Osawa, Mitaka, Tokyo 181-8588, Japan}
\affiliation{Department of Astronomical Science, Graduate University for Advanced Studies (SOKENDAI), 2-21-1 Osawa, Mitaka, Tokyo 181-8588, Japan}

\author[0000-0002-1283-8420]{Michelle Cappellari}
\affiliation{Sub-department of Astrophysics, Department of Physics, University of Oxford, Denys Wilkinson Building, Keble Road, Oxford OX1 3RH, UK}

\author{Jay Strader}
\affiliation{Center for Data Intensive and Time Domain Astronomy, Department of Physics and Astronomy, Michigan State University, 567 Wilson Road, East Lansing, MI 48824, USA}

\author{Laura Chomiuk}
\affiliation{Center for Data Intensive and Time Domain Astronomy, Department of Physics and Astronomy, Michigan State University, 567 Wilson Road, East Lansing, MI 48824, USA}

\author{Evangelia Tremou}
\affiliation{AIM, CEA, CNRS, Universit\'e Paris Diderot, Sorbonne Paris Cite, Universit\'é Paris-Saclay, F-91191 Gif-sur-Yvette, France}

\author[0000-0001-9879-7780]{Fabio Pacucci}
\affiliation{Department of Physics, Yale University, 52 Hillhouse Avenue, New Haven, CT 06511, USA}

\author{Kouichiro Nakanishi}
\affiliation{National Astronomical Observatory of Japan (NAOJ), National Institute of Natural Sciences (NINS), 2-21-1 Osawa, Mitaka, Tokyo 181-8588, Japan}
\affiliation{Department of Astronomical Science, Graduate University for Advanced Studies (SOKENDAI), 2-21-1 Osawa, Mitaka, Tokyo 181-8588, Japan}

\author{Arash Bahramian}
\affiliation{International Centre for Radio Astronomy Research Curtin University, GPO Box U1987, Perth, WA 6845, Australia}

\author{Phuong M. Nguyen}
\affiliation{Department of Physics, Quy Nhon University, 170 An Duong Vuong, Quy Nhon, Vietnam}

\author{Mark den Brok}
\affiliation{Leibniz-Institut f\"ur Astrophysik Potsdam (AIP), An der Sternwarte 16, 14482 Potsdam, Germany}

\author[0000-0002-9852-2258]{Christopher C. Ahn}
\affiliation{Department of Physics and Astronomy, University of Utah, 115 South 1400 East, Salt Lake City, UT 84112, USA}

\author[0000-0001-6215-0950]{Karina T. Voggel}
\affiliation{Department of Physics and Astronomy, University of Utah, 115 South 1400 East, Salt Lake City, UT 84112, USA}

\author{Nikolay Kacharov}
\affiliation{Max Planck Institut f\"ur Astronomie (MPIA), K\"onigstuhl 17, D-69121 Heidelberg, Germany}

\author{Takafumi Tsukui}
\affiliation{National Astronomical Observatory of Japan (NAOJ), National Institute of Natural Sciences (NINS), 2-21-1 Osawa, Mitaka, Tokyo 181-8588, Japan}
\affiliation{Department of Astronomical Science, Graduate University for Advanced Studies (SOKENDAI), 2-21-1 Osawa, Mitaka, Tokyo 181-8588, Japan}

\author{Cuc K. Ly}
\affiliation{Department of Physics, Quy Nhon University, 170 An Duong Vuong, Quy Nhon, Vietnam}

\author{Antoine Dumont}
\affiliation{Department of Physics and Astronomy, University of Utah, 115 South 1400 East, Salt Lake City, UT 84112, USA}

\author{Renuka Pechetti}
\affiliation{Department of Physics and Astronomy, University of Utah, 115 South 1400 East, Salt Lake City, UT 84112, USA}



\begin{abstract}
We improve the dynamical black hole (BH) mass estimates in three nearby low-mass early-type galaxies--NGC 205, NGC 5102, and NGC 5206. We use new \hst/STIS spectroscopy to fit the star formation histories of the nuclei in these galaxies, and use these measurements to create local color--mass-to-light ratio (\ml) relations.  We then create new mass models from \hst~imaging and combined with adaptive optics kinematics, we use Jeans dynamical models to constrain their BH masses.  The masses of the central BHs in NGC 5102 and NGC 5206 are both below one million solar masses and are consistent with our previous estimates, $9.12_{-1.53}^{+1.84}\times10^5$\Msun~and $6.31_{-2.74}^{+1.06}\times10^5$\Msun~(3$\sigma$ errors), respectively. However, for NGC 205, the improved models suggests the presence of a BH for the first time, with a best-fit mass of $6.8_{-6.7}^{+95.6}\times10^3$\Msun~(3$\sigma$ errors).  This is the least massive central BH mass in a galaxy detected using any method.  We discuss the possible systematic errors of this measurement in detail. Using this BH mass, the existing upper limits of both X-ray, and radio emissions in the nucleus of NGC 205 suggest an accretion rate $\lesssim$$10^{-5}$ of the Eddington rate.  We also discuss the color--\mleff~relations in our nuclei and find that the slopes of these vary significantly between nuclei.  Nuclei with significant young stellar populations have steeper color--\mleff~relations than some previously published galaxy color--\mleff~relations. 

\end{abstract}




\section{Introduction}\label{sec:intro}

Observational efforts over the last two decades have revealed that every massive galaxy ($M_{\star}\gtrsim10^{11}$\Msun) contains a central supermassive black hole (SMBH, $M_{\rm SMBH} \gtrsim10^6M_\odot$) at its center \citep[e.g.,][]{Kormendy13, Saglia16}. Empirical surveys have shown the macroscopic properties of massive galaxies (e.g., the bulge velocity dispersion, bulge mass, bulge luminosity) correlate with their SMBHs \citep[e.g.,][]{Kormendy95, Magorrian98, Ferrarese00, Marconi03, Haring04, Gultekin09, Beifiori12, Kormendy13, McConnell13, Saglia16}. These scaling relations suggest SMBHs may play a pivotal role in the growth and evolution of galaxies \citep[e.g.,][]{Schawinski07}. Theoretical work suggests these correlations between the masses of SMBHs and the properties of their hosts can be created by feedback from the central engine of active galactic nuclei (AGN) onto the outer gas reservoirs \citep[e.g.][]{Silk98, DiMatteo08,Fabian12,Netzer15}.

However, the presence of SMBHs and the importance of AGN feedback in lower mass galaxies is less clear. In particular it remains unclear if the scaling relations between SMBH masses and galaxy properties that hold at higher mass break down for lower-mass galaxies. Increased scatter around the relation has been seen for Milky Way like galaxies \citep{Greene16,Lasker16}, while BH masses in the lowest mass galaxies seem to fall below the bulge mass relation seen for higher mass galaxies \citep{Scott13,Graham15, Nguyen17, Chilingarian18, Nguyen18}.  The cause of this change at low masses is still debated: perhaps it is tied to the formation history of the bulge \citep[e.g.,][]{Kormendy12} or perhaps to the star formation history of the galaxy more generally \citep[e.g.,][]{Caplar15, Terrazas17}. 

A population of intermediate mass black holes (IMBHs, $10^{3}$\Msun$<M_{\rm BH}<10^{6}$\Msun) with masses inferred from the velocity widths of their optical broad-line emissions have been found in galaxies with stellar masses $\lesssim$$10^{10}$\Msun, but these systems account for $<$1\% of low-mass galaxies \citep{Barth04, Greene07, Thornton08, Dong12, Reines13, Reines15, Baldassare15, Chilingarian18}. Other accretion signatures are also used to identify IMBHs, including narrow-line emission \citep[e.g.,][]{Moran14}, coronal emission in the mid-infrared \citep[e.g.,][]{Satyapal09}, tidal-disruption events \citep[e.g.,][]{Maksym13}, and hard X-ray emission \citep[e.g.,][]{Gallo08, Desroches09, Gallo10, Miller15, She17}. In addition, ultracompact dwarfs (UCDs) are known as the lowest-mass systems to host central SMBHs, which are likely stripped galaxy nuclei \citep{Mieske13, Seth14, Ahn17, Afanasiev18, Ahn18}.  However, only a few dynamical BH mass measurements have been made at the low-mass, low-dispersion end \citep{denbrok15, Thater17}, including our recent work on nearby early-type galaxies \citep[ETGs;][hereafter N17; N18]{Nguyen17, Nguyen18}.  Due to the small fraction of all low-mass galaxies with identifiable AGN (and the difficulty in measuring the masses of SMBHs in detected AGN), these small number of dynamical measurements in the nearest systems still provide our best information on how BHs populate host galaxies. 

The demographics of BHs in low-mass galaxies can shed light on a number of interesting astrophysical problems. First, the number of low-mass galaxies that host IMBHs--the ``occupation fraction''--is one of the only currently feasible ways to investigate the unknown formation mechanism of BH seeds in the early Universe, which form either from the direct collapse of massive ($\sim$$10^5$\Msun) seeds \citep[e.g.,][]{Lodato06, Bonoli14} or from the lighter remnants of the first stars \citep[e.g., Population III;][]{Volonteri08, vanWassenhove10, Volonteri10, Volonteri12a, Volonteri12b, Volonteri12c, Fiacconi16, Fiacconi17}. The massive seeds scenario predicts a smaller occupation fraction in low-mass galaxies than the Pop III stars scenario \citep[][N18]{Gallo08, Volonteri08, Greene12, Miller15}. Second, the fraction of low-mass galaxies hosting IMBHs is crucial for measuring the BH number density and therefore the expected rate of stellar tidal disruptions \citep{Kochanek16}. Stellar tidal disruptions are used to probe the IMBH populations of low-mass galaxies \citep[e.g.,][]{Law-Smith17, Wevers17} and may eventually provide constraints on occupation fraction \citep[e.g.,][]{Stone16}.  Finally, the occupation fraction of BHs in dwarf galaxies is the key measurement for studies of the number of BHs we expect to find in stripped galaxy nuclei \citep[e.g.,][]{Mieske13, Pfeffer14, Seth14, Ahn17, Afanasiev18, Ahn18, Voggel18}. 

In N17 and N18 we have measured the dynamical masses of five SMBH/IMBHs in a volume complete sample of low-dispersion ($\sigma\sim20-70$ \kms), low-mass ($M_{\star}\sim10^9-10^{10}$\Msun), ETGs within 3.5~Mpc (M32, NGC 205, NGC 404, NGC 5102, and NGC 5206). We created mass models for the central regions of each galaxy using multi-band \emph{Hubble Space Telescope} (\hst) imaging.  We then compared Jeans anisotropic models \cite[JAM;][]{Cappellari08} to stellar kinematic measurements from Gemini/NIFS or VLT/SINFONI observations to constrain the BH and NSC masses. We found the mass of the SMBH in M32 consistent with previous measurements \citep[$M_{\rm BH} = 2.5\times10^6$\Msun;][]{Verolme02, vandenBosch10} and measured for the first time the masses of two sub-million Solar masses IMBHs in NGC 5102 ($M_{\rm BH} = 8.8^{+4.2}_{-6.6}\times10^5$\Msun) and NGC 5206 ($M_{\rm BH} = 4.5^{+2.3}_{-3.4}\times10^5$\Msun). We obtained an upper limit on the BH mass in NGC 205 of $M_{\rm BH} < 7\times10^4$\Msun, a factor of two larger than the upper limit estimated by \citet{valluri05}.  This work has added up to 50\% of numbers of dynamical sub-million Solar masses IMBHs that have been constrained so far, and resulted in an estimate of 80\% for the BH occupation fraction of ETGs between $M_{\star}\sim10^9-10^{10}$\Msun.  

One of the primary challenges in finding BHs in low-mass galaxies is that their nuclei typically have spatially varying stellar populations.  These varying stellar populations make it challenging to transform luminosity models into mass models.  In N17, we used STIS data to measure the star formation history (SFH) of NGC 404, and then used this to construct a color--mass-to-light ratio (\ml) relationships. The derived relationships differed in both slope and normalization with previously published relations \citep{Bell03, Roediger15}. However, the results of our BH mass upper limit changed very little regardless of the color--\ml~relation used.  This is possibly due to the heavily dust extinction on the northeast side in the NGC 404 nucleus and the dust is mixed with the stellar population on the line-of sight direction. Our spectroscopic fitting method although can be able to disentangle the stellar population simply assuming the dust located in front of the population and does not account for the dust and population mix attribution. In N18, for the dynamical modeling of the SMBH/IMBHs in M32, NGC 205, NGC 5102, and NGC 5206 we compared models with a constant \ml, and those based on the color--\ml~relations of \citet{Bell03} and \citet{Roediger15}, using the latter as our default models.  The BH mass results were somewhat sensitive to the assumed color--\ml~relation (especially in NGC 5206).  

This work presents new constraints on dynamical mass estimates of the BHs in three nearby low-mass ETGs (NGC 205, NGC 5102, and NGC 5206), which were previously presented in N18.  We use new \hst/STIS spectroscopy and ACS/HRC (NGC 205), WFC3 (NGC 5102), and WFPC2 (NGC 5206) imaging to quantify the spatial variations in their nuclei \mleff~based on colors and specific star formation histories (SFHs) throughout their nuclei and improve their BH mass estimates using the Gemini/NIFS (NGC 205) and VLT/SINFONI (NGC 5102 and NGC 5206) kinematic data. The method we use in this work was developed and presented in N17.
 
This paper is organized into seven Sections. In Section \ref{sec:data}, we present the observations and data reduction. The \hst/STIS spectroscopic color--\ml~relations and their new mass maps and mass models for all three galaxies are constructed in Section \ref{sec:mlvary}. In Section \ref{sec:relationinsight}, we provide an insight of our color--\ml~relations into these nuclei stellar populations and a guidance to apply them to measure the \ml~variability and mass map in the nuclei, which lack of the stellar spectroscopic information. We model the new BH mass constraints and their uncertainties via Jeans models using the new mass models from Section \ref{sec:mlvary} and the kinematic measurements from N18 in Section \ref{sec:steldyn}. We discuss our results and conclude in Section \ref{sec:dis} and Section \ref{sec:cons}, respectively.

\section{Data and Data Reduction}\label{sec:data}

\subsection{\emph{HST} Imaging}\label{ssec:images}

The imaging data we use here for NGC 205 and NGC 5206 is presented in detail in N18, while new \hst~imaging data was obtained for NGC 5102.  To briefly summarize, for NGC~205, we use \hst/ACS/HRC data in the F555W and F814W filters, while for NGC 5206 we use WFPC2/PC data in the F555W and F814W filters. 

Our new data for NGC 5102 includes WFC3/UVIS data in F336W, F547M, and F814W images.  These data were observed in the UVIS2-C512C-SUB aperture and obtained contemporaneously with the STIS spectroscopic observation (see Section \ref{ssec:stisspec}). Details are given in Table  \ref{hst_data_tab}.  We downloaded the WFC3/UVIS flat-fielded images from the \hst/The Barbara A. Mikulski Archive for Space Telescopes (MAST) and combined the images in each filter using drizzlepac/Astrodrizzle \citep{Avila12}.

The astrometry for the NGC~205 nucleus was a particular challenge to calculate due to the lack of cataloged point sources around the galaxy in 2MASS and SDSS. We correct the astrometry from the \hst/F814W and F555W ACS HRC images using Gaia astrometry. 

For all three galaxies, we correct the central positions of their nuclei to align images from all \hst~filters to the F814W data (after applying the astrometric correction for NGC 205). The sky backgrounds of these images are determined by comparing them to ground-based data. These ground-based data include the $I$-band for NGC 205  \citep{valluri05} and the Carnegie-Irvine Galaxy Survey \citep[CGS;][]{Ho11, Li11, Huang13} for NGC 5102 and NGC 5206. Additional details are given in N18.

We create point-spread functions (PSFs) for the new WFC3 images of NGC 5102 in the same way presented in \citet{denbrok15} and N18. The PSF model for each WFC3 exposure is created using the \texttt{Tiny Tim} routine to insert a PSF into each of the four individual \texttt{flt} exposures. We apply the \texttt{tiny3} task to model these PSFs, which takes into account the frame distortion and charge diffusion kernel of a PSF. The position of the nucleus in each individual exposure is then transferred to these mock \texttt{flt} images to simulate our observations. The final PSF for each filter is the stack of the four PSFs at each of the four dither positions using the \texttt{Astrodrizzle} package. The PSFs of the ACS/HRC (NGC 205) and WFPC2/PC (NGC 5206) images are taken from N18.  The imaging in each galaxy is used to create our color--\ml~relations and make our final mass maps below.

\begin{table*}[!ht]
\scriptsize
\caption{\hst/WFPC2 PC,  WFC3/UVIS, ACS HRC Images, and STIS Spectroscopies}    
\hspace{-34mm}
\begin{tabular}{ccccccccccccccc}
 \hline\hline
Object       &$\alpha$(J2000)&$\delta$(J2000)&       Camera      &         Aperture       &     UT Date  &   PID  &  Filter  &     Exptime     &Pixelscale&Zeropoint$\tablenotemark{a}$&A$_{\rm \lambda}\tablenotemark{b}$\\
                 &    (h~m~s)         &(\deg~\mm~\se)&                         &                              &                     &           &            &           (s)       &  (\se/pix)  &                        (mag)              &               (mag)                   
                    \\
  (1)           &           (2)           &           (3)         &            (4)         &               (5)           &       (6)         &    (7)   &    (8)   &           (9)        &      (10)    &                         (11)                &               (12)    			  \\	                
\hline
\multirow{3}{*}{NGC 205}      &                         &                         &  ACS/HRC  &              HRC         &2002 Sep 08 & 9448  &F555W& $4\times640$   &  0.0300   &                    25.262             &    	        0.047 \\
   &  00:40:22.054$\tablenotemark{c}$  &   41:41:07.50$\tablenotemark{c}$   &  ACS/HRC  &              HRC         &2002 Sep 08 & 9448  &F814W& $8\times305$   &  0.0300   &                    24.861             &    	        0.026 \\[2mm]
                                              &                         &                         &  STIS/CCD &   52$\times$0.1      &2017 Jul 21-23&14742&G430L & $5\times946$  &  0.0500  &                        --                  &   		   --     \\  
\hline
\multirow{3}{*}{NGC 5102}    &                         &                         &WFC3/UVIS&UVIS2-C512C-SUB&2017 Jun 01&14742&F336W& $1\times1232$ &  0.0400   &                   23.481                 &   		0.075  \\
                                              &   13:21:55.96   &  $-$36:38:13.0 &WFC3/UVIS&UVIS2-C512C-SUB&2017 Jun 01&14742&F547M& $1\times524$   &  0.0400   &                   24.748                 &   		0.050  \\  
                                              &                         &                         &WFC3/UVIS&UVIS2-C512C-SUB&2017 Jun 01&14742&F814W& $1\times464$   &  0.0400   &                   24.686                 &   		0.026  \\[2mm]                                                 
                                              &                         &                         &  STIS/CCD &   52$\times$0.1      &2017 Jun 01&14742&G430L & $11\times933$ &  0.0500   &                        --                     &   		   --      \\                 
\hline
\multirow{3}{*}{NGC 5206}   &                         &                          &    WFPC2   &      PC1-FIX            &1996 May 11  & 6814&F555W& $6\times350$   &  0.0445   &                    24.664              &   		0.047  \\
                                             &   13:33:43.92   &  $-$48:09:05.0  &    WFPC2   &      PC1-FIX            &1996 May 11  & 6814&F814W& $6\times295$   &  0.0445   &                    23.758              &   		0.026  \\[2mm]
                                             &                         &                          &  STIS/CCD &   52$\times$0.1      &2017 Sep 21-22&14742&G430L & $7\times1924$ &  0.0500&                        --                  &   		    --     \\ 
\hline
\end{tabular}
\tablenotemark{}
\tablecomments{Column 1: galaxy name. Columns 2 and 3: position (R.A. and Decl.) of the galaxy from \hst/HLA data.   Columns 4 and 5: the camera and the aperture in which the data were taken. Column 6: date when the observations were performed. Column 7: the principle investigator identification numbers. Column 8: filter. Column 9: the exposure times of the observations. Column 10: the pixel-scale of each camera.  Columns 11 and 12: the photometric zero point and extinction value in each filter.\\ 
$^{\rm a}$The photometric zero points were based on Vega System.\\
$^{\rm b}$The extinction values A$_{\lambda}$ were obtained from \citet{Schlafly11} with interstellar extinction law from UV to near-infrared \citep[NIR;][]{Cardelli89}.\\
$^{\rm c}$Astrometrically corrected using the nucleus position observed from Gaia; the \hst/HLA position is at (00:40:22.00, 41:41:07.10).}
\label{hst_data_tab}
\end{table*}


\subsection{\emph{HST/STIS} Spectroscopic Data}\label{ssec:stisspec}

The STIS spectroscopic observations (PID: 14742, PI: Nguyen) were taken on 2017 September 22-23 for NGC 205, 2017 June 01 for NGC 5102, and 2017 July 21-22 for NGC 5206.  All data was taken with the G430L grating and the $52\arcsec0\times0\arcsec.1$ slit. This provides spectra over a wavelength range of 2900~\AA~to 5700~\AA~with a pixel size of 2.73 \AA, and spectral resolving power of R $\sim$ 530 -- 1040. The specific exposure times per frame and total exposure times for each galaxy is detailed in Table \ref{hst_data_tab}. However, only nine exposures of NGC 5102 are used in this analysis, as the last two are unusable due to dithering off the chip.  For each target, the source was dithered along the slit and centered near the E1 aperture position to minimize charge-transfer inefficiency losses. 

Reduced and rectified spectroscopic exposures (\texttt{x2d} files) were downloaded from \hst/MAST. Details of the data reduction follow N17.  First, median frames were constructed from the undithered combined data in each galaxy and subtracted to remove hot pixels.  Each dither was separated by 30 (NGC 205), 13 (NGC 5102), and 20 (NGC 5206) pixels, thus, some background galaxy light will be included in their median images. However, this effect is very small; the maximum row-averaged fluxes of their median images are $<$10\% (NGC 205), $<$5\% (NGC 5102), and $<$9\% (NGC 5206) of that at the outermost radii we analyze for each galaxy.  After median combining, we combine the dithered images including rejection of cosmic rays and bad pixel masking. 

We examine the quality of the reduced spectroscopic data by looking at the signal-to-noise (S/N) of the central pixel in the combined spectra at 3700 \AA~and 5000 \AA~for each galaxy and get S/N (3700 \AA, 5000 \AA) =  (73, 94) for NGC 205, (178, 200) for NGC 5102, and (52, 79) for NGC 5206. At larger radii, we bin pixels together to obtain S/N $\gtrsim$ 20 at 5000 \AA.  The outermost bins used are at $\pm$(1\farcs15--0\farcs85) for NGC 205, $\pm$(2\farcs15--1\farcs85) for NGC 5102, and $\pm$(1\farcs45--1\farcs15) for NGC 5206. We note that the $\pm$ signs indicate bins on either side of the galactic center.

We match the astrometry of our \hst/STIS spectroscopy to the F814W image as described in Section 2.5 of N17.  This astrometric alignment is an important factor in measuring the color--\ml~correlation and creating the mass maps.  Because of dramatic drop in the S/N in the STIS image at large radius, the fitting of the one dimensional (1D) images was performed at radii $<$1\farcs2 (NGC 205), $<$2\farcs2 (NGC 5102), and $<$1\farcs5 (NGC 5206).

\subsection{Integral Field Spectroscopic Data}\label{ssec:ifuspec}

The kinematic data used here are identical to that presented in N18.  The integral field unit (IFU) spectroscopic data of NGC 205 was obtained with Gemini/NIFS using the Altair tip-tilt laser guide star system, while NGC 5102 and NGC 5206 were observed with SINFONI \citep{Eisenhauer03, Bonnet04} on the UT4 (Yepun) of the European Southern Observatory's (ESO) VLT at Cerro Paranal, Chile. Details are given in Table 2 and Section 2.2 of N18. The stellar kinematics are derived from the CO band-head absorption lines and are shown in Figure 5 and Section 5 of N18. We note that in NGC~205, subtraction of bright individual stars was done to obtain measurements of the smoother component of the galaxy kinematics using PampelMuse \citep{Kamann18}.

\begin{figure*}[!ht]  
\vspace{-2mm}
\centering
   	\includegraphics[scale=0.155]{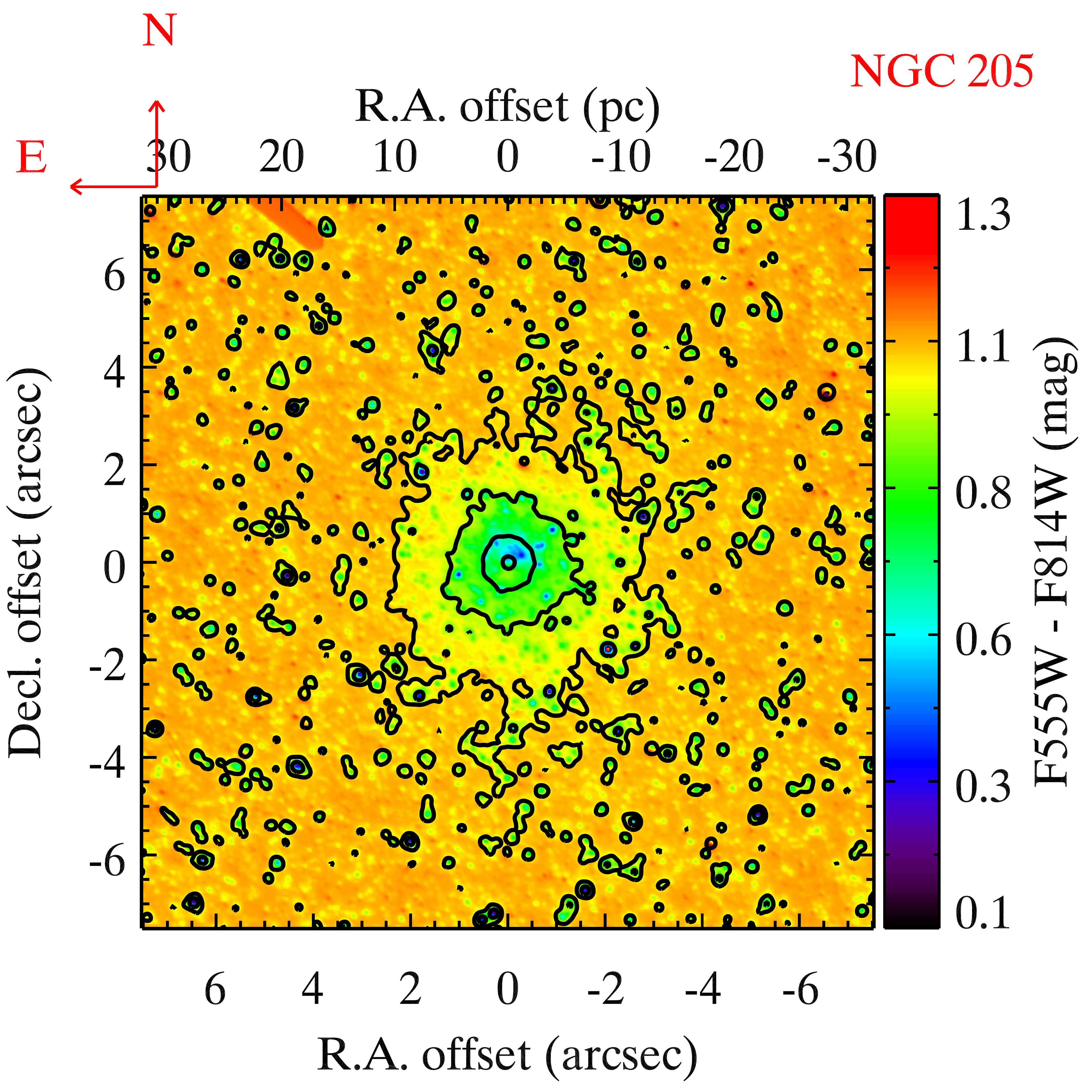}
   	\includegraphics[scale=0.155]{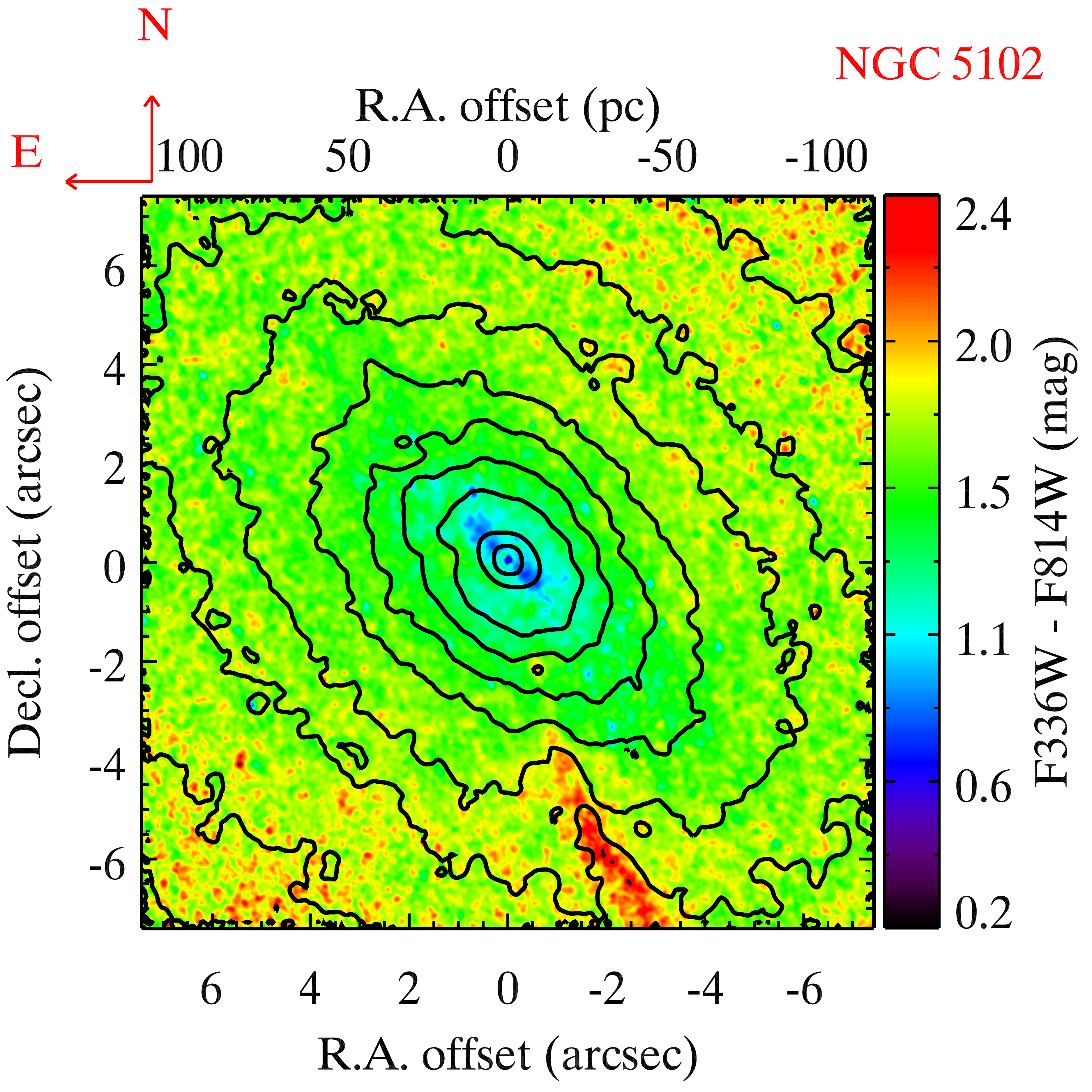}	
   	\includegraphics[scale=0.155]{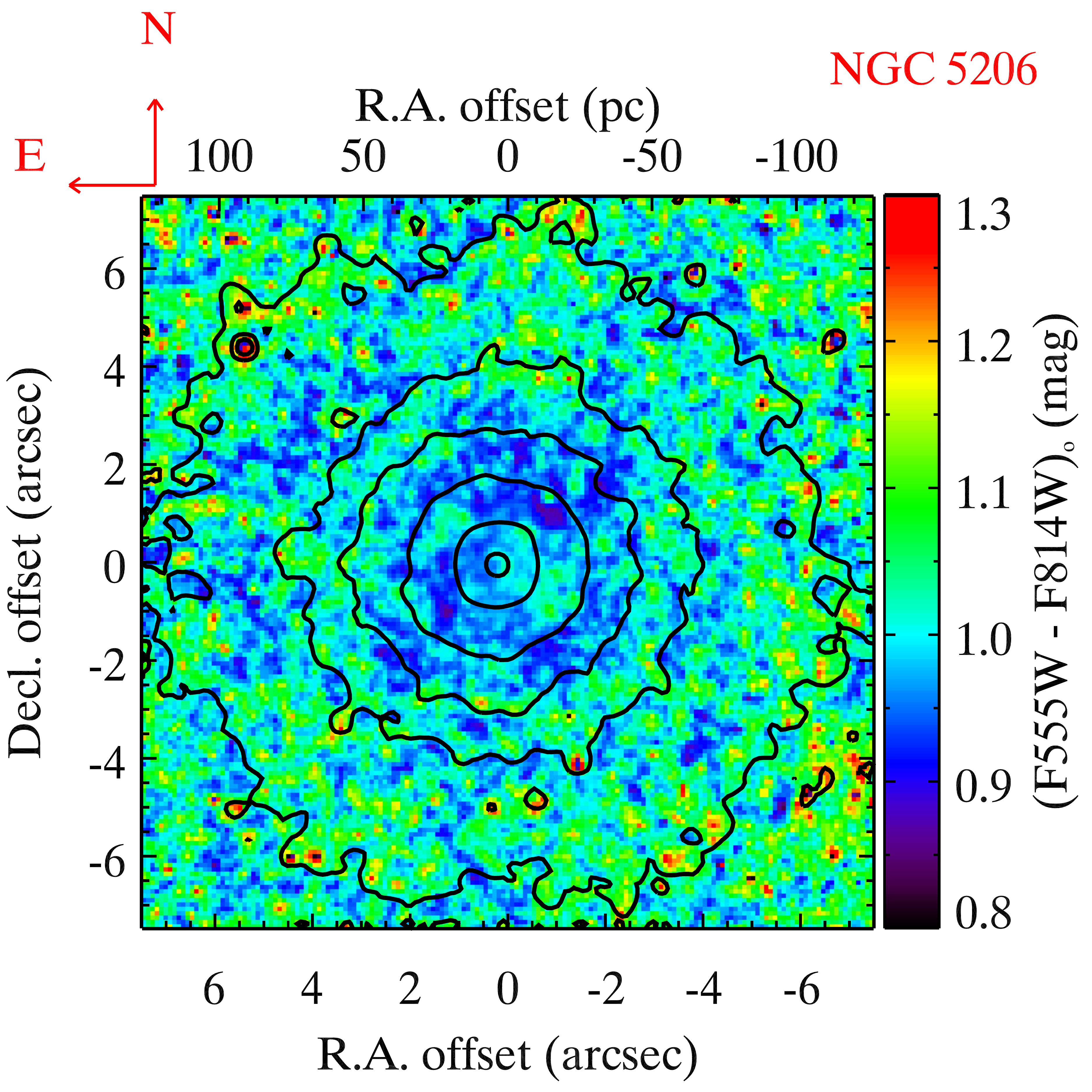} 
\caption{\small{Color maps of the nuclei of NGC 205 (F555W--F814W), NGC 5102 (F336W--F814W), and NGC 5206 (F555W--F814W). These color maps are made from \hst~imaging, cross-convolved to match the PSFs. The contours show the F814W surface brightness at $\mu_{\rm F814W}$ of (13.0, 15.0, 16.5, 17.5) mag arcsec$^{-2}$ for NGC 205, (12.2, 13.0, 13.3, 13.7, 14.0, 15.4, 16.0, 16.5, 17.0) mag arcsec$^{-2}$ for NGC 5102, and (15.0, 16.0, 17.0, 18.0, 18.5, 19.0) mag arcsec$^{-2}$ for NGC 5206. The centers of the nuclei of NGC 205, NGC 5102, and NGC 5206 are represented as (0$\arcsec$, 0$\arcsec$) on these maps (their real Equatorial J2000 coordinates are presented in columns 2 and 3 of Table \ref{hst_data_tab}).}}   
\label{colormap}   
\end{figure*}

In this work, we use these stellar kinematics in combination with updated mass maps (see Section \ref{ssec:maps}) to fit Jeans models to these three galaxies, and present the improved and more accurate estimates of their central BH masses in Section \ref{ssec:jeans}. We note that the dynamical modeling requires the astrometric alignment of these IFU data to their corresponding \hst~images as well.  As was done with the NGC~205 and NGC~5206, our new data was aligned to the F814W image.  These alignments are quite trivial due to the fact that our galaxies have minimal dust and internal extinction in the area around the nucleus as seen in Figure \ref{colormap}.  The PSFs of the IFU are presented in  Section 2.3 and Table 3 of N18. 

\begin{figure} 
\centering
	\includegraphics[scale=0.17]{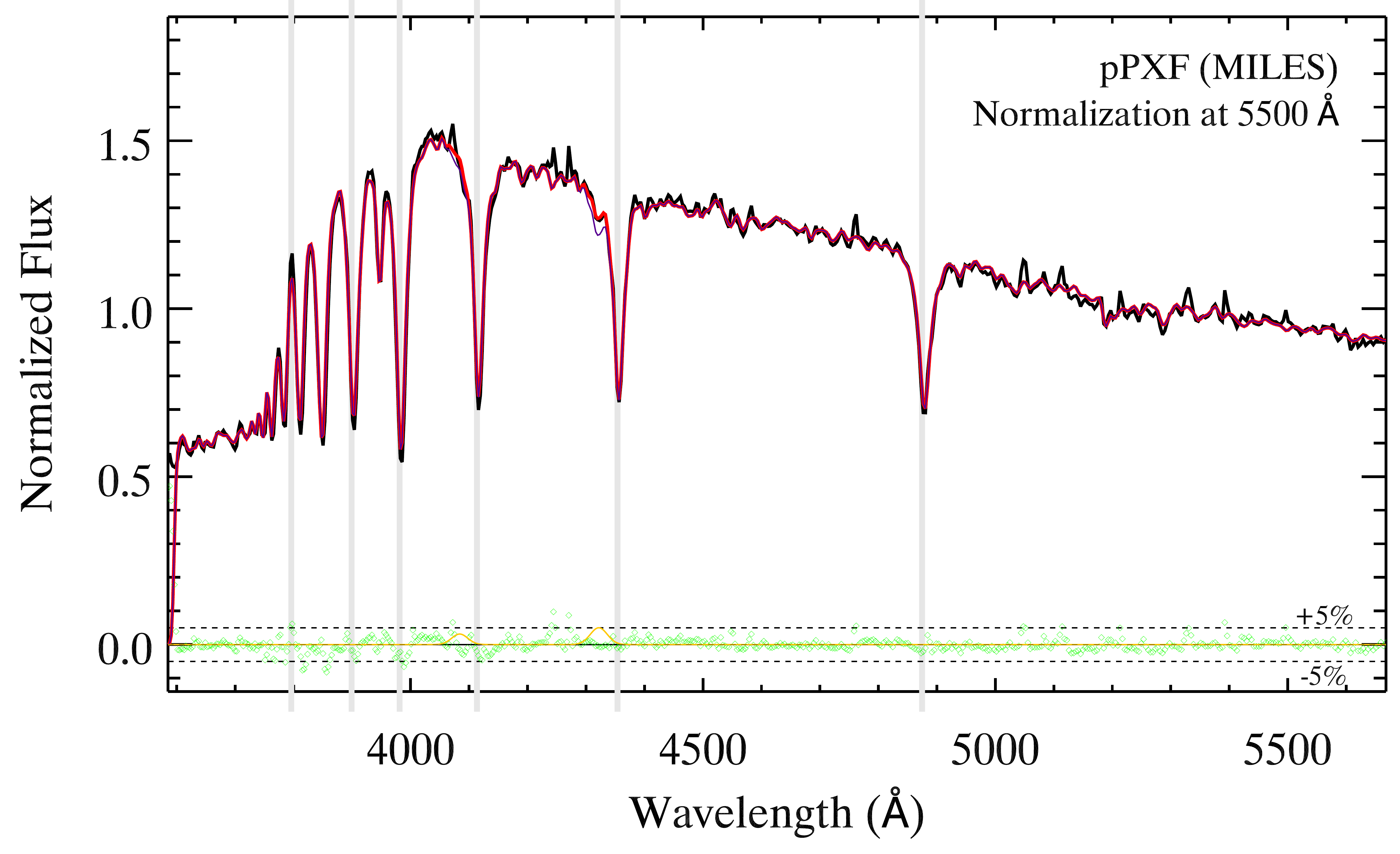}
\caption{\small{The central \hst/STIS spectrum of NGC 5102 (from its central pixel) is shown in black, while the best-fitting stellar population synthesis model fit is shown in red. Vertical gray lines show the position of possible emission line regions. In this pPXF fit we include these emission lines (yellow, embedded in the residuals), and fit them with their stellar component (purple) at the same time. Green data points are the fractional residuals (\texttt{(Data-Model)/Data}) between data and model. The spectrum is normalized at a wavelength of $\lambda = 5500$\AA.}}  
\label{stis}   
\end{figure}

\section{Mass-to-Light Ratio Variations}\label{sec:mlvary}

Significant color variations are seen in all three nuclei (see Figure 1 and also Section 4, N18).  These color variations suggest spatial variations in stellar populations (or extinction) that will create a variable \ml. Here, we combine \hst~imaging and STIS spectroscopic data to fit the color--\ml~relation in each nucleus based on its distinctive stellar populations.  These color--\ml~correlations are the core ingredient in creating the new accurate mass maps enabling better constraints on the dynamical mass estimates of the central SMBHs. 

\subsection{Nuclear Color Variations}\label{ssec:colorvary}

We examine the color variations in all three nuclei using \hst~images (Table \ref{hst_data_tab}). We create the F555W--F814W color map using ACS/HRC images for NGC 205 and WFPC2 PC1 images for NGC 5206, while for NGC 5102, we create a F336W--F814W color map with our new WFC3 images. To create these maps, first, we use the astrometrically aligned image pairs (Section~\ref{ssec:images}). We then cross convolve each image with the PSF of the other filter (e.g., for the F555W--F814W color map, the F555W image was convolved with the F814W PSF and vice versa). The cross-convolution is applied to eliminate spurious gradients near the center of the galaxies that can be caused caused by the different widths of PSFs. Second, we estimate the background level on each image in an annulus at the maximum radius available in each observation, which varies in the range of 10$\farcs$0--12$\farcs$0 away from the nucleus and subtract it off. We then create color images using the Vega-based zero points and correct for foreground extinction as listed in Column 12 of Table \ref{hst_data_tab}.

Figure \ref{colormap} shows these color maps within a field-of-view (FOV) of 14$\farcs$0. The color maps of NGC~205 and NGC~5102 show the signatures of two distinct and dominant stellar populations with young stars concentrated in the nucleus and older populations at larger radii \citep[][N18]{valluri05, Davidge15, Mitzkus17, Kacharov18}.  In addition, NGC 205's nucleus is contaminated by individual young stars or clumps at larger radii \citep{Cappellari99}, while outside the radius of 4$\farcs$0, there are dust clumps and a dust lane in the southern part of NGC 5102's nucleus \citep[][N18]{Davidge15}. We note that we also create F336W--F547M and F547M--F814W color maps for NGC 5102 for use in testing our new mass models. The nucleus of NGC 5206 shows a smaller color range, as expected given its older stellar population \citep{Kacharov18}.

\subsection{Color Correlations with Spectroscopic M/Ls}\label{ssec:correlation}

We use the STIS spectroscopy to measure the star formation history (SFH) and \ml~along the major axis across the nucleus in each galaxy.  We follow the fitting described in N17.  Briefly, we use the penalized pixel-fitting (pPXF) code\footnote{Specifically, we use the IDL version of the code, available at http://purl.org/cappellari/software} \citep{Cappellari04, Cappellari17} to fit the spectra to a set of MILES stellar population templates \citep{Vazdekis10, Vazdekis12} with Chabrier initial mass function (IMF) with massive stars segment logarithmic slope of 1.3 and BaSTI isochrones \citep{Girardi00}. We determine the spectroscopic \ml~based on the single-stellar population (SSP) model weights produced by the pPXF fitting \citep{Mitzkus17, Kacharov18}, and these SSP mass and light predictions are obtained from the MILES website \footnote{http://www.iac.es/proyecto/miles/pages/predicted-masses-and-photometric-observables-based-on-photometric-libraries.php}. We use a set of population models covering the age range from 0.03 to 14.0 Gyr spaced into 53 logarithmically steps, [$\alpha$/Fe] = +0.00, and 12 metallicites Z ([M/H] = $-$2.70, $-$1.79, $-$1.49, $-$1.26, $-$0.96, $-$0.66, $-$0.35, $-$0.25, +0.06, +0.15, +0.26, +0.40). All template spectra are scaled with one scalar to have a median value of 1 at 5,500 \AA~as we do the same for the galaxy spectrum. The gas emission lines are added to fit simultaneously with the stellar spectral templates without masking them.  We did not include an AGN continuum component as used in N17 during the fit because of the lack of AGN signatures within these galaxies.  We note that the pPXF fitting method uses the description of \citet{Calzetti00} to fit for the extinction by default and does not fit $\tau_V$ directly, it fits a \texttt{reddening} = E(B--V) as a variable parameter which relates to dust extinction as $\tau_V=(3.1\times{\texttt{reddening}})/1.086$. Here, we have changed the function. The dust extinction distributions can be separated using the prescription of \citet{Charlot00}, which is represented by the parameter $\tau_V$ in the models.  The translations of this extinction in $V$-band into other bands depend on their effective wavelengths as $\tau_\lambda=\tau_V\times(\lambda/5500$~\AA)$^{-0.7}$ \citep{Charlot00}.  The details of the SFH will be presented in another work (A. Dumont et al., {\em in prep}), while here we just discuss the derived \ml~values, and use them to improve our mass models and refine our BH mass estimates.  We show the spectrum and its SSP fit at the central bin (one pixel) of NGC 5102 as an example in Figure \ref{stis}.  We use these fits to measure the effective mass-to-light ratio (\mleff), which includes both \ml~variations due to stellar population and dust extinction.  Errors on the \mleff are determined via Monte Carlo refitting of the spectra.

\begin{figure*}[!htb]  
\centering
	\includegraphics[scale=0.21]{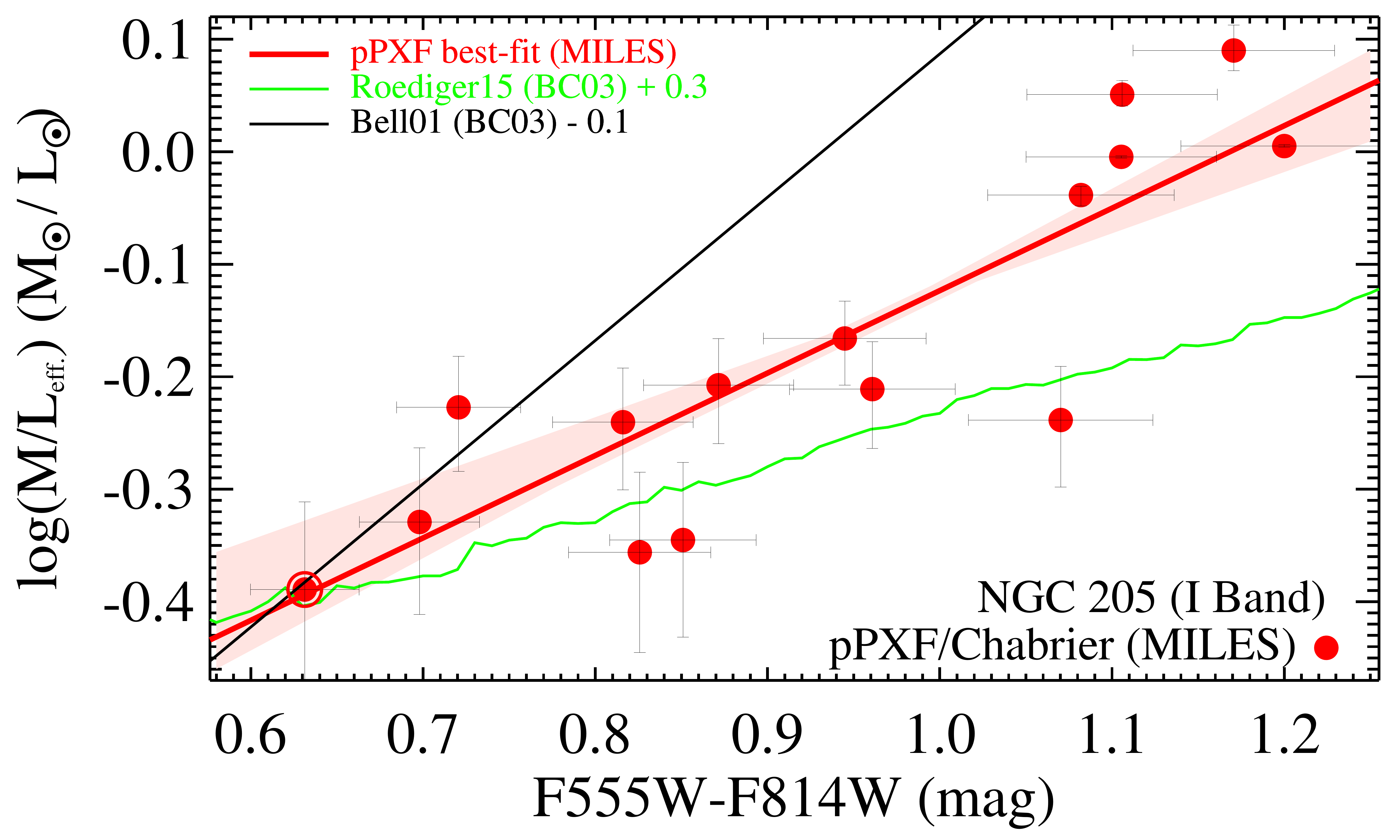}
  	\includegraphics[scale=0.21]{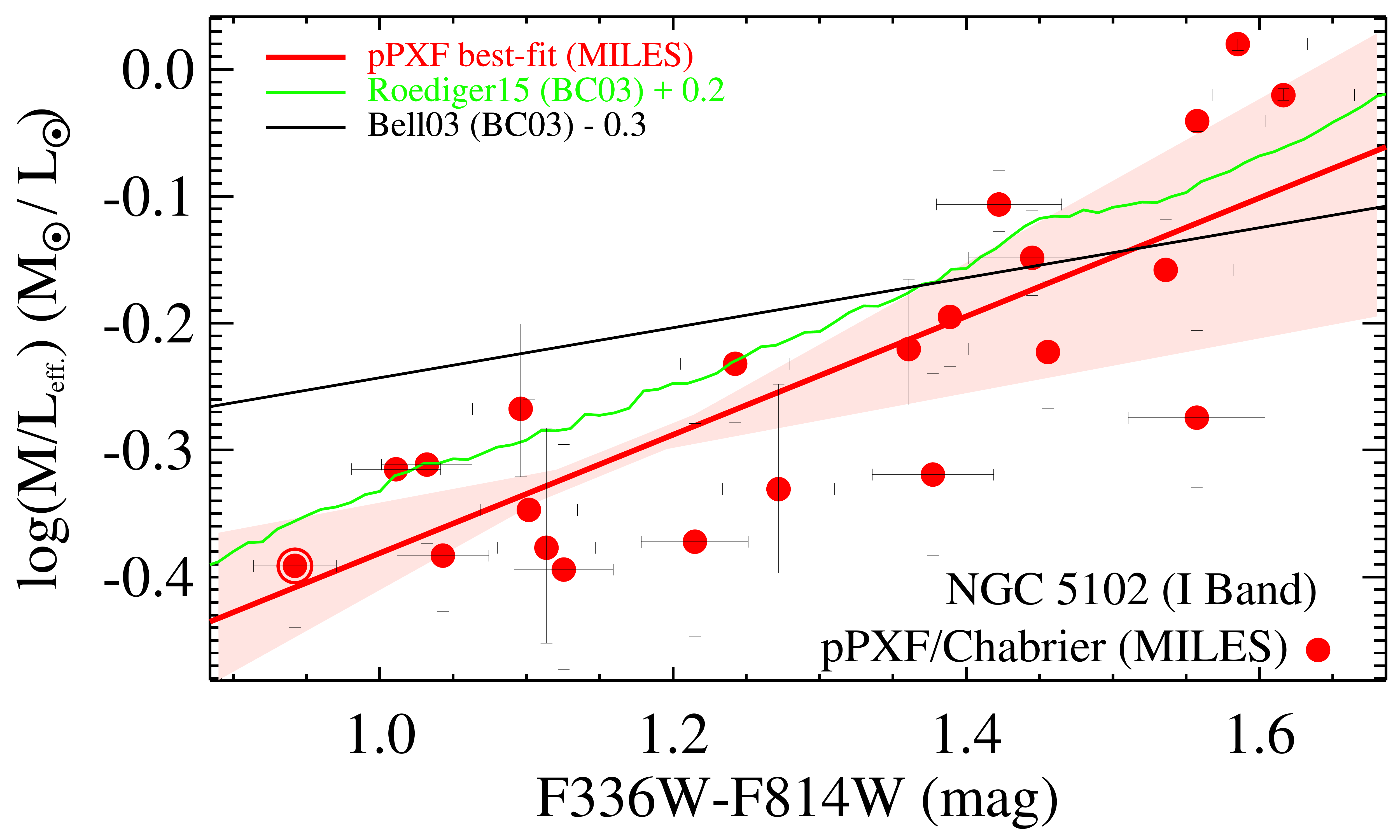}
	\includegraphics[scale=0.21]{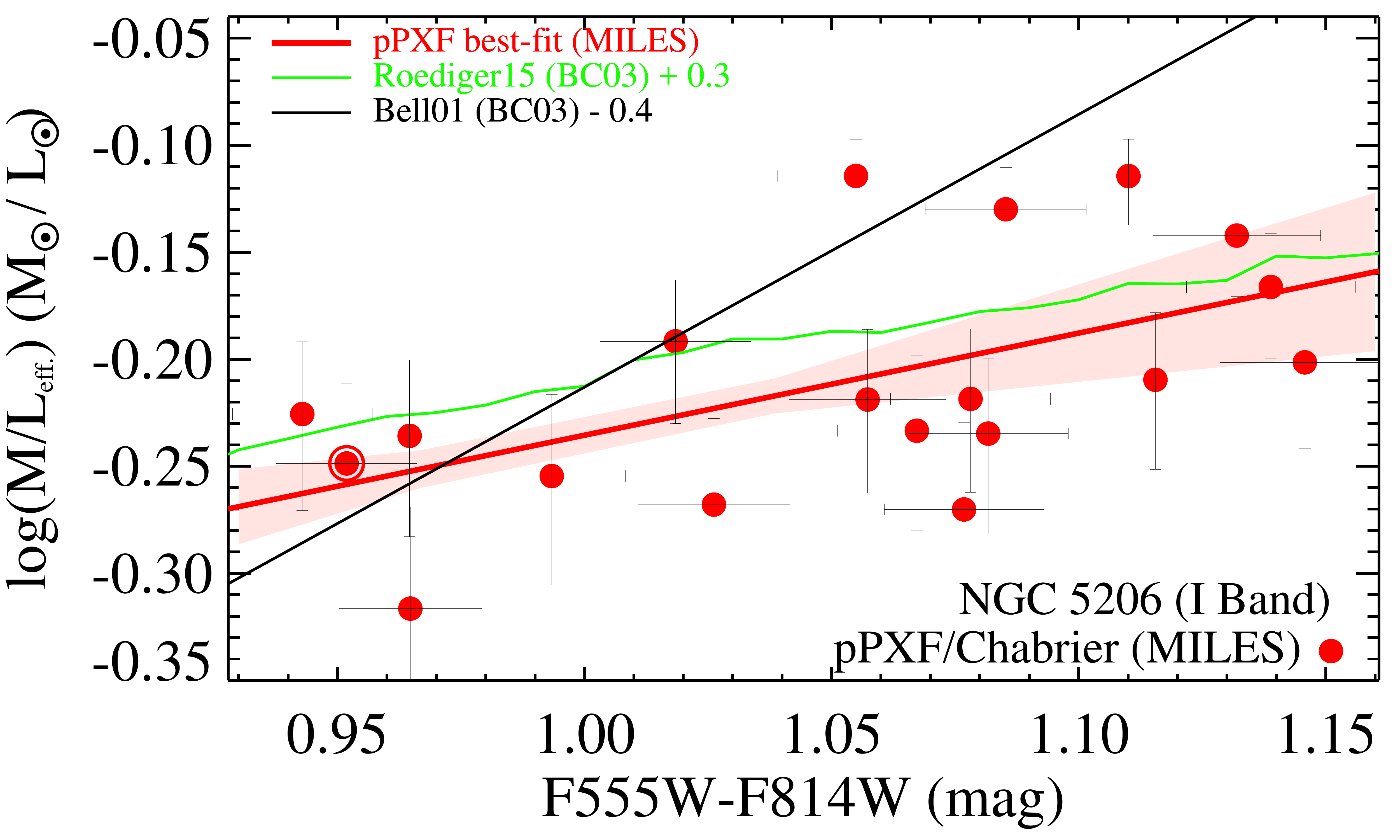} 
        \caption{\small{The effective mass-to-light ratio (\mleff) -- color relations for NGC 205 (top), NGC 5102 (middle), and NGC 5206 (bottom). The horizontal axes show either the F336W--F814W color determined from WFC3 imaging (NGC 5102) or F555W--F814W colors determined from ACS HRC (NGC 205) and WFPC2 PC1 (NGC 5206) images, while the vertical axes show the \mleff~in F814W determined from stellar population fits to STIS spectroscopy of their nuclei. Red points and lines illustrate the data from the stellar population fits using the MILES models and the best-fit linear relations to these data. The circled red points are the central bins (fits to single STIS pixels). The error bars in log(\mleff) were determined via a Monte Carlo analysis of the stellar population fits and these errors are the dominant ones in the best fits of the log(\mleff)--color relations. The black and green solid lines are the predicted color--\ml~correlations from the \citet{Bell01} or \citet{Bell03} and from the \citep{Roediger15} relation; these have been shifted as indicated in the legends in each panel.   The shaded-pink regions are the uncertainties of the best-fit linear relations taking into account both the $\pm1\sigma$ uncertainties in the slopes and intercepts of these best-fit linear relations from the fit as well as boot-strapping uncertainties.}} 
\label{color_m2l_relation}   
\end{figure*}

By aligning the STIS spectroscopy with our color images, we can look at the correlations between the integrated color and spectroscopic \mleff~in each spectroscopic bin along the major axis. The region where the STIS data is useful (S/N $\gtrsim$ 20) covers a wide enough range of integrated color in each galaxy to provide useful constraints on the color--\ml~relations, with the largest variations seen in NGC 205 (0.6~mag in F555W--F814W) and in NGC 5102 (0.7~mag in F336W--F814W), with a smaller variation in NGC 5206 (0.3~mag in F555W--F814W). We present the $I$-band \mleff~along the STIS slit for each galaxy as the red data points in Figure~\ref{color_m2l_relation} with the horizontal axis show the colors determined from the \hst-based color images, and the vertical axis shows the spectroscopic STIS \mleff~values at the same astrometric positions, respectively. We find strong linear correlations between these colors and the logarithm of the \mleff~in all three galaxies as expected based on previous work \citep{Bell01, Bell03, Zibetti09, Roediger15}.  The best linear fits to these data are presented as red solid lines.  

Similar to N17, we determine the errors in these relations using both (1) Monte Carlo errors based on propagating the errors in the spectroscopic \mleff, and (2) bootstrap errors using random replacement sampling.  In both cases, errors in the slope and intercept were estimated by taking the standard deviation of the resulting fits. We first estimate Monte Carlo errors by adding  random noise to the \mleff~measurements of each galaxy and generate new data sets of  \mleff~measurements via Monte Carlo simulation. We then repeat the linear fit  in log scale of \mleff~vs. color and loop this process 100 times. The errors in these relations are measured as the 1$\sigma$ deviation of these best-fit values in terms of \mleff s and intercepts. The  bootstrapping error for each galaxy is determined later during the linear fit of the best-fit color--\mleff~correlation from the whole measurements along the STIS slit. The Monte Carlo \mleff~errors are larger than the bootstrap errors, but both are represented in the shaded pink 1$\sigma$ confidence regions shown in each panel of Figure~\ref{color_m2l_relation}.  We note that for NGC 5102 and NGC 404, we also fit the F336W--F547M and F547M--F814W color--\mleff~relations for more direct comparison between the four relations (Section~\ref{sec:relationinsight}). 
 
\begin{figure*}[!ht]  
\centering
    \includegraphics[scale=0.155]{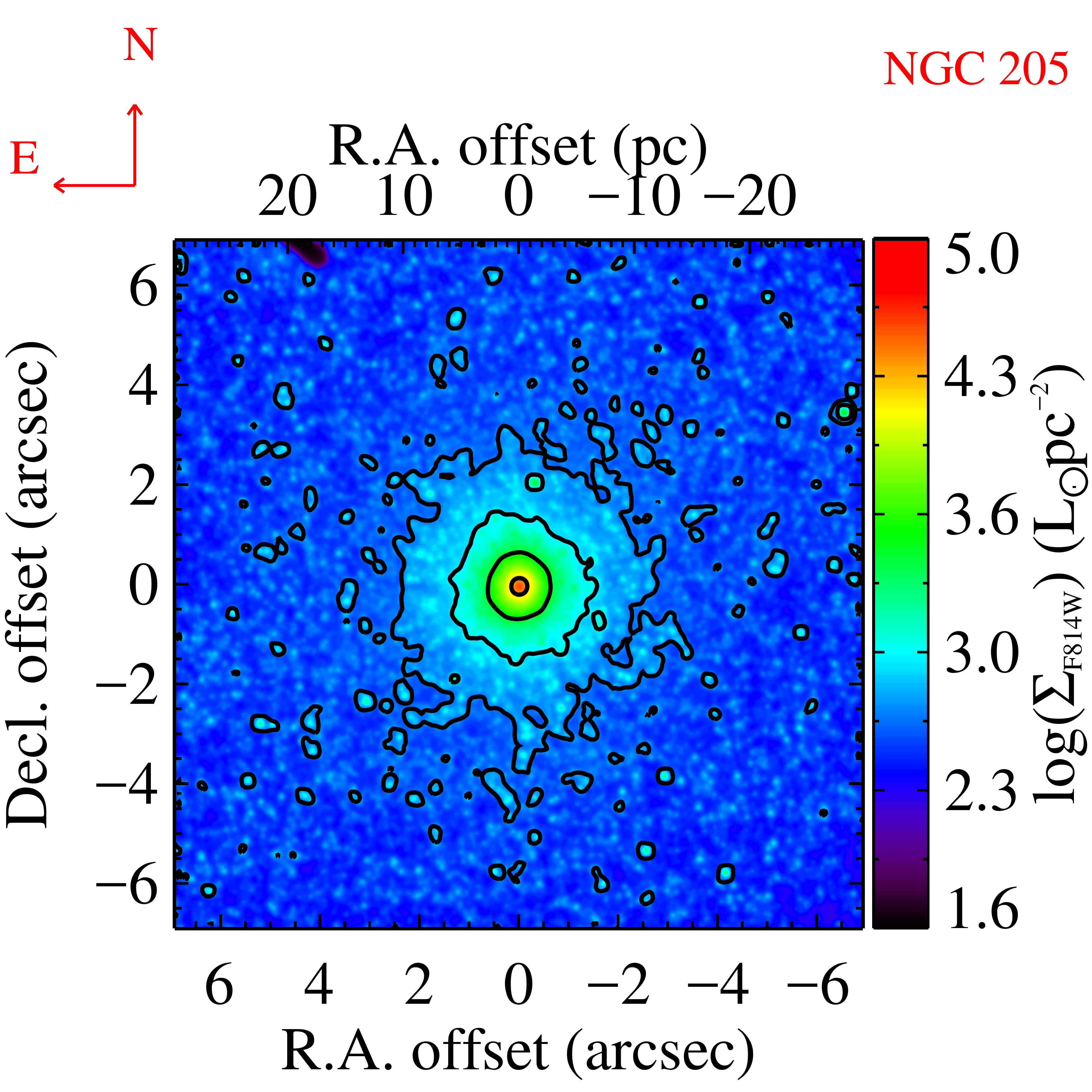}
    \includegraphics[scale=0.155]{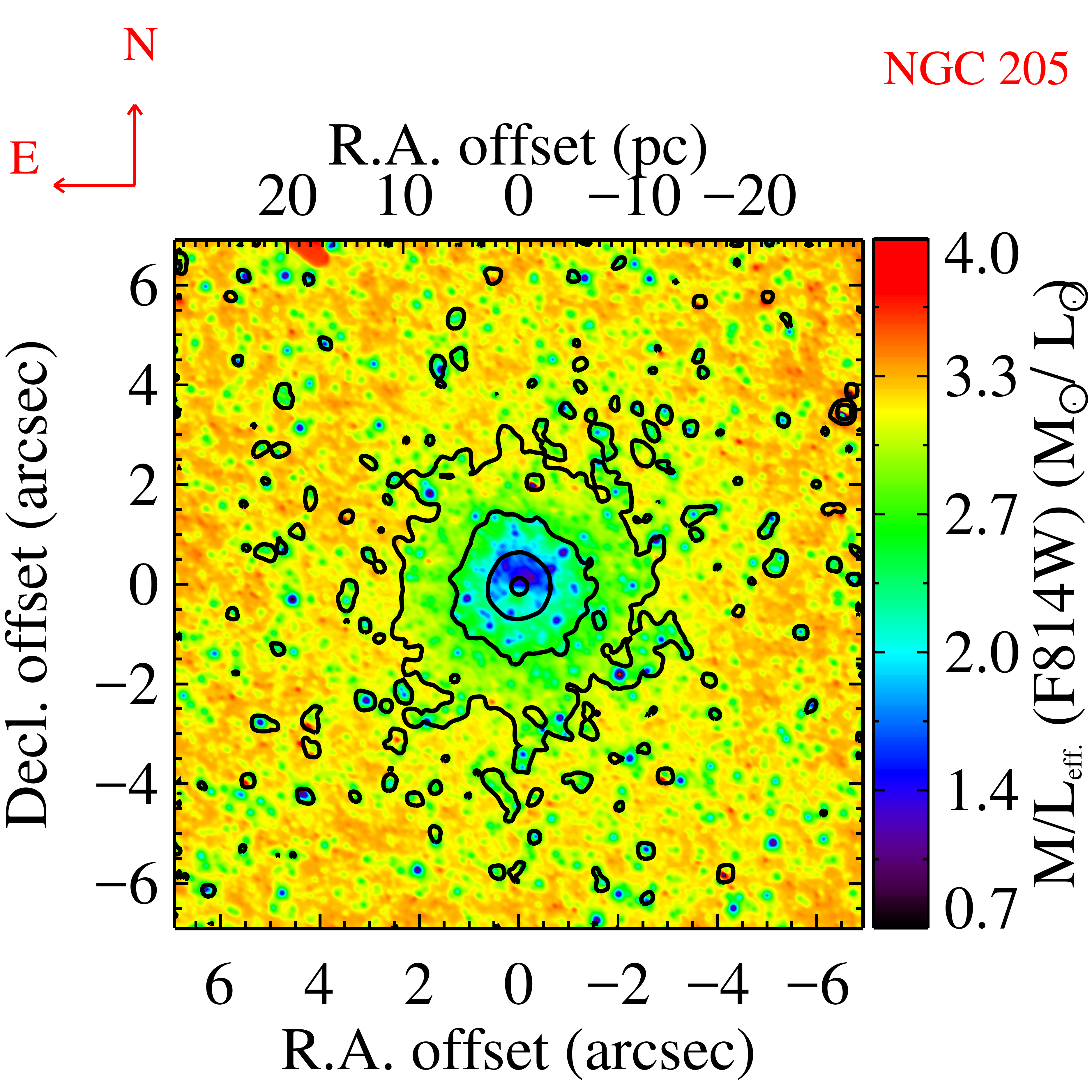}	
    \includegraphics[scale=0.155]{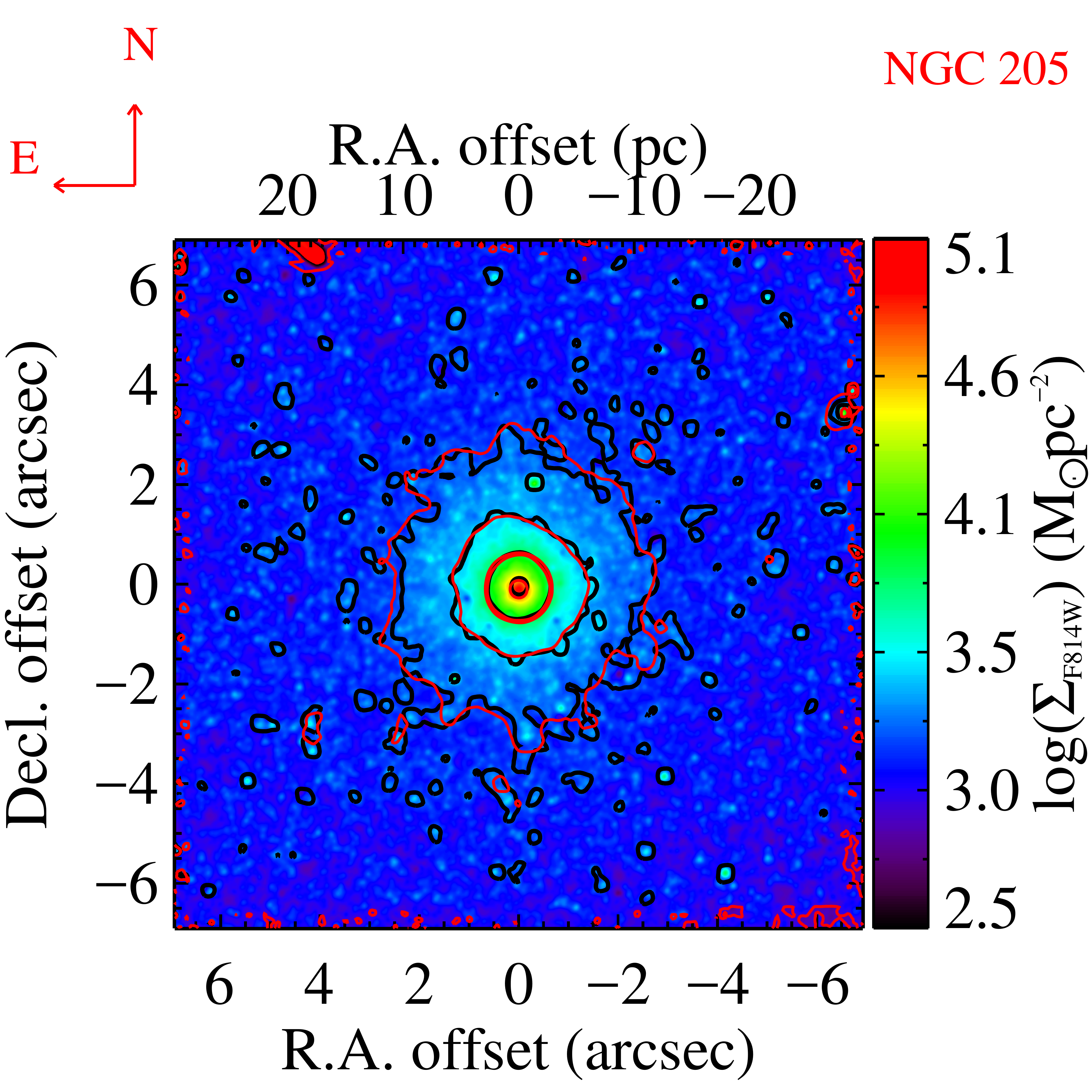} 
        
	\includegraphics[scale=0.155]{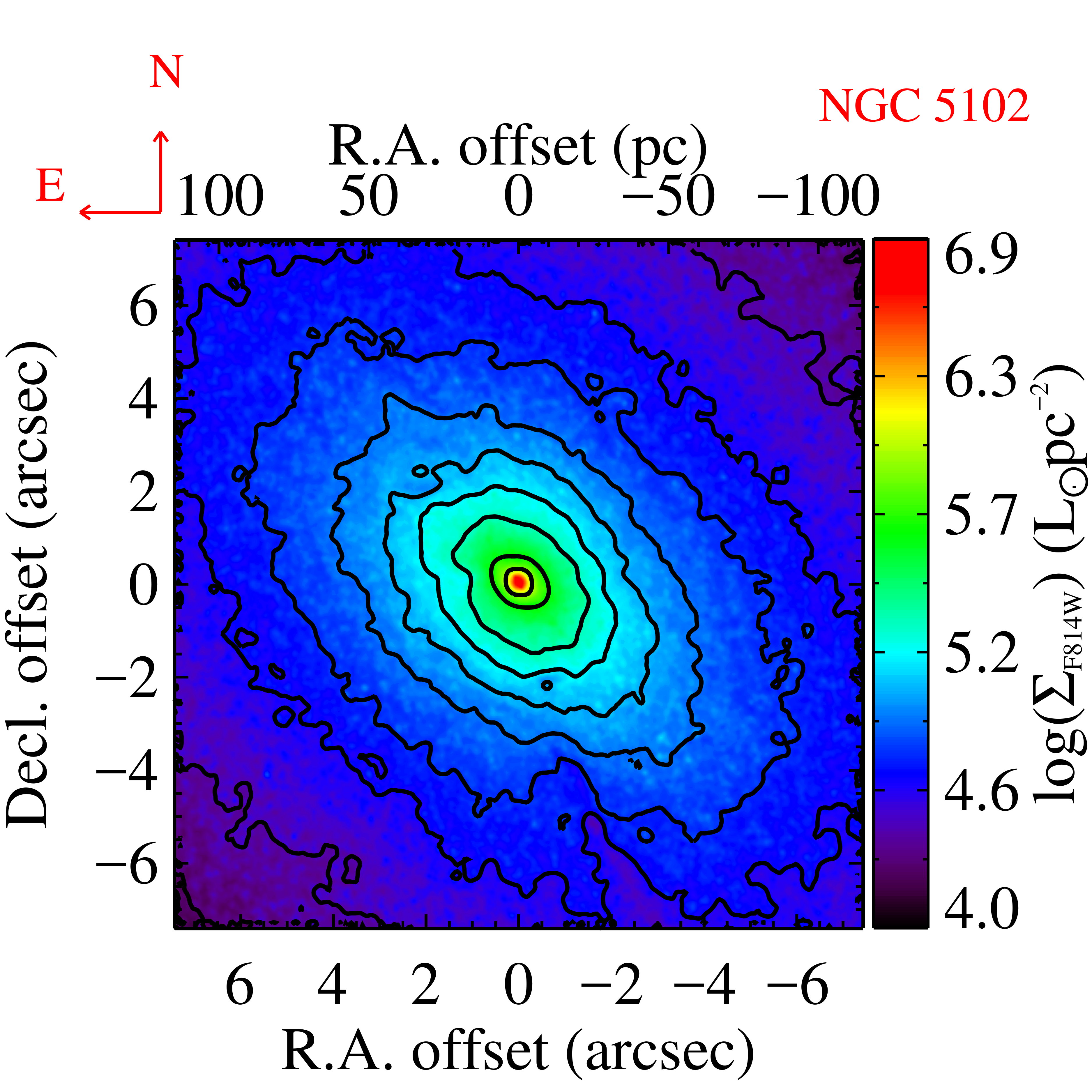}
    \includegraphics[scale=0.155]{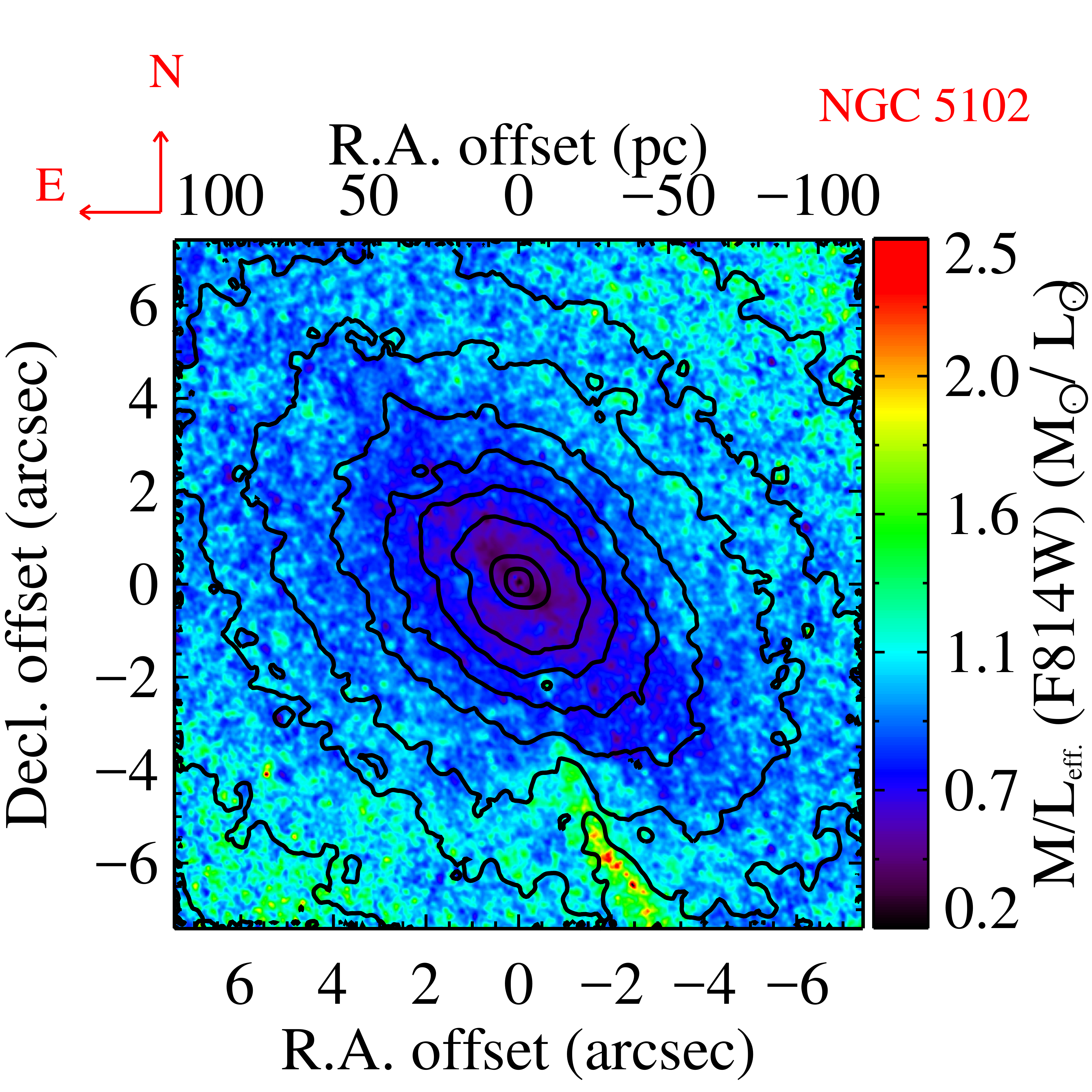}	
    \includegraphics[scale=0.155]{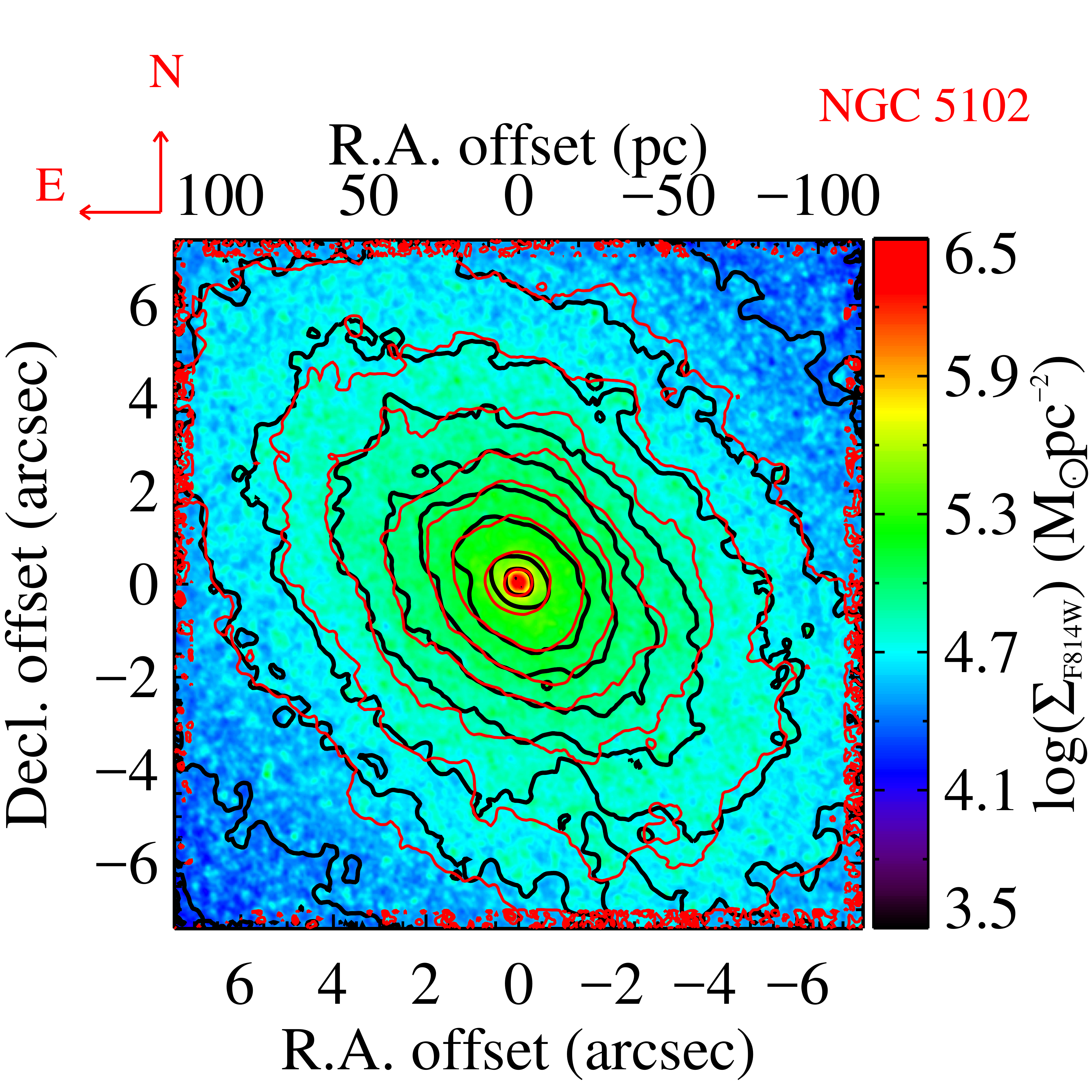} 
         
         
   \includegraphics[scale=0.155]{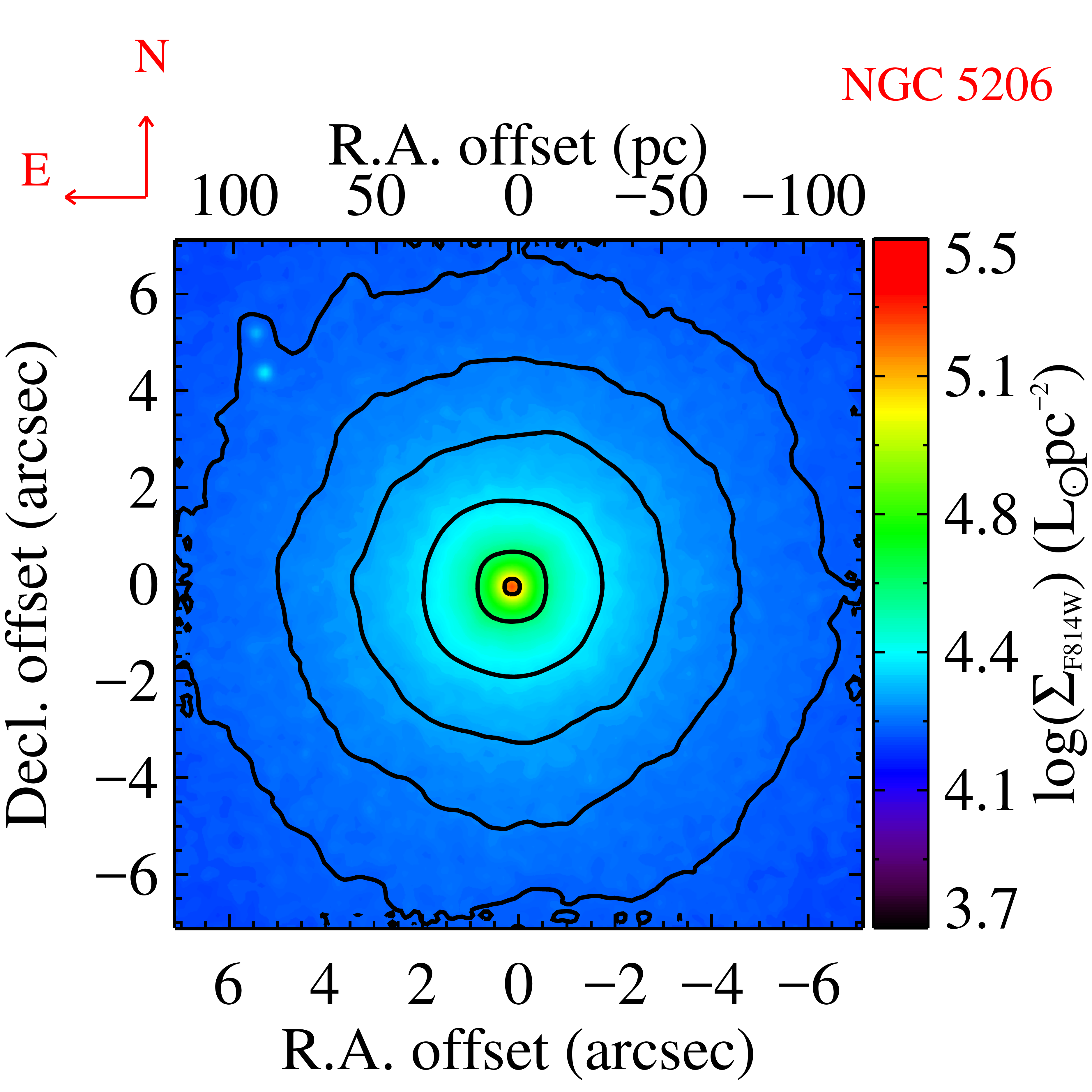}
   \includegraphics[scale=0.155]{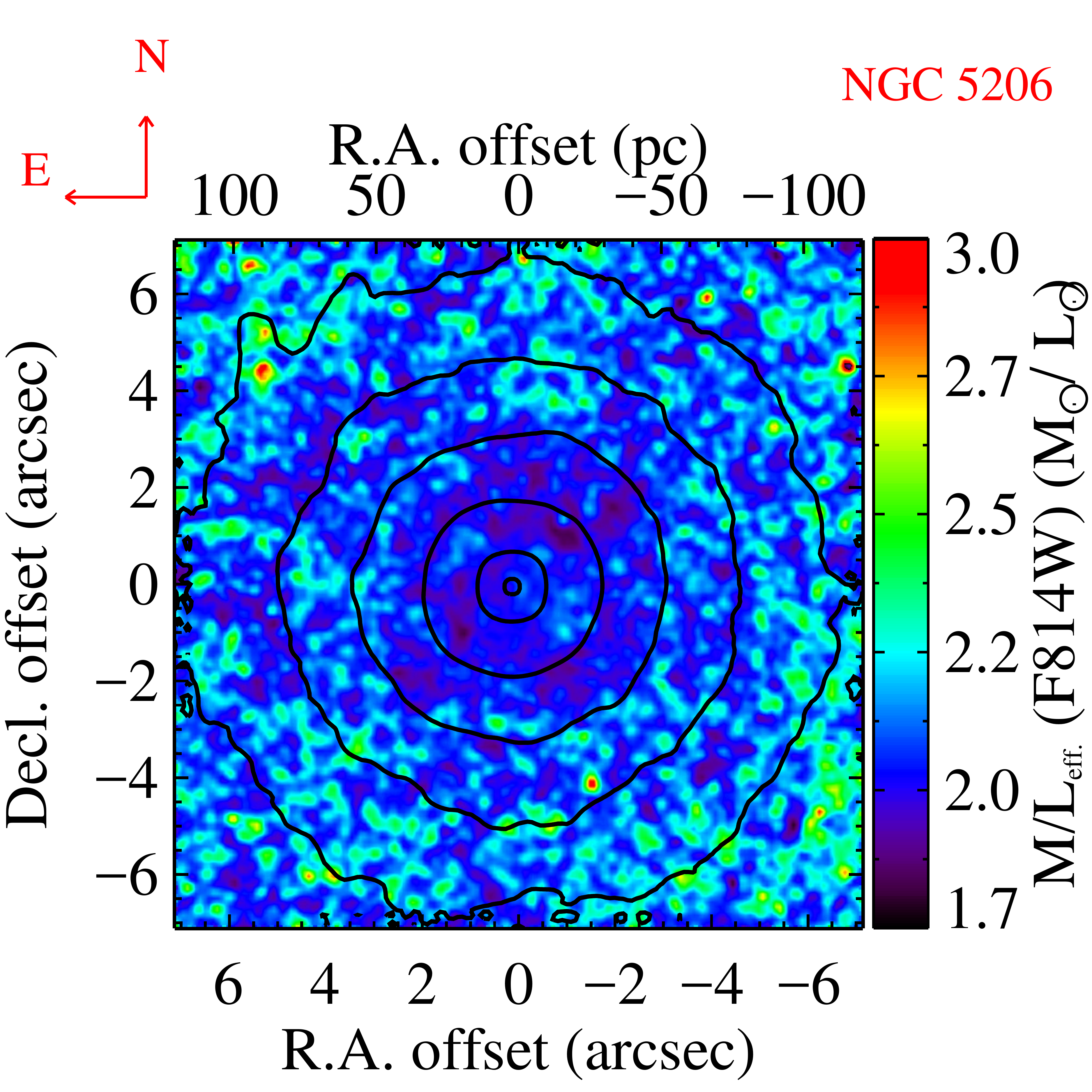}	
   \includegraphics[scale=0.155]{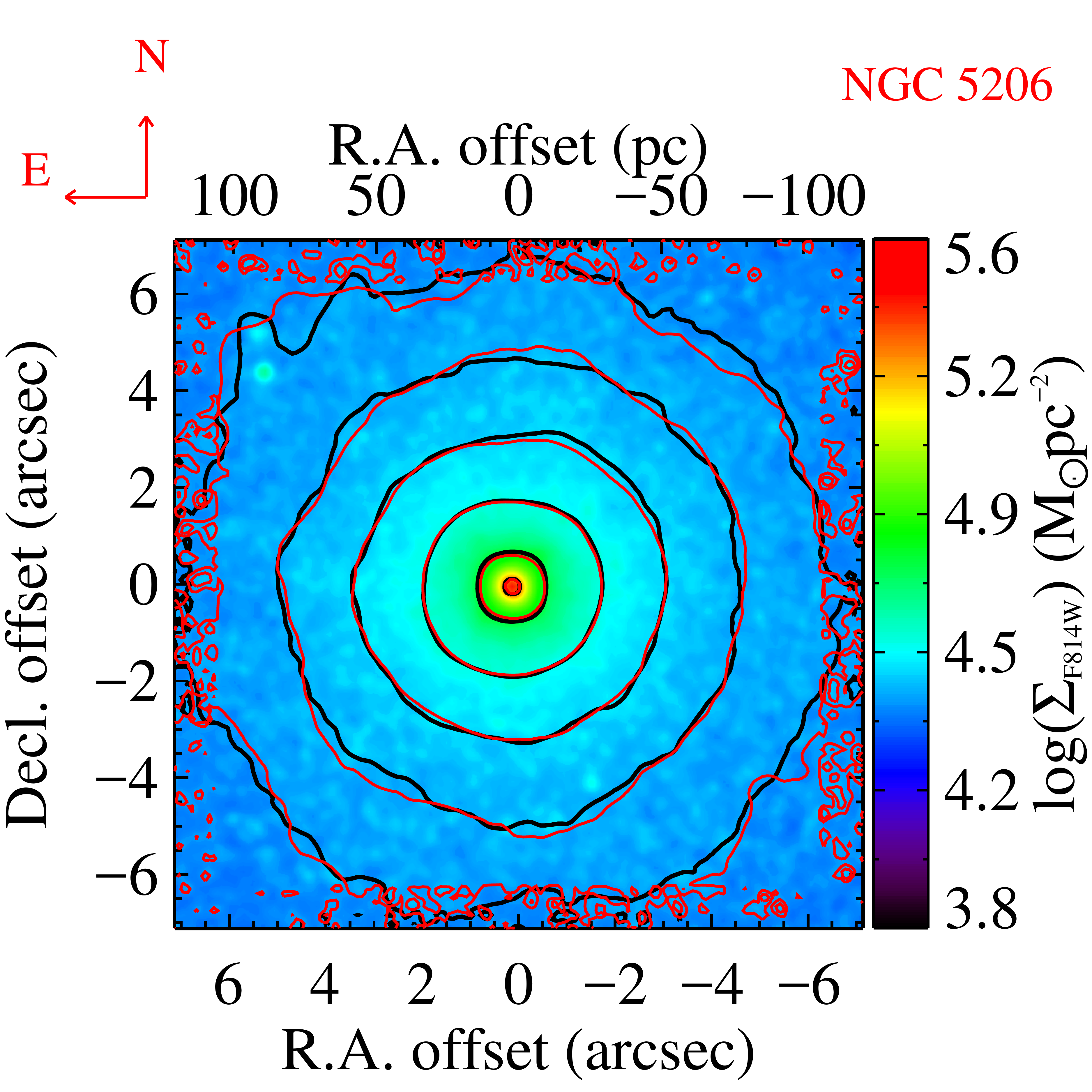}
\caption{\small{The luminosity density (left), \mleff~(middle), and mass density (right) maps built from the ACS HRC images for NGC 205 (top panels), WFC3 UVIS2 images for NGC 5102 (middle panels), and WFPC2 PC1 images for NGC 5206 (bottom panels). The black contours are the same as in Figure \ref{colormap} corresponding to each galaxy, while the red contours show the mass densities created using the F814W filter ($I$-band).  We use contours with log scale of (4.8, 4.2, 3.5, 3.1) \Msun~pc$^{-2}$ for NGC 205, (6.2, 5.9, 5.6, 5.4, 5.2, 5.1, 4.8, 4.6, 4.5) \Msun~pc$^{-2}$ for NGC 5102, and (5.4, 4.9, 4.6, 4.5, 4.4, 4.3) \Msun~pc$^{-2}$ for NGC 5206, respectively.}}  
\label{lum_m2l_mass_map}   
\end{figure*}

\subsection{Creating New Mass Maps and Mass Models}\label{ssec:maps}

We describe our new mass maps and models for the nuclei of NGC 205, NGC 5102, and NGC 5206 in this section. We first apply each galaxies' color--\mleff~relations to their color maps. This step yields their nuclear \mleff~maps in F814W, which are shown in the middle column panels of Figure \ref{lum_m2l_mass_map}. Specifically, these \mleff~maps were created using the fitted correlations of the F814W \mleff~versus F555W--F814W for NGC 205 (ACS/HRC) and NGC 5206 (WFPC2 PC1) or F336W--F814W for NGC 5102 (WFC3).

To obtain the mass maps for these nuclei, we simply multiply the \mleff~maps and the F814W luminosity maps pixel by pixel. These luminosity maps are plotted in the left panels of Figure \ref{lum_m2l_mass_map}. The right panels of Figure \ref{lum_m2l_mass_map} show the new mass maps of the nuclei of NGC 205, NGC 5102, and NGC 5206. The comparison of our new mass maps (red contours) with the F814W luminosity maps (black contours) at the same radii shows that the mass distributions in these nuclei are more symmetric than their corresponding F814W light emission profiles. This results in larger values of the axis ratio ($q=a/b$) of the mass multiple Gaussian expansions (MGEs, Table \ref{tab_mges}) than their luminosity based MGEs presented in Appendix B of N18. These results are similar to what we found in NGC 404 (N17, Figures 8 and 9). The astrophysical reason behind this is that the \mleff~maps account for both dust extinction and young stellar regions; because the dust and young stars are distributed less symmetrically, the mass profiles more axisymmetric than the light profiles.  
The increased symmetry in the mass maps indicates the success of our color--\ml~relations in modeling the true mass distribution of these galactic nuclei.

\begin{figure*}[!ht] \centering
\includegraphics[scale=0.6]{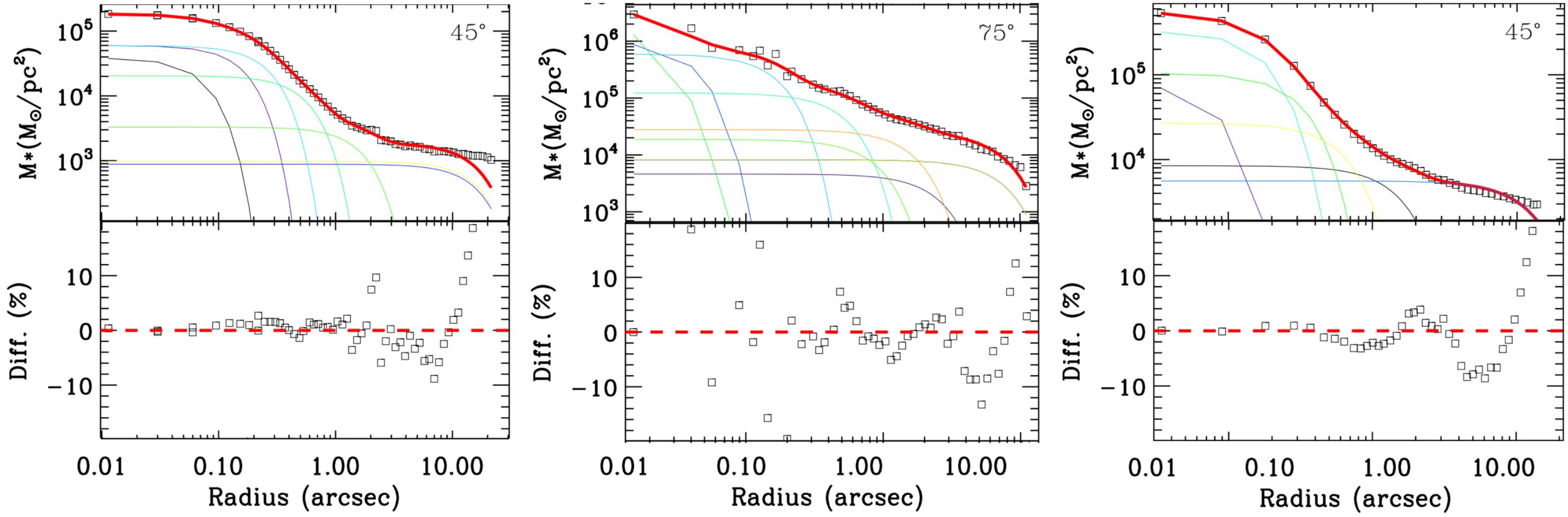}
\includegraphics[scale=0.27]{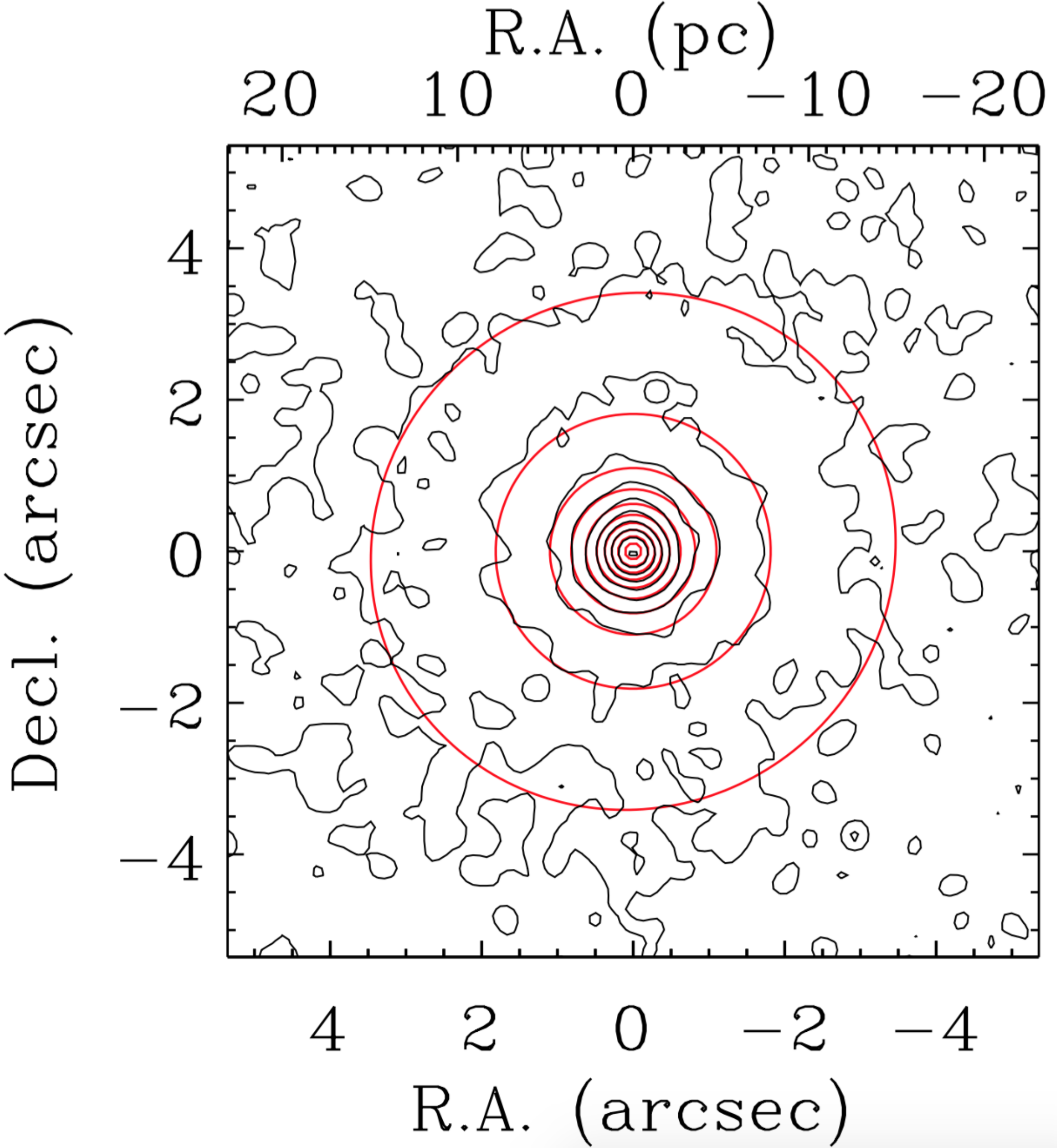}
\hspace{3mm}\includegraphics[scale=0.31]{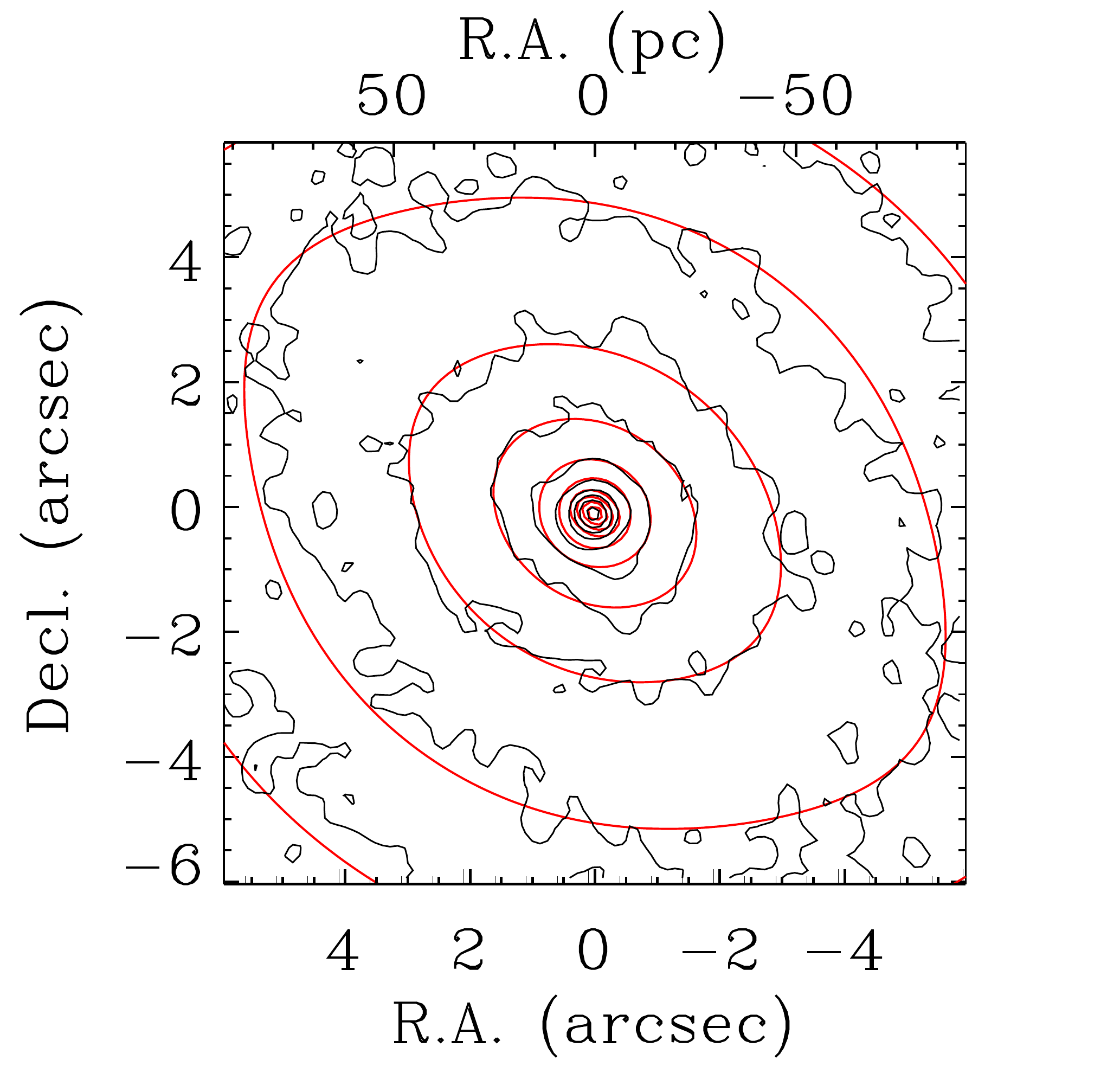}	
\hspace{-5mm}\includegraphics[scale=0.265]{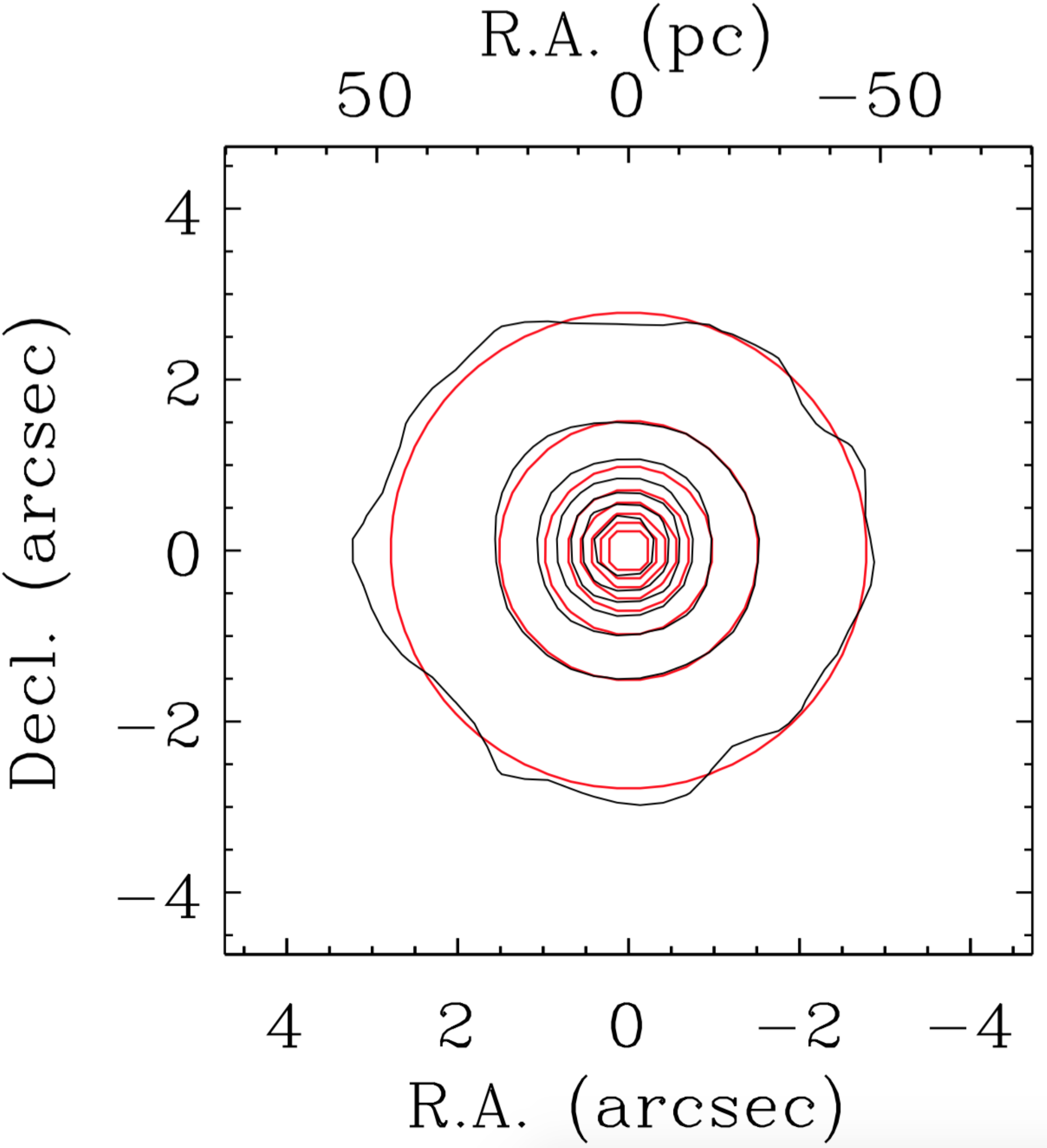} 
\caption{\small{Upper Panels: a comparison between the \hst~photometry of NGC 205 (left), NGC 5102 (middle), and NGC 5206 (right) in F814W (open squares) and their corresponding best-fit mass MGE models (red solid lines), which are the sum of multiple Gaussians (color thin lines). These best-fit models are projected along the sectors with their corresponding inclination angle in the top-right label. The fractional residuals \texttt{(Data-Model)/Data} are shown in the corresponding lower panels. Lower Panels: comparison between the F814W mass maps in Figure \ref{lum_m2l_mass_map} and their best-fit MGE models in the form of the contours of the mass surface densities for NGC 205 (left), NGC 5102 (middle), and NGC 5206 (right) within the central ($5\arcsec\times5\arcsec$). Black contours show the data, while the red contours show the models.}} 
\label{mge}   
\end{figure*}

We create the mass models for these nuclei by utilizing MGE models \citep{Emsellem94a, Cappellari02} to decompose the mass surface densities into individuals MGE components. We use the \texttt{mge\_fit\_sectors IDL} code\footnote{version 4.14, http:purl.org/cappellari/software} \citep{Cappellari02} to fit the mass maps and deconvolve the effects of the PSF.  We first parametrize the PSF using MGEs and then use them to fit the 2D mass map directly. These MGEs PSF models are tabulated in Table~\ref{tab_psfmges} in Appendix \ref{sec:mgepsf}. Here, we use the ACS/HRC, WFC3, and WFPC2 PC F814W PSFs for the mass maps of NGC 205, NGC 5102, and NGC 5206 because their mass maps are the weighted versions of their F814W luminosity maps (N17).  We show the mass surface densities profiles of these galactic nuclei (data (open black squares) vs. models (red solid lines)) in the top-upper panels of Figure~\ref{mge}, while the fractional residuals, which indicates the agreement between our best-fit models to the data along the radii are shown in the lower panels. The 2D surface mass densities in the F814W bands of the three galaxies (black contours) are also plotted  with their MGE models (red contours) in the three bottom panels of Figure \ref{mge}. The differences between the best-fit models and the data are $<$15\% across the mass maps' FOV ($\sim$10$\arcsec$). All three panels show the agreements of the mass surface density maps and their models at the same radii and contours levels to highlight the consistency between the data and models. 

Although our mass maps and mass models extend out to the radii of $\sim$10$\farcs$0, this limited range has minimal impact on our dynamical models, as our kinematics only extend out to $\sim$1$\farcs$0.  In N17, we found that our dynamical models were unchanged when the mass map was larger than 6$\farcs$0.  We list the parameters of these mass MGE models in Table \ref{tab_mges}.

The errors in the color--\mleff~relations could affect the central dynamical BH mass estimates due to the uncertainties they leave on their mass map models. The discussions of this effect can be found in N17. Here, we assess these uncertainties will have on our dynamical models by creating the mass maps for other filters, such as F555W for NGC 205 and NGC 5206, and F336W and F547M for NGC 5102. For NGC 5102 due to the availability of its F547M data, we examine its other color--\mleff~relations with different color bases (i.e., F336W--F547M and F547M--F814W).  Another impact that these color--\mleff~relations may have on our mass maps are the uncertainties in their best-fit slopes. We also create mass maps using the 1$\sigma$ uncertainties on the color--\mleff~relations. We will discuss the effects these uncertainties have on our JAM dynamical models in Section \ref{ssec:masserror}, and their results are listed in Table \ref{full_fittable} of Appendix \ref{sec:fulltable}.

\begin{table}
\caption{New Mass MGE Models}
\centering
\begin{tabular}{cccc}
\hline\hline   
$j$  &$\log$(Mass Density)&$\sigma$&  $a/b$ \\
       &        (\Mppc) &     (arcsec)     &            \\
 (1) &      (2)           &       (3)            &    (4)    \\
\hline
      &                      &{\bf NGC 205}&             \\
\hline
1    & 4.598       &     0.061         &   0.837 \\
2    & 4.785       &     0.120         &   0.999 \\
3    & 4.781       &     0.198         &   0.994 \\
4    & 4.316       &     0.412         &   0.987 \\
5    & 3.524       &     1.220         &   0.984 \\
6    & 2.951       &     13.11         &   0.850 \\
7    & 2.996       &     13.11         &   0.851 \\
\hline
      &                     &{\bf NGC 5102}&           \\
\hline
1   & 5.848   &    0.015         &  0.997 \\
2   & 6.606   &    0.033         &  0.886 \\
3   & 5.448   &    0.101         &  0.661 \\
4   & 4.706   &    0.362         &  0.997 \\
5   & 4.026   &    0.865         &  0.695 \\
6   & 3.589   &    1.272         &  0.884 \\
7   & 3.473   &    5.922         &  0.887 \\
8   & 3.038   &    6.132         &  0.611 \\
9   & 2.627   &    7.945         &  0.884 \\
10 & 2.520   &    15.93         &  0.884 \\                                        
\hline                    
     &                      &{\bf NGC  5206}&       \\
\hline       
1   &  5.107   &      0.061         &  0.837 \\
2   &  5.689   &      0.135         &  0.996 \\
3   &  5.182   &      0.236         &  0.995 \\
4   &  4.587   &      0.472         &  0.999 \\
5   &  4.083   &      1.148         &  0.998 \\
6   &  3.797   &      9.389         &  0.999 \\
7   &  2.970   &      20.45         &  0.850 \\
8   &  2.367   &      32.390       &  0.850 \\
\hline  
\end{tabular}
\tablenotemark{}  
\tablecomments{\small{MGE models used in JAM model fits (see Section~\ref{ssec:jeans}).  Column 1: Gaussian component number. Column 2:    the MGE models which represented for the mass models of the galaxies.  Column 3: the Gaussian width along the major axis.  Column 4: the axial ratios.}}
\label{tab_mges}
\end{table}

\begin{figure}
\centering
	\includegraphics[scale=0.14]{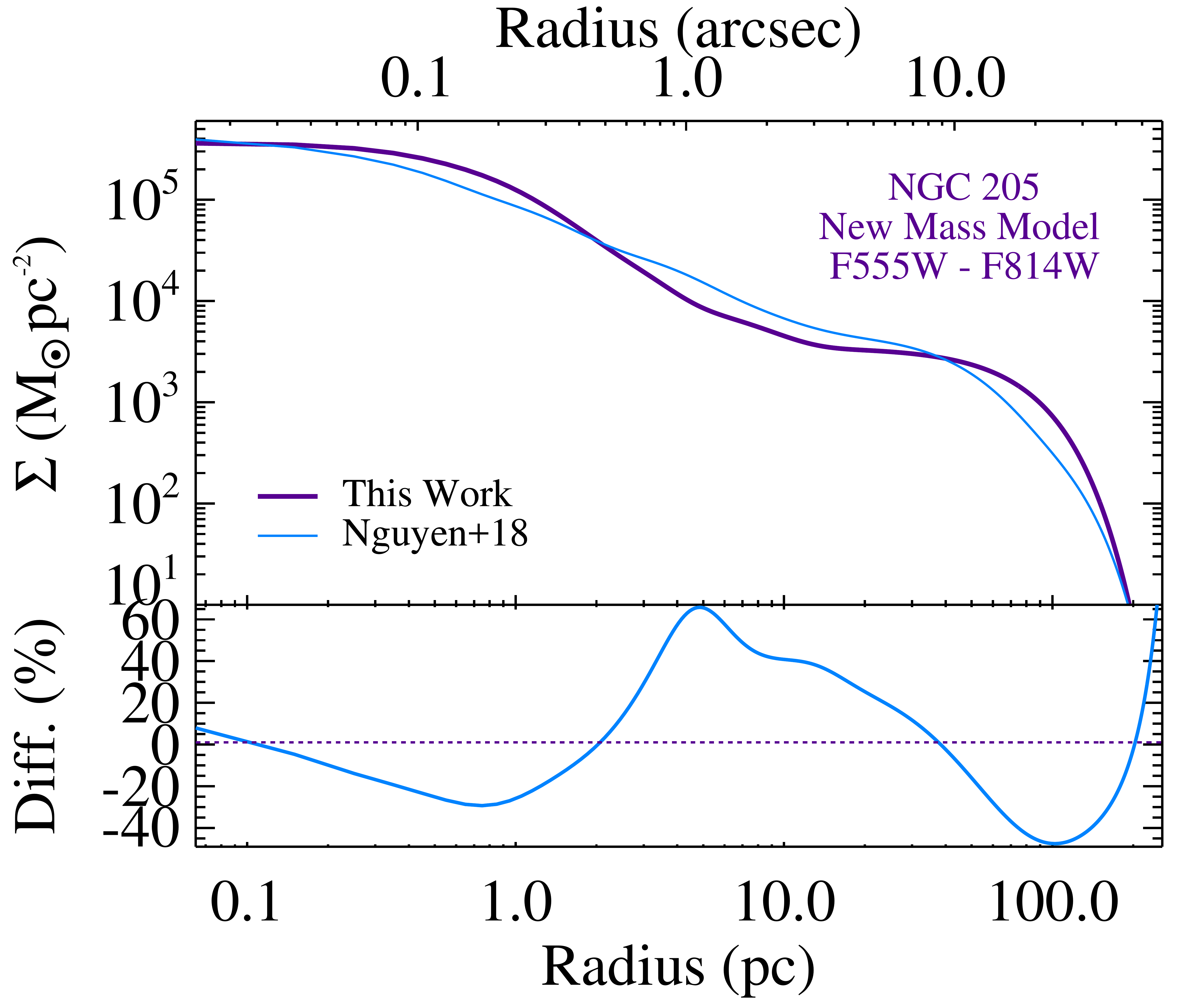}
	\includegraphics[scale=0.14]{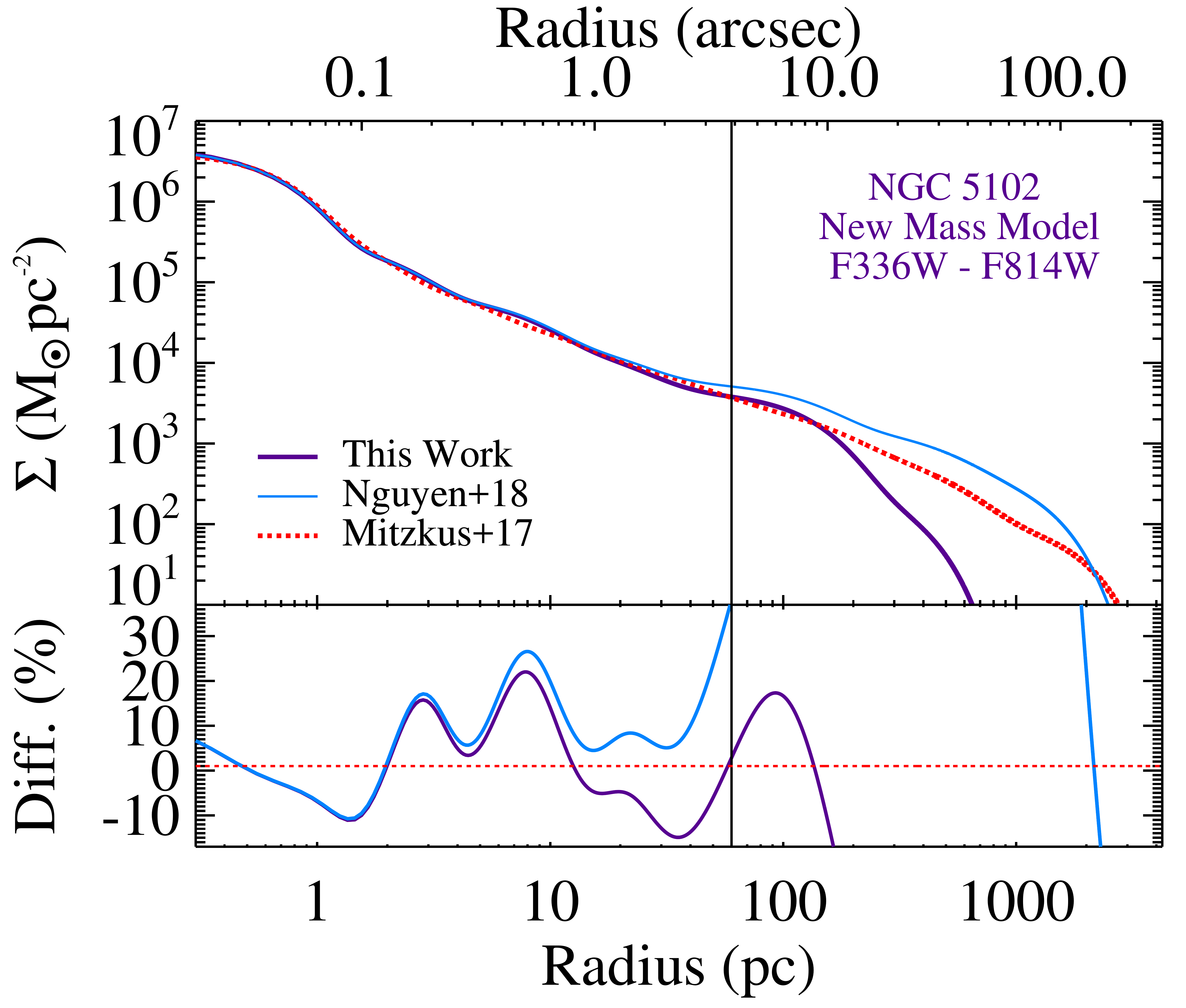}
	\includegraphics[scale=0.14]{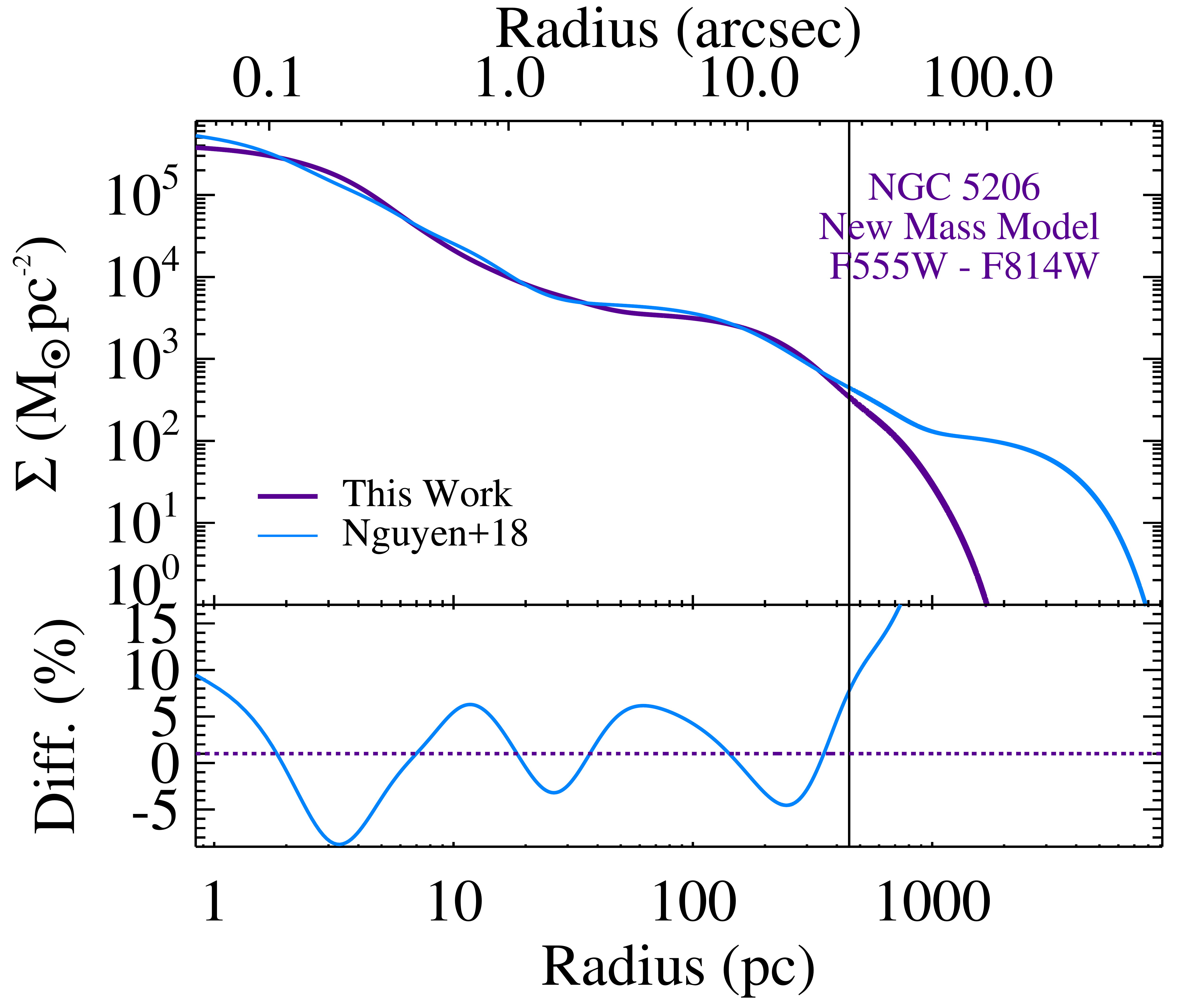}
\caption{\small{Upper part of the panels: comparison of MGE models of the NGC 205 (top), NGC 5102 (middle), and NGC 5206 (bottom) mass density models (purple lines) constructed from our new nuclear mass maps (Figure \ref{lum_m2l_mass_map}) to the MGE mass models from \citet{Mitzkus17} (red line) and N18 (blue lines). Lower part of the panels: relative comparison of three different mass profiles. We plot their fractional differences relatively to our mass density models (purple dashed line) for NGC 205 and NGC 5206. For NGC 5102, we plot our updated mass model in this work and N18 relative to the mass model of \citet{Mitzkus17} (red dashed line). The black vertical lines indicate the radii at which our new mass map models (which only cover the central regions) and their previous version start to have large discrepancies; we highlight, though, that these differences have little effect on our dynamical models of the central regions.}}  
\label{compare_mass_n5102}   
\end{figure}

We compare our new mass models of these nuclei to their previous works including N18 for all three galaxies.  We also include a comparison to the mass model with variable \ml~of \citet{Mitzkus17}, which is based on their modeling of Multi Unit Spectroscopic Explorer integral field unit (MUSE IFU) spectroscopic data for NGC 5102.  We plot these new MGE mass surface density models of NGC 205 (top), NGC 5102 (middle), and NGC 5206 (bottom) and compare them to the MGE models created in N18 (using the default models based on the \citet{Roediger15} color--\ml~relation) and \citet{Mitzkus17} in Figure \ref{compare_mass_n5102}. Our new mass model of NGC 205 predicts roughly $\sim$20\% more mass than that prediction of N18 at the radius less than 0$\farcs$8 (3.4 pc); however, from this radius out to 10$\farcs$0 (43 pc), this ratio is reversed by 38\%. In NGC 5102, our new mass model calculation is very similar to the estimates of N18 and \citet{Mitzkus17} $<$20\% within 4$\farcs$0 (64 pc); outside this radius these mass models start to diverge dramatically. Out to the maximum radii of our spectroscopic population fitting ($\sim$2$\farcs$0), our model is in good agreement with the \citet{Mitzkus17} model, but falls below the N18 model. For NGC 5206, our new mass model produces more/less mass ($<$18\% and $>\pm$8\%) than the mass model of N18 across the nucleus until 28$\farcs$0 (476 pc). We should note that these differences of our new mass models compare to that of N18 will have large impacts on our dynamical mass estimates for BH masses in Section \ref{sec:steldyn}, especially for the BH in NGC~205 since its BH is small (upper limit $<$$7\times10^4$\Msun, N17). We found in N17 the best-fit BH mass of NGC 404 ($<$$1.5\times10^5$\Msun) was vary significantly if the mass model show variations within $6\arcsec$; a larger fitting area made the BH mass changed negligibly small. We will demonstrate how do the new mass model of NGC 205 improve its BH mass over the N18 mass model in Section~\ref{ssec:n205}, and the results can be seen clearly in Figures \ref{posterior_n205}, \ref{rms1d_mcut_oldnewmass}, \ref{rms2d_oldnewmass}, \ref{chi2_oldnewmass},  \ref{n205_bhsoi}, and \ref{posterior_n205_n18mges}.

\section{Dependence of the Color--M/L Relations on the Nuclei's Stellar Populations}\label{sec:relationinsight} 

Before considering the impact of our new color--\ml~relations on our BH mass estimates, we consider the slopes and normalizations of these relations to previous relations of \citet{Bell01, Bell03, Roediger15}. We note that the stellar populations and color--\mleff~relation in the nucleus of NGC~404 are presented in detail in N17. 

\begin{figure*}  
\centering
	\includegraphics[scale=0.07]{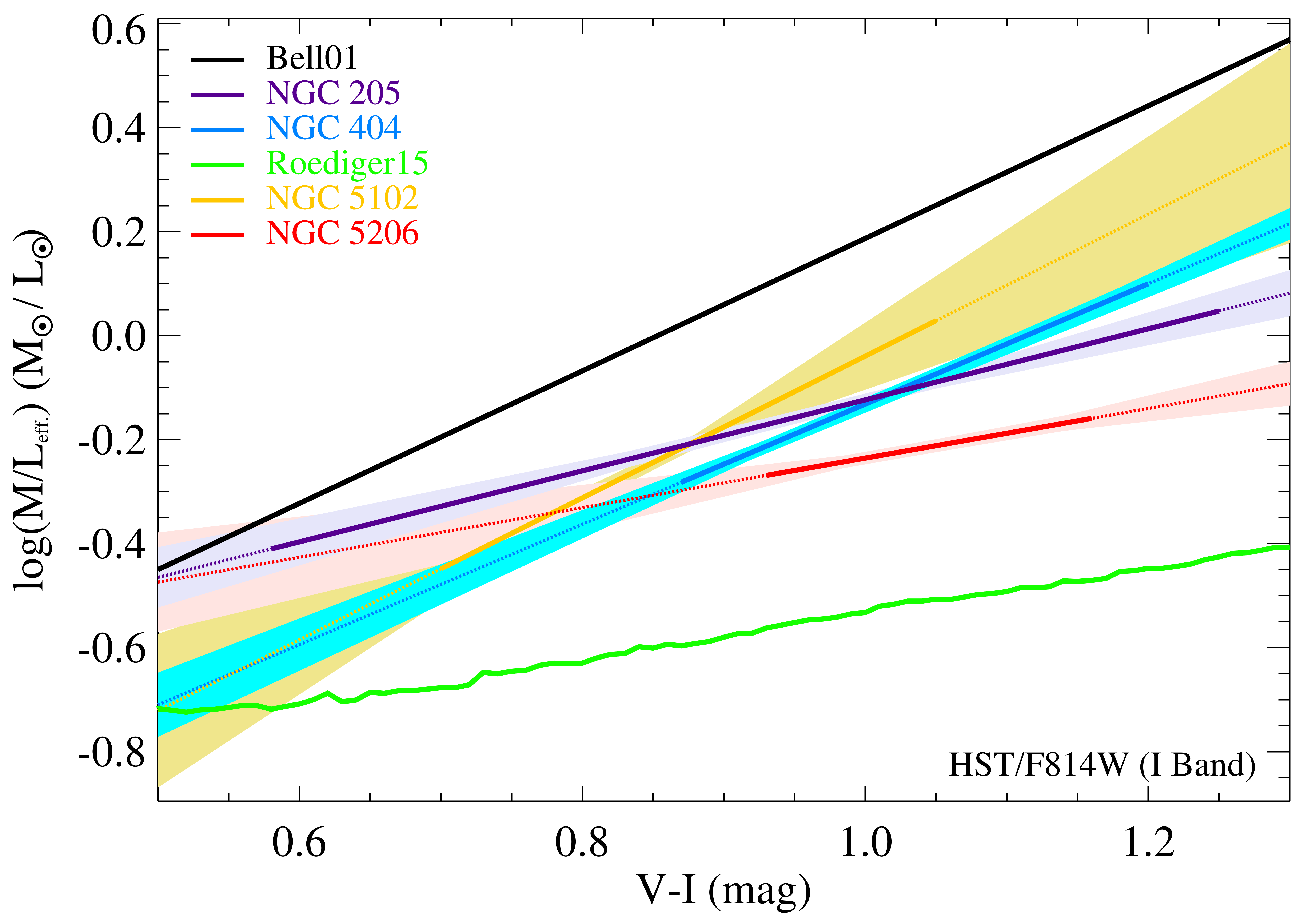}
	\includegraphics[scale=0.07]{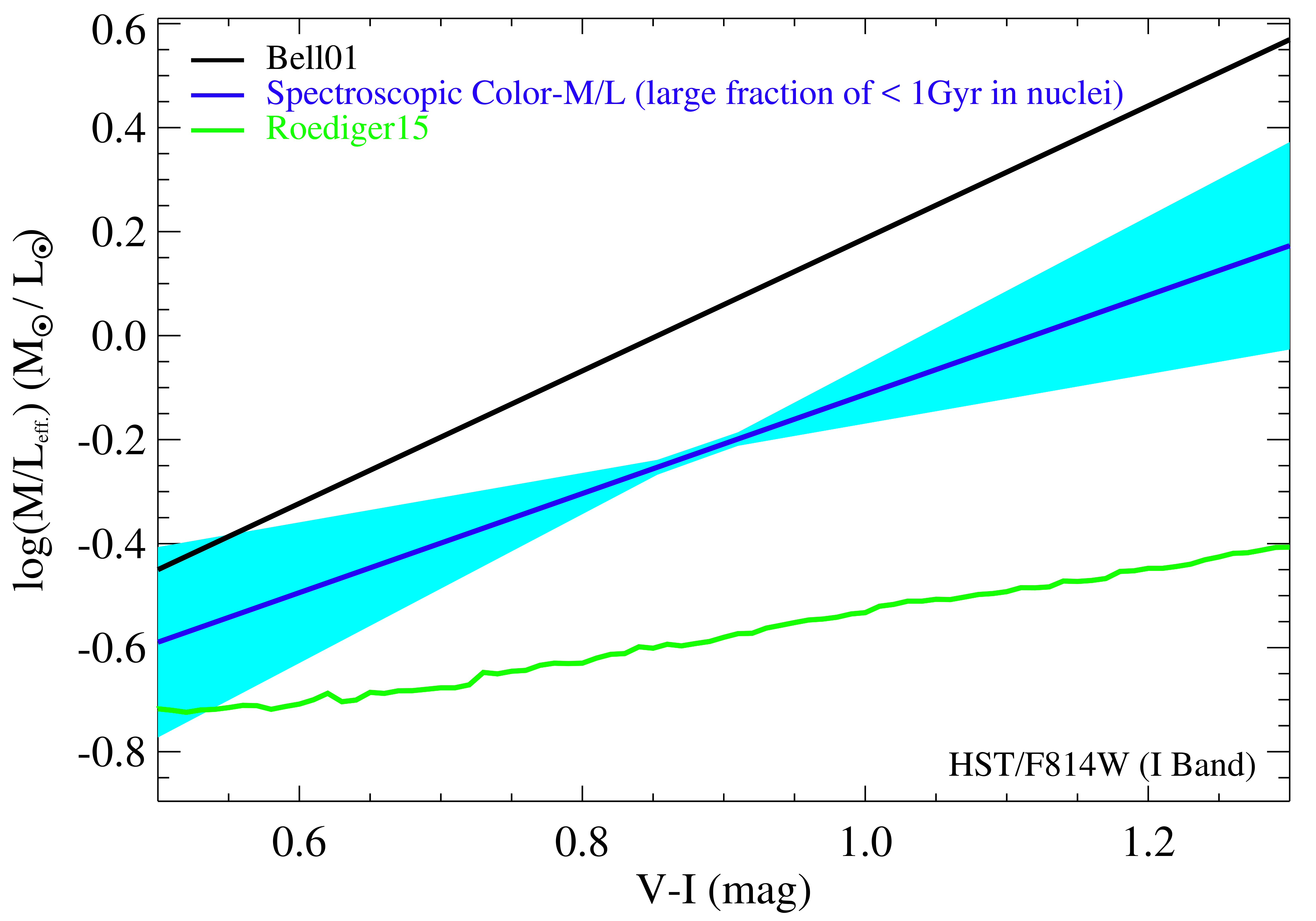}
\caption{\small{Left panel: Comparison of our color--\ml~correlations in $I$-band based on the specific stellar populations in the nuclei of NGC 404 (N17), NGC 205, NGC 5102, and NGC 5206 (this work) to the color--\mleff~correlations available in literature, including \citet{Bell01} and \citet{Roediger15}. Specific color--\mleff~relations are reported in the legend. The same color region around each best-fit line are their specific $1\sigma$ uncertainty as described in Figure \ref{color_m2l_relation}. Here, the real color ranges for the galaxies are plotted with thick solid lines and their extrapolations toward the smaller and  higher colors are shown with dotted lines. Right panel: Comparison of our new spectroscopic color--\ml~relation derived from the combined data of NGC 205, NGC 404, and NGC 5102; these nuclei host large factions of young stellar populations ($<$1 Gyr).}}
\label{all_color_m2l_relation}   
\end{figure*}

We compare our best-fit color--\ml~relations of these galactic nuclei to those of  \citet{Bell01} and \citet{Roediger15}, which are plotted as the black and green solid lines in Figures \ref{color_m2l_relation} and \ref{all_color_m2l_relation}. To consistently compare all the relations, we have to transform our models into predictions of \mleff~in $I$ band vs.~$V-I$~color. We use the $F547M$ (NGC~404 and NGC~5102) or $F555W$ (NGC~205 and NGC~5206) as our $V$ band filter and the $F814W$ filter as our $I$ band filter.  We use filter transformations from \citep{Sirianni05}, assuming the WFPC2 and WFC3 filters have similar transformations.

The slope of the color--\mleff~relations can depend strongly on the stellar population properties. For this reason, we briefly summarize what we know about the stellar content of the nuclei of these four galaxies.
\begin{itemize}
\setlength\itemsep{0em}
	\item {\it NGC 205} contains in its nucleus two distinctive stellar populations including young bright blue stars with age $<$0.1 Gyr, metallicities of [Fe/H] $\sim-0.5$ \citep{Monaco09} at the center ($r<0\farcs6$) and older stars with age $\sim$1--5 Gyr \citep{Cappellari99, Davidge03}
	\item {\it NGC 5102} harbors young stellar populations with ages of (0.3--0.7) Gyr with most of its stars formed less than 1 Gyr ago  with a very bursty and stochastic SFH (\citet{Davidge15, Mitzkus17, Kacharov18}. The mean age of these stellar populations are increasing gradually with radius  \citep{Kraft05, Davidge08, Davidge15, Mitzkus17}.
	\item {\it NGC 5206}'s center has a wide range of stellar populations with ages ranging from 1 to 10 Gyr, and a continuous SFH with gradual metallicity enrichment \citep{Kacharov18}; no stars younger than 1 Gyr appear to be present.
	\item  {\it NGC 404}'s nucleus is dominated by a large fraction ($\sim$70\%) of 1 Gyr population within the radius of $\sim$0$\farcs$5, with an additional contribution of even younger populations (N17). We note that the north-east side of the nucleus of NGC 404 has significant dust extinction, which has a large affect on the steep slope of its color--\ml~relation (N17), instead of the relation being largely dominated by stellar population variations as seen in the three nuclei studied here.
\end{itemize} 

Based on their stellar contents and color--\mleff~relations in Figure~\ref{all_color_m2l_relation}, we can give some guidance on understanding the color--\ml~relations in low-mass galaxies that are lacking spectroscopic data for measuring \mleff~accurately.  The shallowest slope is seen in NGC~5206, which lacks a significant young ($<$1~Gyr) population; for this galaxy, the \citet{Roediger15} relation has a very similar slope.  For the purposes of dynamical modeling, this will result in similar results apart from a global scaling in \ml.  However, the presence of younger populations appears to steepen the  $V-I$ vs. \mleff~slopes. We combine the relations derived in the three nuclei with $<$1~Gyr populations to create the empirical color-\mleff~relation that is shown in the right panel of Fig.~\ref{all_color_m2l_relation}.  This can be used to model the mass distributions of similar nuclei without resolved spectroscopy in the future. The details of these relations are given in Table \ref{tablecolorm2l}.

\begin{table}[h]
\caption{Coefficients of color--\mleff~relation fits in $I$-band with $V-I$ color base.}
\centering
\begin{tabular}{cccc}
\hline\hline   
   Galaxy   &  $a$   &    $b$     & references\\
        (1)     &   (2)   &       (3)    &  (4)\\
\hline
NGC 205  & 0.683 & $-$0.806 & This work\\
NGC 404  & 1.157 & $-$1.288 &  N17        \\
NGC 5102& 1.364 & $-$1.403 & This work\\
NGC 5206& 0.477 & $-$0.712 & This work\\
\hline
Spectroscopic Color--\ml$^\star$&1.067&$-$1.182&This work\\
\hline
\end{tabular}
\tablenotemark{}  
\tablecomments{\small{The color--\mleff~relation are calculated in the form: $\log$(\mleff) = $a\times(V-I)+b$. Column 1: The galaxy's name.  Columns 2 and 3: The slope and intercept of the linear color--\mleff~relation in log scale. Column 4: The references where the relations are determined.  $^\star$This relationship combines NGC~205, NGC~404, and NGC~5102, and is appropriate for objects with significant young ($<$1 Gyr) stellar populations}}.
\label{tablecolorm2l}
\end{table}

\section{Stellar Dynamical Modeling}\label{sec:steldyn}

\subsection{Jeans Anisotropic Models}\label{ssec:jeans}

\begin{figure*}
\centering\vspace{-10mm}
	\includegraphics[scale=0.75]{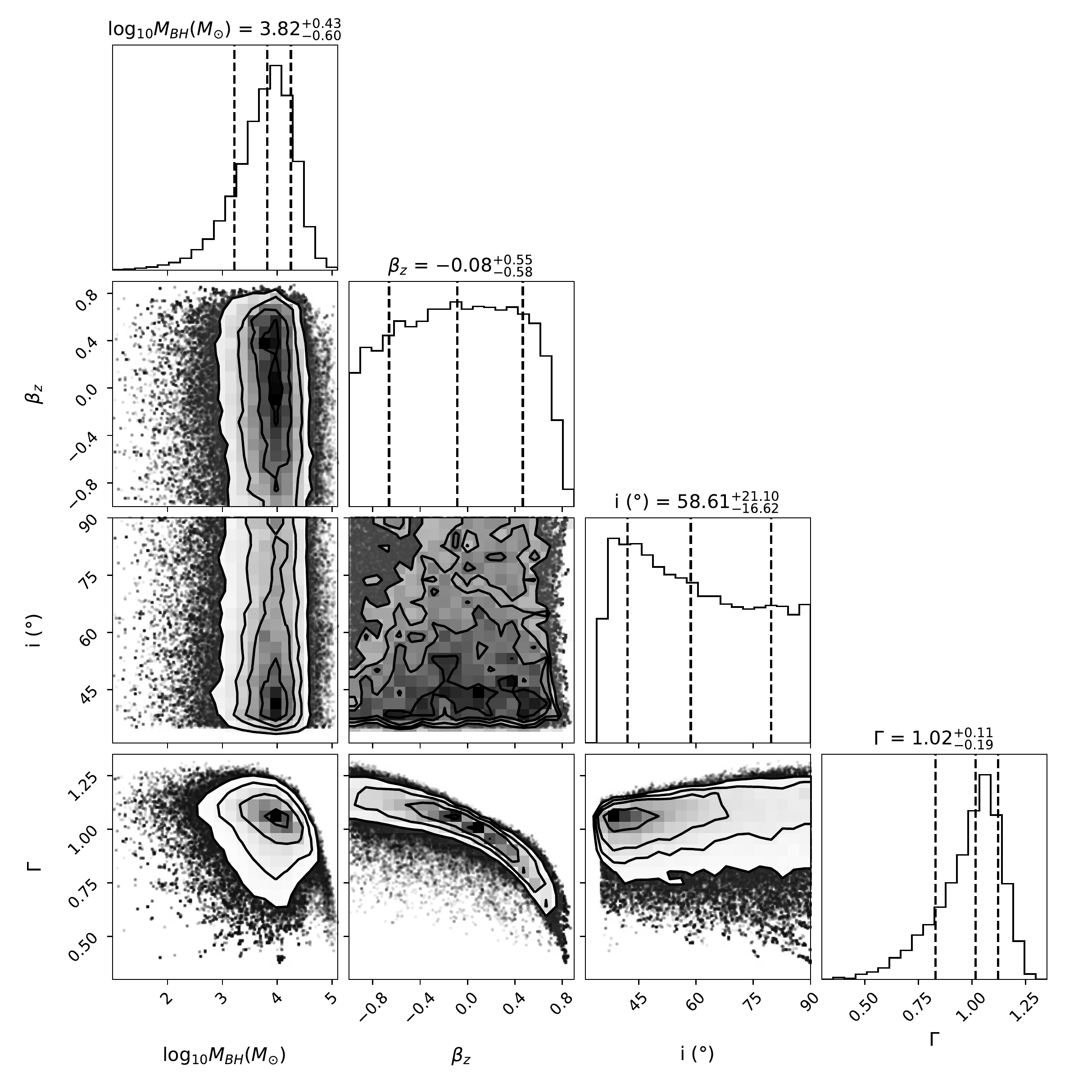}
	\vspace{-7mm}
        \caption{\small{The MCMC posterior distribution of the parameter space that we explored with the JAM dynamical models for the central BH in NGC 205. Each panel shows the projected 2D distributions for a pair of parameters after marginalizing over the other two. See Table \ref{fittable} for a quantitative description of the range of the priors and the likelihoods of all fitting parameters and their best fits. In the top panel of each column we report a 1D histogram distribution of that parameters with their best fits and $1\sigma$ errors. Our model explores the black hole mass \Mbh, anisotropy $\beta_z$, mass scaling factor $\Gamma_{\rm F814W}$, and inclination $i$.}}  
\label{posterior_n205}   
\end{figure*}

\begin{figure}[!h]  
\centering
   	\includegraphics[scale=0.17]{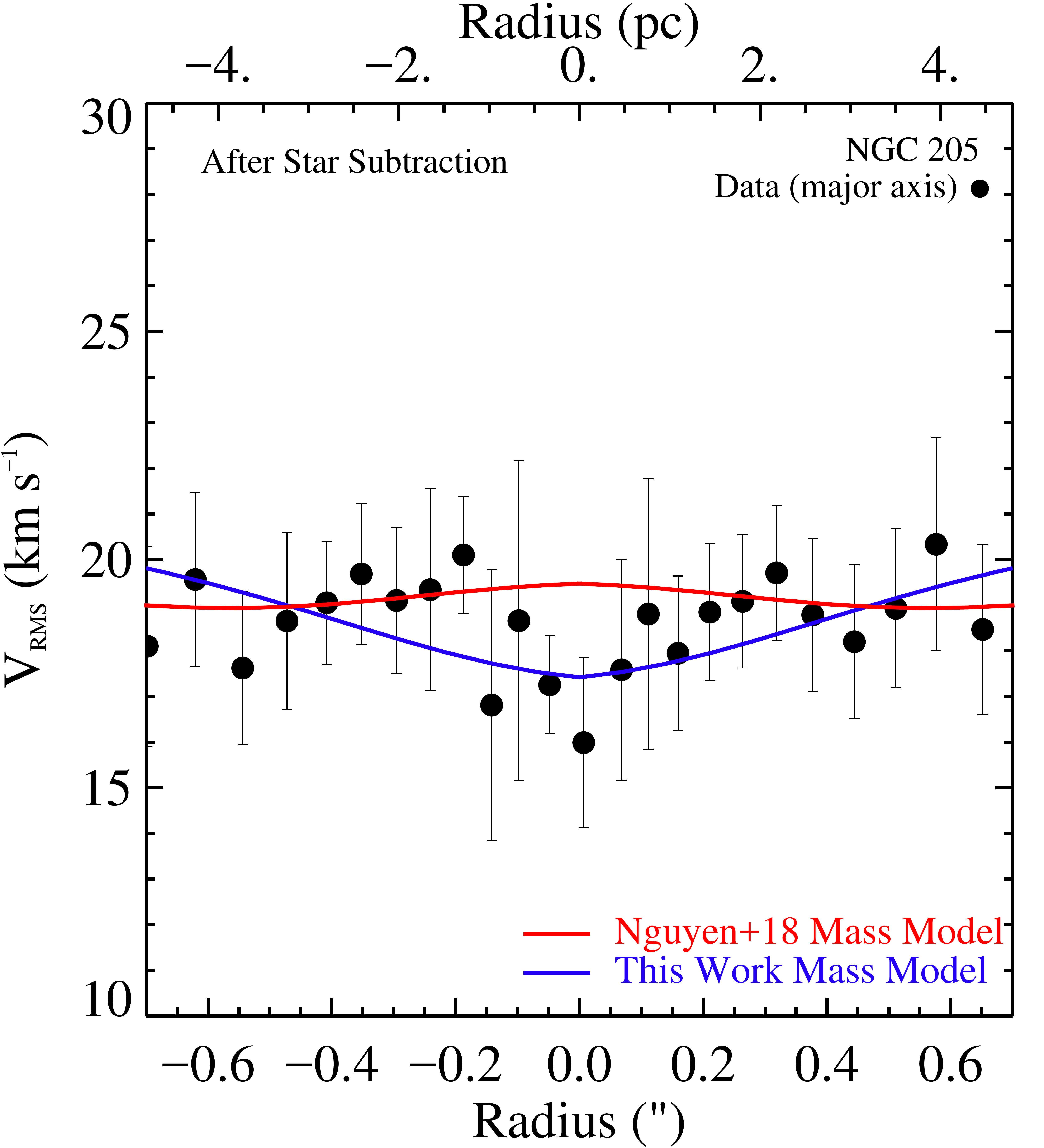}
\caption{\small{$V_{\rm rms}$ comparison between the best-fit JAM models using the N18 mass model (red line) and this work's mass model (blue line) for NGC 205.  Our new mass model of NGC 205 produces a best-fit JAM model that better traces the central kinematic drop where the best-fit JAM model of the N18 mass model failed. The measured data points are individual points with 0$\farcs$05 of the major axis; note that the models were fit to the full 2D kinematic maps in Figure \ref{rms2d_oldnewmass}}.}  
\label{rms1d_mcut_oldnewmass}   
\end{figure}

\begin{figure*}  
\hspace{-8mm}
   	\includegraphics[scale=0.15]{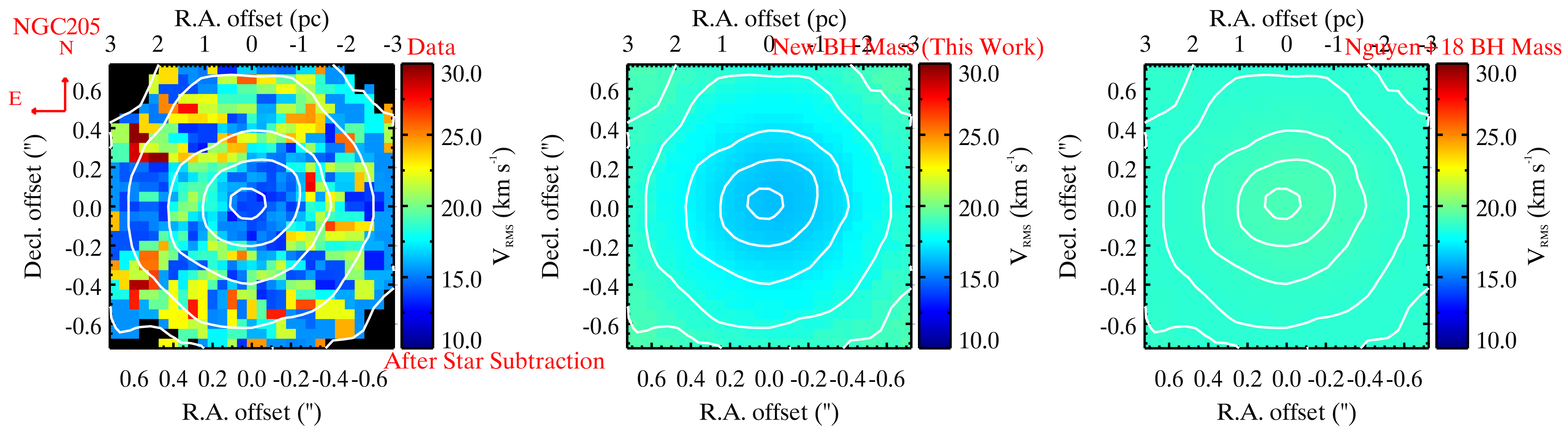}
\caption{\small{A 2D data-model comparison of the best-fit JAM BH masses between the new mass model (middle) and the N18 mass model for NGC 205 (right). The left panel shows the map of $V_{\rm rms}$ data. These models are the same as those shown in 1D in Figures \ref{rms1d_mcut_oldnewmass} and \ref{rms1d_mcut}. The white contours show the continuum, the red arrows indicate the N-E orientation of the field of view.}}  
\label{rms2d_oldnewmass}   
\end{figure*}

\begin{figure}  
\centering
   	\includegraphics[scale=0.17]{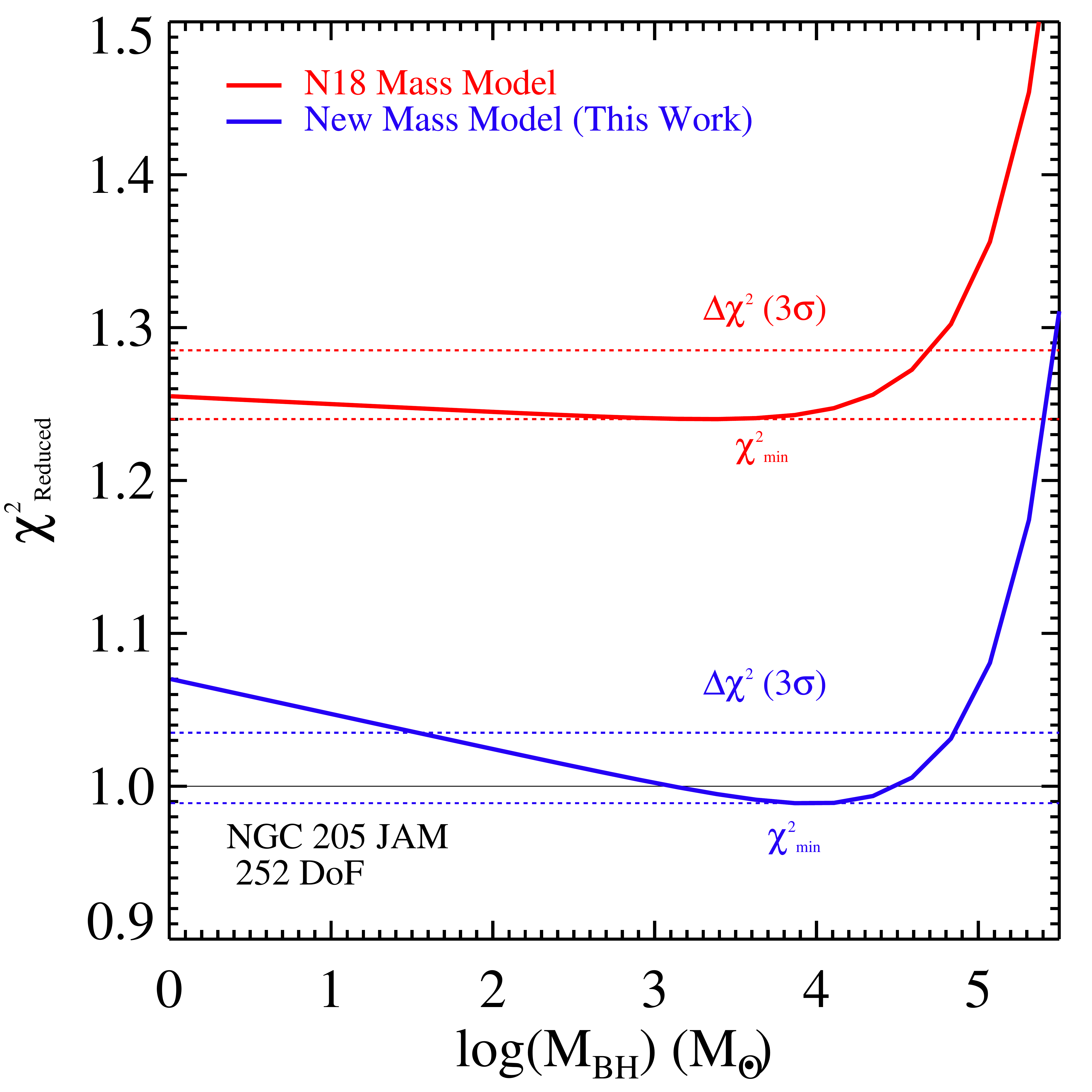}
\caption{\small{The reduced $\chi^2_{\rm reduced}$ distributions of the best-fit JAM models for the kinematics of NGC 205 are plotted as functions of BH mass using the N18 mass model (black line) and our new spectroscopic mass model in this work (red line).  The latter distribution shows that the BH mass can be measured within $3\sigma$, while the former indicates it is only estimated at $3\sigma$ upper limit (N18). Dashed lines with corresponding colors show the minimum $\chi^2_{\rm reduced,\; min}$ values and the $\Delta\chi^2_{\rm reduced}=\chi^2_{\rm reduced}-\chi^2_{\rm reduced,\; min}=0.06$ with 252 degree of freedom (DoF) (or $\Delta\chi^2=\chi^2-\chi^2_{\rm min}=11.3$) at the $3\sigma$ confidence level. Note that we fix $\beta_z$, $\gamma$, and $i$ as their best-fit values  for the new and N18 mass model, respectively, and calculate their $\chi^2_{\rm reduced}$ over a range of BH mass.}}  
\label{chi2_oldnewmass}   
\end{figure}

\begin{figure}  
\centering
	\includegraphics[scale=0.135]{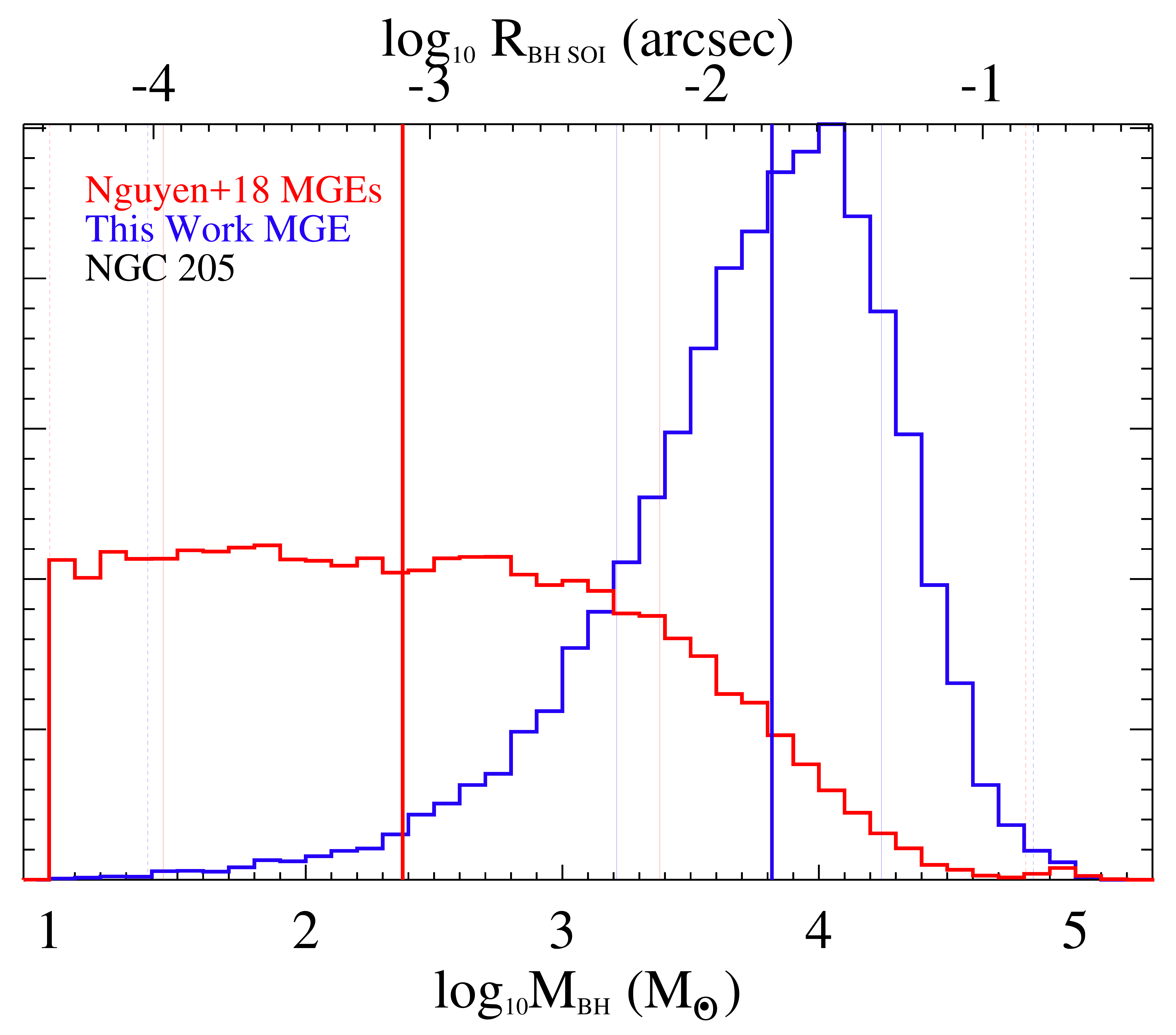}
\caption{\small{Comparison of the BH mass posterior distribution in our new mass model (blue) vs.~the N18 one (red).  The central vertical thick solid lines are the median BH mass estimates (50\% of the cumulative likelihood of \Mbh~from the JAM modeling), while the two thin solid lines and the two dashed lines are their PDF at $1\sigma$ and $3\sigma$ errors, respectively. The BH mass is shown in the lower horizontal axis, while the upper horizontal axis shows the sphere of influence of the BH in arcsec.}}  
\label{n205_bhsoi}   
\end{figure}

\begin{figure}  
\centering
	\includegraphics[scale=0.15]{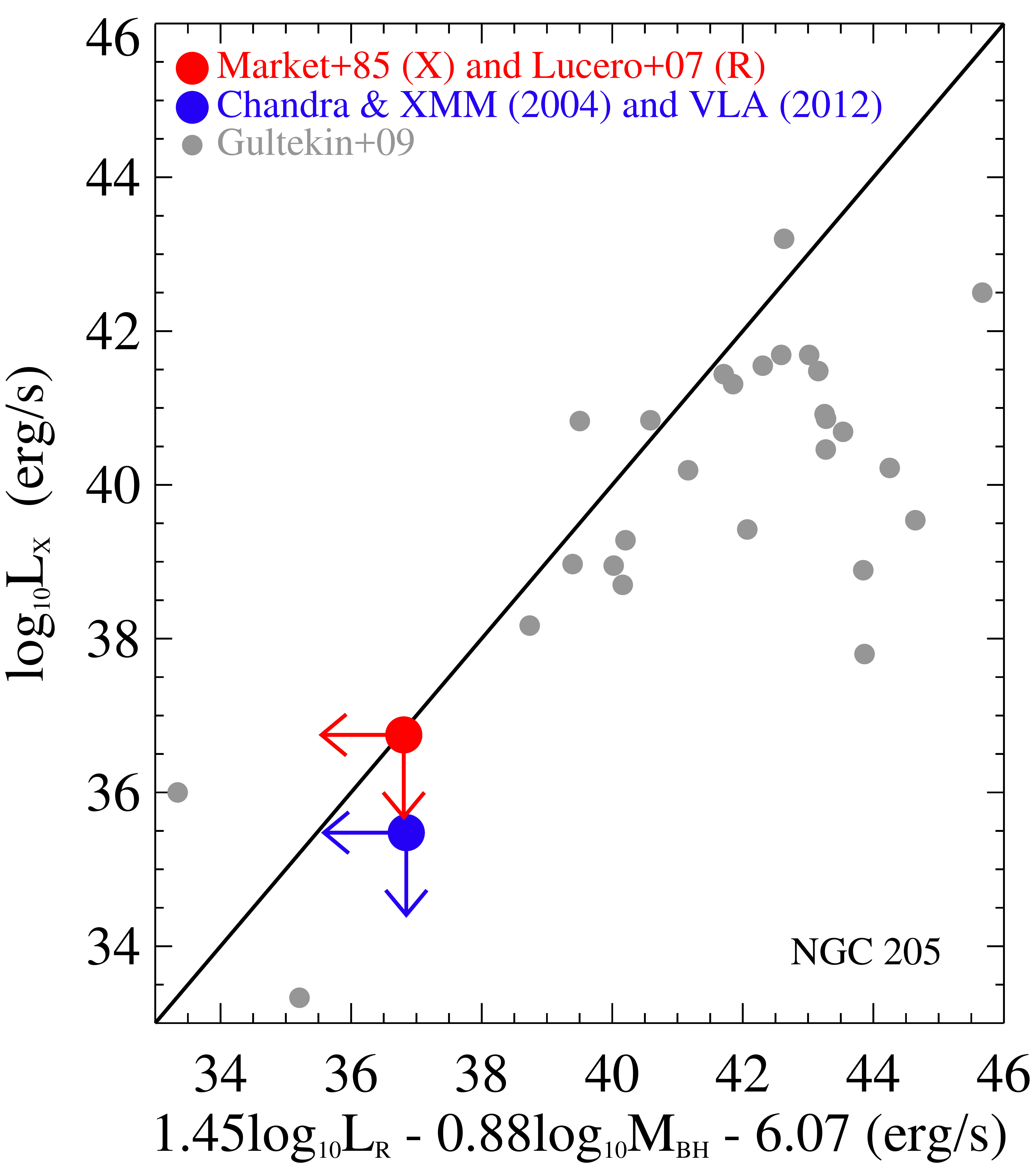}
        \caption{\small{The NGC 205 BH emission measurements from X-ray \citep{Markert85} and radio \citep{Lucero07} projected onto the fundamental plane of \citet[][the solid line]{Plotkin12} using our median BH mass measurement, which is plotted as a red dot. The blue dot is based on deeper 2004 observations with  \emph{Chandra/XMM} and 2012 observations with the VLA. The arrows indicate their upper limit estimates for radio and X-ray.  Additional data are plotted from \citet{Gultekin09b}  with gray dots.}}
\label{n205_fundamental_plane}   
\end{figure}

\begin{figure*}
\centering\vspace{-3mm}
	\includegraphics[scale=0.75]{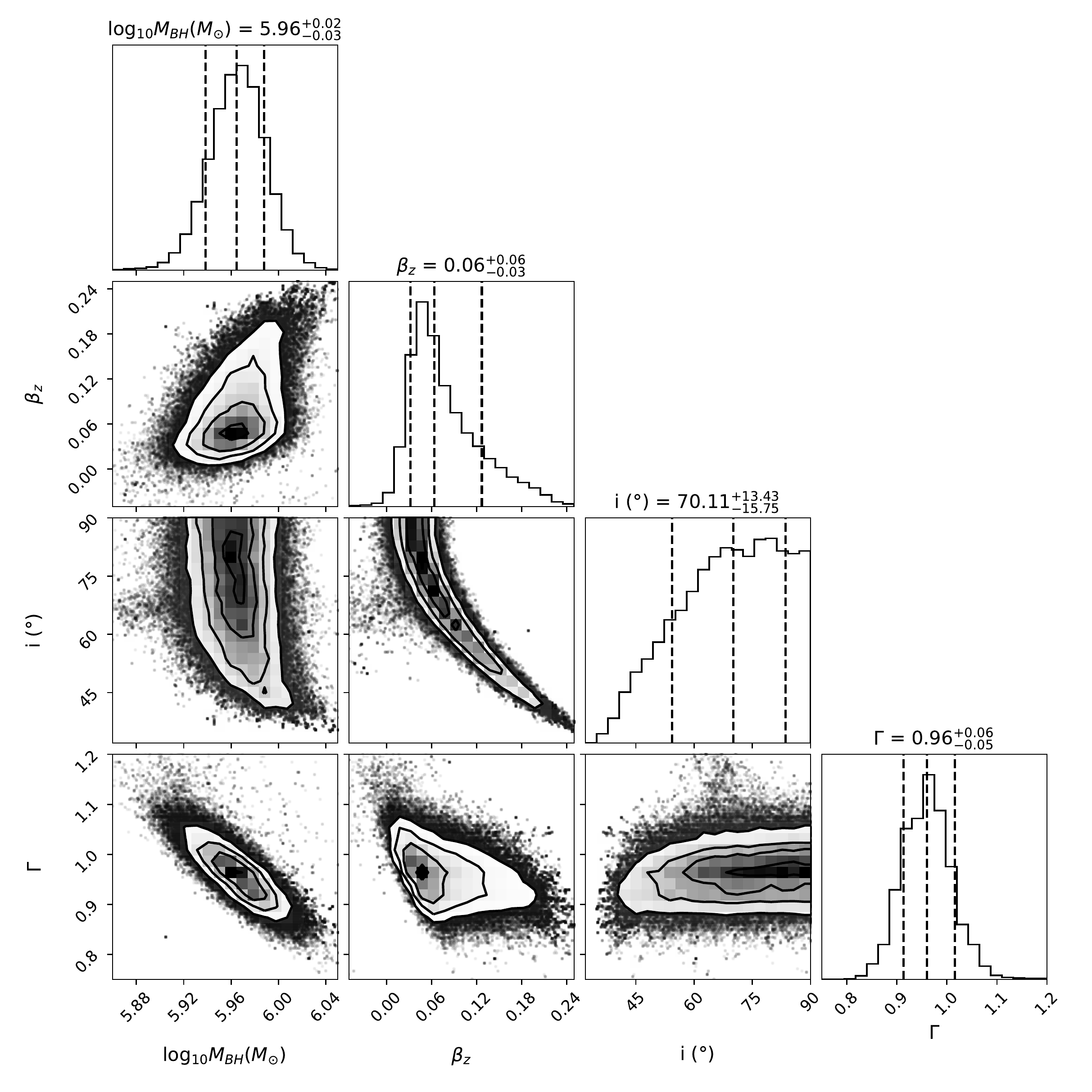} 
	\vspace{-7mm}
\caption{\small{The MCMC posterior distribution of the parameter space that we explored with the JAM dynamical models for the central BH in NGC 5102. See the caption of Figure \ref{posterior_n205} for details.}}     
\label{posterior_n5102}   
\end{figure*}

\begin{figure*}
\centering\vspace{-3mm}
	\includegraphics[scale=0.75]{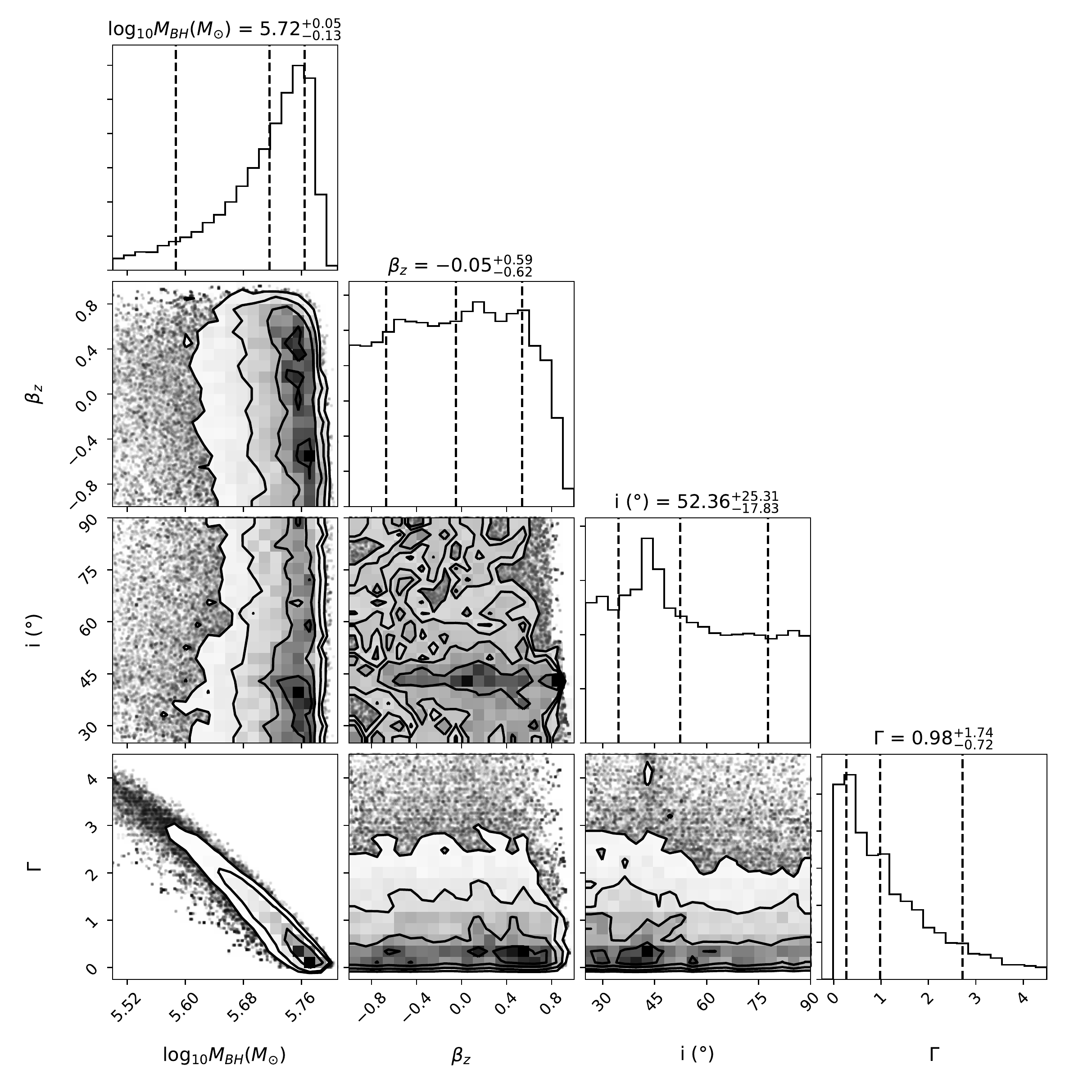}   
	\vspace{-7mm}
\caption{\small{The MCMC posterior distribution of the parameter space that we explored with the JAM dynamical models for the central BH in NGC 5206. See the caption of Figure \ref{posterior_n205} for details.}}  
\label{posterior_n5206}    
\end{figure*}
 
\begin{figure*}[!ht]  
\hspace{-3mm}
   	\includegraphics[scale=0.12]{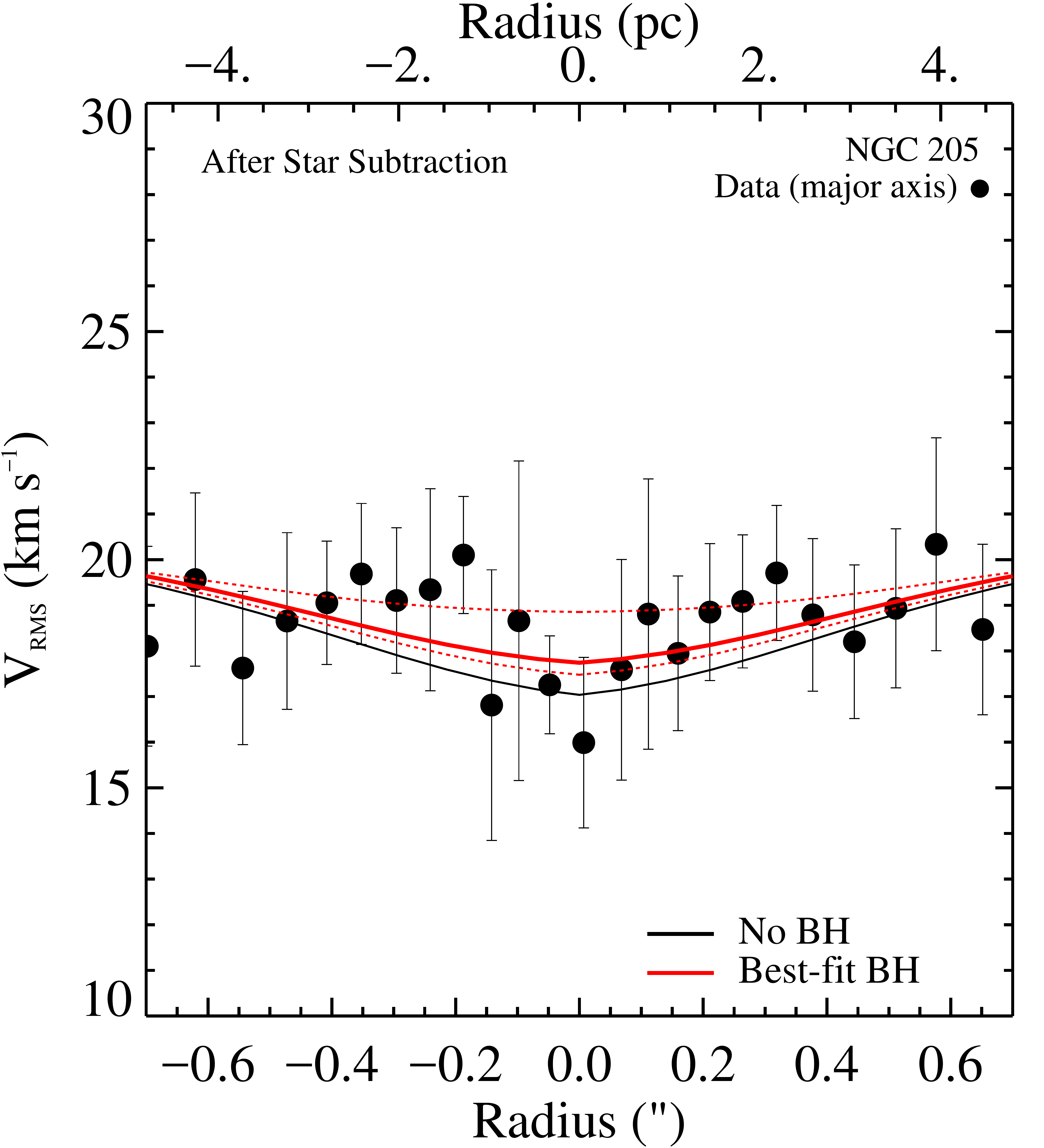}
   	\includegraphics[scale=0.12]{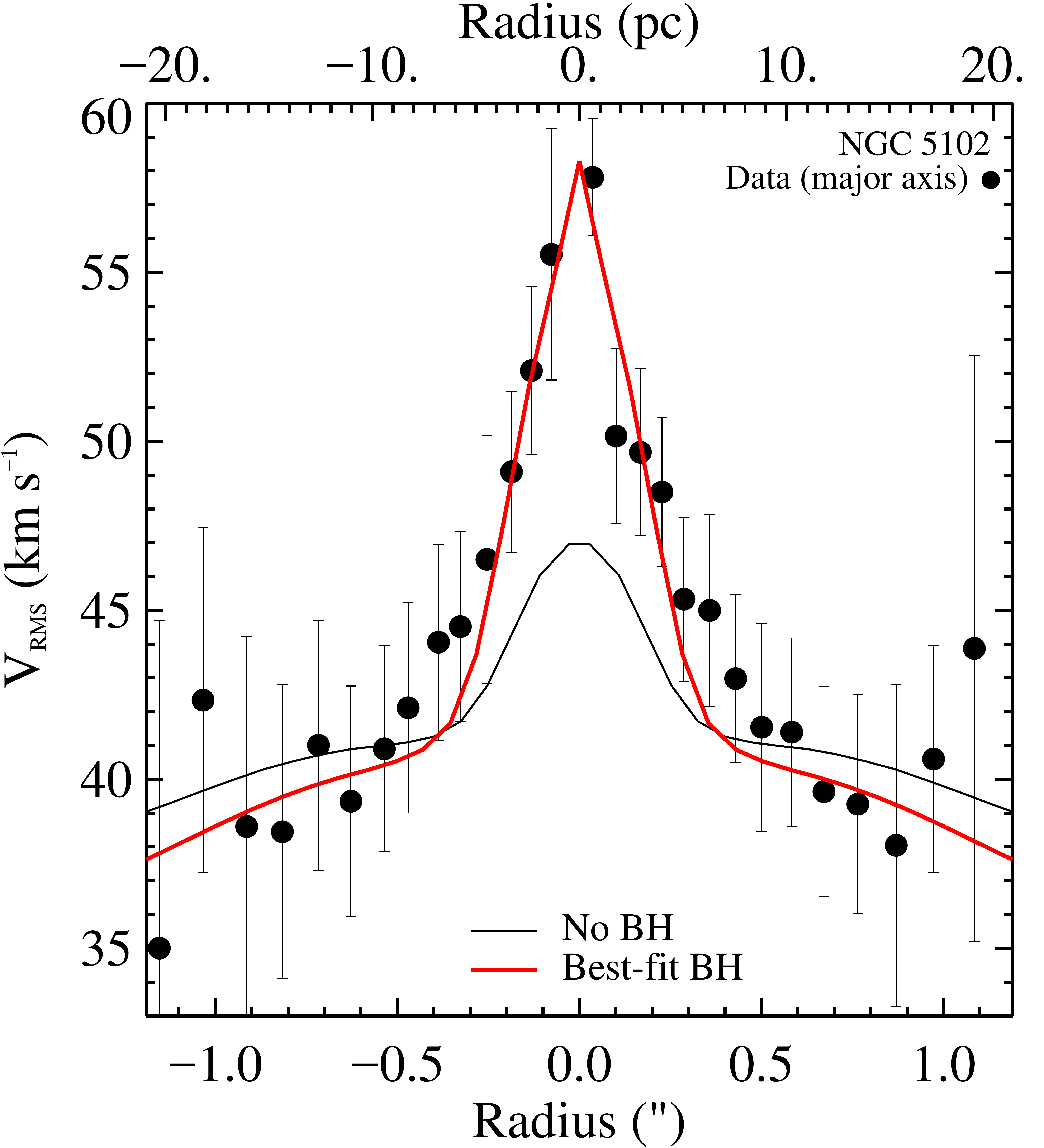}
   	\includegraphics[scale=0.12]{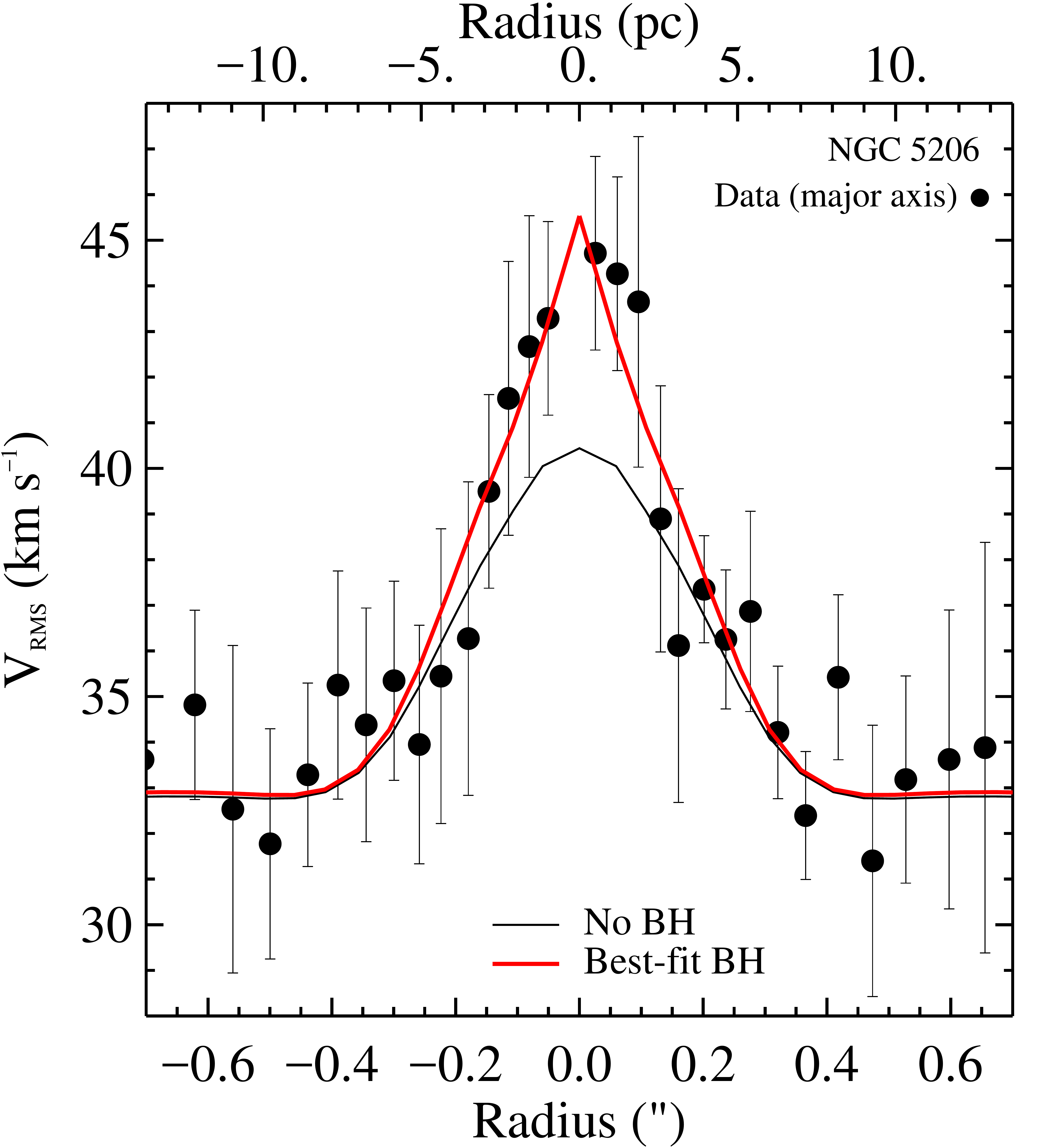} 
\caption{\small{$V_{\rm rms}$ comparison between the data (black points), the best-fit JAM model (red solid lines), and the corresponding 3$\sigma$ BH masses upper/lower limit (red dashed lines) for each galaxy.  Other parameters are fixed as their corresponding best-fit listed in Table \ref{fittable}. We plot the data points extracted in an elongated rectangular aperture one pixel wide ($0\farcs05$) along the semi-major axis \citep{Ahn18, Krajnovic18}. We also add the same $V_{\rm rms}$ 1D profile cut for the case of no BH (black lines), in order to visualize the imprint of the central BHs in these nuclei. We note that the best-fit models were fitted to the full 2D kinematics.}}   
\label{rms1d_mcut}   
\end{figure*}
 
In this section, we present our JAM dynamical modeling using the new mass constraints for the central SMBHs in NGC 205, NGC 5102, and NGC 5206, and compare these to our previous results (N18).  The two main improvements are: (1) the new mass models presented in Section \ref{ssec:correlation} and (2) the use of a Markov chain Monte Carlo (MCMC) sampler to fit the BH masses with JAM\footnote{We use the Python version of the JAM code (JamPy package), available at http://purl.org/cappellari/software.}, along with the \texttt{emcee} code\footnote{https://github.com/dfm/emcee} \citep{Foreman-Mackey13}, which utilizes an affine-invariant ensemble sampler \citep{Goodman10} to explore the parameter space with a set of walkers. At each step, the relative likelihoods for each walker determine their motion through parameter space. A few previous studies have shown that this MCMC mode  \citep[e.g.,][]{Leung17, Poci17, Ahn18, Krajnovic18} provides consistent results in estimating BH masses compared to the orbit based methods of triaxial Schwarzschild modeling and axisymmetric Schwarzschild modeling \citep[e.g.,][]{Verolme02, Cappellari10, vandenBosch10, Seth14, Drehmer15, Feldmeier-Krause17}.  Using MCMC is also more computationally  efficient for comparing JAM models with data than a grid search when the number of free parameters is $\geq$3.

The ingredients fed into the JAM models are: (1) The new stellar mass MGE models derived from using our \hst/STIS spectroscopic color--\mleff~relations in Section \ref{ssec:correlation} to parameterize the stellar mass components. (2) The synthetic $K$-band MGE models to parameterize the tracer population, which are taken from N18. (3) The nuclear stellar kinematic measurements from Gemini/NIFS (NGC 205) and VLT/SINFONI (NGC 5102 and NGC 5206); more details are given in Section 5 of N18. (4) The best-fit kinematic PSF functions parametrized from \hst~images and spectroscopic data in form of a Gauss + Moffat function for NGC 205 and double-S\'ersic functions for NGC 5102 and NGC 5206. These PSF functions are described in Section 3.2 of N18.   

Our JAM models have  four free parameters including BH mass (\Mbh), anisotropy ($\beta_z=1-\sigma_z^2/\sigma_R^2$), inclination ($i$), and the mass-scaling factor ($\Gamma=((M/L)_{\rm dyn.}/(M/L)_{\rm pop.}$) which parametrizes the best fit dynamical mass relative to that predicted in our stellar-population based mass maps; $\Gamma=1$ indicates a stellar mass map similar to what is expected from stellar population models (including the assumed IMF).  The anisotropy parameter ($\beta_z$) relates the velocity dispersion in the radial direction ($\sigma_R$) and z-direction ($\sigma_z$) assuming the velocity ellipsoid is aligned with cylindrical coordinates ($R, z, \phi$). The inclination ($i$) parameterizes the intrinsic axial ratio ($q=a/b$) in form of $q=\frac{\sqrt{q'^2-\cos^2(i)}}{\sin(i)}$, where $a$ and $b$ are the semi-minor and major axis; and $q'$ is the flattest axis ratio of the observed mass/light MGEs and the axis ratio parameter of the model \citep[see][for a detailed discussion]{Cappellari08}.  With these inputs and parameters, the JAM model calculates the projected $V_{\rm rms}$ ($V_{\rm rms}=\sqrt{\sigma^2+V^2}$, where $V$ is the radial velocity relative to the systemic velocity and $\sigma$ is the line-of-sight (LOS) velocity dispersion) of the observations.  The likelihood of each model is then determined from the $\chi^2$ differences of data and model, assuming Gaussian errors, which results in the likelihood $\mathcal{L} \propto e^{-\chi^2/2}$.  

To find the best-fit parameters with the \texttt{emcee} MCMC model of JAM, we first perform an initial MCMC JAM model run to explore the parameter space with an initial guess of the model parameters as the best-fit values from N18. This test will give us a sense of how each parameter behaves over the entire parameter space. Next, we create a follow up MCMC JAM model run with uniform priors on $\log(M_{\rm BH})$, $\beta_z$, $\Gamma$, and $i$ as given in Table \ref{fittable}.  The ``best-fit" parameters found in the initial test are used as the initial guess to maximize the sampling distribution around the best-fit model in this final MCMC JAM model run.  We set an MCMC chain with 100 walkers for a total of 300,000 steps. We consider the first 300 steps of each walker as the burn-in phase.

Figures \ref{posterior_n205}, \ref{posterior_n5102}, \ref{posterior_n5206} show the post burn-in phase distributions for the models of NGC~205, NGC~5102, and NGC~5206, respectively. In each of these figures, the contours illustrate their 0.5$\sigma$, 1$\sigma$, 2$\sigma$, and 3$\sigma$ confidence level. The histograms show the 1D probability distribution functions (PDFs) for each parameter, while vertical dashed lines from left to right are the quantiles of 0.16, 0.5, and 0.84, which are corresponding to 16\%, 50\%, and 84\% of the sampler distribution or 1$\sigma$ of the confidence level. We use the 1D distributions to calculate the best-fit values and their corresponding uncertainties. Their best-fit parameter values and 1$\sigma$ confidence level for each galaxy are shown at the top of each corresponding 1D PDF histogram. However, following the discussions in \citet{Seth14} and N17, we quote our final uncertainties on the BH masses and other parameters at the 3$\sigma$ confidence level (Table \ref{fittable}). We chose to quote 3$\sigma$ levels due to the fact that Jeans models have a restricted orbital freedom compared to e.g., Schwarzschild models. We should  also note that the way we interpret the best-fit models in this work is quite different to previous work \citep[e.g.,][N17; N18]{Seth14} where they identified the best-fit models are corresponding to minimum $\chi^2$ values; and the 1$\sigma$, 2$\sigma$, and 3$\sigma$ uncertainties are determined based on the differences of their $\chi^2$ values to the minimum $\chi^2$ at $\Delta\chi^2=2.3, 6.8, 11.2$, respectively. We take the median-likelihood model as our best-fit model,while the 1$\sigma$ and 3$\sigma$ uncertainties are estimated from all models within (16\% and 84\%) and (0.2\% and 99.8\%) of the PDFs, respectively.  However, we also checked for the choice of the maximum-likelihood (minimum $\chi^2$) from the PDF.  This test reveals that these highest probability ``best-fit" parameters are within our reported 1$\sigma$ uncertainties of using the above median distributions of the PDF ``best-fit" values except for the inclination of NGC 205, which is somewhat smaller (outside the lower 1$\sigma$ boundary).  Moreover,  as clearly see in Figures \ref{posterior_n205}, \ref{posterior_n5102}, and \ref{posterior_n5206}, our dynamical models provide poor constraints on inclinations. We therefore prefer to quote our ``best-fit" models using the median PDF for each parameter instead of using the highest probability values. 
We present our best-fit Jeans models with default color--\mleff~relations for the three galaxies in Table \ref{fittable}, and list the results from other filter/color combinations in Table~\ref{full_fittable} in the Appendix \ref{sec:fulltable}.

We also show the best-fit parameters for each nucleus in Figure~\ref{rms1d_mcut} in which the model $V_{\rm rms}$ values are compared to the data near the centers. Our best-fit models are sensitive to the presence of central BHs; especially for the case of NGC 205, the new mass map model allows us to replicate the drop of the $V_{\rm rms}$ towards the center where we failed with the previous mass map (N18) as shown in Figure~\ref{rms1d_mcut_oldnewmass}. These 1D $V_{\rm rms}$ profiles are extracted along an elongated rectangular aperture with the width of one pixel ($0\farcs05$) along the semi-major axis for both data and model; we note that the models were fit to the 2D data, not the 1D slices shown.
 
\begin{table*}
\caption{Best-fit Model Parameters and Statistical Uncertainties for Default Color Maps and Color--\ml~Relations}
\centering 
\begin{tabular}{c|cccc|ccc} 
\hline\hline       
          Parameter                & \multicolumn{3}{c}{Search Range} &   step  & Best Fit& $1\sigma$ Error (68\% conf.) &$3\sigma$ Error (99.7\% conf.)\\
                       (1)               &    (2)    &    (3)                &    (4)      &    (5)     &   (6)      &                   (7)    &            (8)     \\  
\hline                
{\bf NGC 205}      &           &{\bf F555W--F814W} Color Map&    &             &             &  {\bf F814W}  Mass Model&       \\    
\hline
$\log M_{\rm BH}$ (\Msun)&    0.0  & $\longrightarrow$&      6.0  &    0.1    &    3.83    &  $-$0.60, +0.43  & $-$1.84, +1.18  \\
$\beta_z$                           &$-$1.0 & $\longrightarrow$&     +1.0 &  0.05    & $-$0.08  &  $-$0.58, +0.55 & $-$0.92, +0.78  \\
$\Gamma$                         &  0.2    & $\longrightarrow$&     1.5   &  0.05    &    1.02    &  $-$0.19, +0.11  & $-$0.52, +0.23  \\
$i$ (\deg)                           &   30    & $\longrightarrow$&      90    &  1.0      &    58.6    &  $-$16.6, +21.1 & $-$38.6, +41.4  \\
\hline
{\bf NGC 5102}      &           &{\bf F336W--F814W} Color Map&            &             &        &     {\bf F814W} Mass Model& \\       
\hline
$\log M_{\rm BH}$ (\Msun)&  4.0    & $\longrightarrow$&  7.0      &    0.1    &    5.96    &  $-$0.03, +0.02 & $-$0.05, +0.04 \\
$\beta_z$                           &$-$1.0 & $\longrightarrow$& +1.0     &   0.05  &    0.06    &  $-$0.03, +0.06 & $-$0.08, +0.10  \\
$\Gamma$                        &    0.5   & $\longrightarrow$&  1.5      &   0.05   &    0.96   &  $-$0.05, +0.06 & $-$0.14, +0.16  \\
$i$ (\deg)                           &    30    & $\longrightarrow$&  90       &   1.0    &    70.1    &  $-$15.8, +13.4 & $-$28.4, +19.9 \\
\hline
{\bf NGC 5206}       &             &{\bf F555W--F814W} Color Map&         &          &           &{\bf F814W} Mass Model&     \\       
\hline
$\log M_{\rm BH}$ (\Msun)&   3.0    & $\longrightarrow$  &  7.0     &   0.1    &   5.72    &  $-$0.13, +0.05 & $-$0.30, +0.06  \\
$\beta_z$                           & $-$1.0 & $\longrightarrow$  & +1.0    &   0.05  & $-$0.05 &  $-$0.62, +0.59 & $-$0.95 +0.60  \\
$\Gamma$                         &   0.5    & $\longrightarrow$  & 15.0    &   0.05  &   0.98    &  $-$0.72, +1.74 & $-$0.05, +4.02  \\
$i$ (\deg)                           &    25     & $\longrightarrow$  & 90       &   1.0    &  52.4     &  $-$17.8, +25.3 & $-$27.4, +37.6  \\
\hline
\end{tabular}
\tablenotemark{}
\tablecomments{\small{Column 1: The list of the fitted model parameters for each galaxy. Columns 2, 3, and 4: The parameter search ranges in uniform linear space for each galaxy; the mass of BHs are linear in log scale. Column 5: The deviation of each parameter from their previous values expected in next step. Columns 6--8: The best-fit value of each parameter and their uncertainties at 1$\sigma$ and 3$\sigma$ confident levels in in F814W mass map models which are created from their corresponding default color maps (Figure \ref{colormap}) and color--\mleff~relations (Figure \ref{color_m2l_relation}).}}
\label{fittable}
\end{table*}

\subsection{Detection of a Massive Black Hole in NGC 205}\label{sec:n205}

\subsubsection{JAM for the BH Mass in NGC 205}\label{ssec:n205} 

The best-fit Jeans model of NGC 205 gives $M_{\rm BH} = 6.8_{-6.7}^{+95.6}\times10^3$\Msun, $\beta_z=-0.08_{-0.92}^{+0.78}$, $\Gamma_{\rm F814W}=1.02_{-0.52}^{+0.23}$ ($M/L_{\rm F814W, dyn.}=1.84_{-0.94}^{+0.41}$ (\Msun/\Lsun)), and $i=58.6$\deg$_{-38.6}^{+41.4}$. \citet{valluri05} put a 3$\sigma$ upper limit on this BH mass ($M_{\rm BH} < 3.8\times10^4$\Msun) using Schwarzschild modeling of \hst/STIS spectroscopic-kinematic measurements; for their \ml, they fit a single value for the nucleus and the galaxy from subsets of their data. N18 also measured an 3$\sigma$ upper limit mass of this BH of $M_{\rm BH} < 7\times10^4$\Msun~(or $\log_{\rm 10} M_{\rm BH} = 4.85$\Msun) using the same kinematic measurements that we use in this work.

We show the comparisons between the best-fit JAM models using the new mass model and the N18 mass model in form of $V_{\rm RMS}$ radial profiles in Figure~\ref{rms1d_mcut_oldnewmass} and 2D maps in Figure~\ref{rms2d_oldnewmass}, respectively.   Our new mass model produces a best-fit JAM model that matches the central $V_{\rm RMS}$ drop at the center of NGC 205, while the N18 mass model predicts a relatively flat profile, with a slight upturn at the center.  This suggests our new spectroscopic mass model provides a significant improvement over the previous one.  
  
This change in the $V_{\rm RMS}$ profile is due to the leveling off of the density profile within the central $\sim$0$\farcs$3 relative to the previous mass models steeper surface density slope; so while the projected central density in the two models is very similar, the predicted $V_{\rm RMS}$ is very different even at the same \ml.  This change in the mass model accounts for the different BH mass result.

The $\chi^2_{\rm reduced}$ resulting from both mass models provide further evidence for the improvement in our new mass model -- Figure \ref{chi2_oldnewmass} compares the $\chi^2_{\rm reduced}$ from the new and old mass models as a function of BH mass. In this plot, we fix $\beta_z$, $\Gamma$, and $i$ as their best-fit values in Table~\ref{fittable} and N18 for the new and N18 mass model, respectively, and calculate their $\chi^2_{\rm reduced}$ over a range of BH mass.

It is clear that the new mass model illustrates the BH mass in NGC 205 can be measured within the $\Delta \chi^2$ corresponding to the $3\sigma$ confidence level (CL), while the same $\Delta \chi^2$ interval with the N18 mass model produces include the without a BH ($M_{\rm BH}=0$). The new mass model also gives a minimum $\chi^2_{\rm reduced}\sim0.99$ for the best-fit JAM model, a big improvement over the N18 mass model, which has a minimum $\chi^2_{\rm reduced}\sim1.24$. This is related to the improved central dispersion profile in the new mass model seen in Figures~\ref{rms1d_mcut_oldnewmass} and \ref{rms2d_oldnewmass}.

Our work is the first one to place a lower limit on the BH mass in this galaxy of $\gtrsim$$5\times10^3$\Msun. This constitutes the first detection of a central BH in NGC 205 and at its current mass estimate makes it the lowest mass BH ever dynamically detected in a galaxy center.

If correct, the detection of a central BH in NGC 205 would be quite exciting, as it is the lowest mass BH ever dynamically detected in a galaxy center.  It is also lower mass than any BHs mass inferred through broad-line emission from an AGN in dwarf galaxies \citep{Baldassare15, Chilingarian18} and the accreting source HLX-1 \citep[e.g.,][]{Servillat11} via modeling the X-ray spectrum and light curve located off the plane of an edge-on lenticular galaxy ESO 243-49. The only dynamical measurements of BHs with comparable mass are those in globular clusters \citep[e.g.,][]{Gebhardt05, Noyola10, Lutzgendorf15, Baumgardt17, Kiziltan17}, which remain controversial \citep[e.g.,][]{vanderMarel10, Lanzoni13, Gieles18, Tremou18},   Because of this significance, we critically examine the possible systematic errors on the BH mass measurement in detail below, and then discuss X-ray and radio observations of the NGC 205 nucleus.

 \subsubsection{Possible Mass Model Errors}

We first focus on how the present mass models are different (and better) than those presented in N18.  In both cases, the colors were used to infer the varying \ml, but in N18, a 1D color and model was used for this purpose, while here we create a 2D mass map, and then fit our MGE directly to this map. Also in N18, we were restricted to use the color--\ml~relations of \citet{Bell03, Roediger15}, while here we derive this based on stellar populations fits to our STIS data.  Despite this improvement, NGC 205 is in the semi-resolved regime (as is visible in its color map, Figure~\ref{colormap}), and thus it is not clear that the color--\ml~relation we derive applies well to all pixels in the 2D image.  In particular, pixels with significant contributions from bright AGB stars could be given artificially high \ml~due to their red colors.  However, because we are fitting azimuthally symmetric MGE models to these data we do not expect that this will significantly impact our MGE models. We test this by applying two approaches: (1) we exclude the obvious stars in the central 1$\arcsec$ in mass model and interpolate over them; and (2) we find the ellipse averaged color from the color maps. The JAM model derives a BH of $M_{\rm BH} = 7.3\times10^3$\Msun~and $\beta_z=-0.38$ for the former mass MGE model, while the latter mass MGE model gives a BH of $M_{\rm BH} = 8.0\times10^3$\Msun~and $\beta_z=-0.82$.  These differences are not significant, although the $\beta_z$ values are low, with values that are further from isotropic.  

Since we are using a new MCMC method, we verify that our original mass model gives similar results with the MCMC approach.  Specifically, we take the N18 mass and light MGEs and run our MCMC code on them with identical priors to our original run; the results are shown in the Figure \ref{posterior_n205_n18mges}. These results are consistent with what we found in N18 except for the inclination angles, which are less constrained in our MCMC models (see Table \ref{fittable}). The cumulative likelihood for the new mass MGE and N18 default MGE \citep{Roediger15} of NGC 205 is shown in Figure \ref{n205_bhsoi}. The black hole mass posterior distribution in our new mass model (shown in blue) is compared to the one with the N18 model (red). The vertical lines (in the same colors) indicate the best fit BH mass (thick solid line) along with the 1$\sigma$ (thin solid line) and 3$\sigma$ (dashed line) uncertainties.

The best-fit BH mass of NGC 205 is $M_{\rm BH} = 6.8\times10^3$\Msun~corresponds to a sphere of influence (SOI) of radius 0$\farcs$018 ($\sim$0.07 pc), while the maximum BH mass within 1$\sigma$ suggests a SOI of 0$\farcs$03 ($\sim$0.12 pc).  
While our stellar population and kinematic data is at lower resolution (core PSF with HWHM$\sim$0$\farcs$05), our mass model was made with ACS/HRC images at resolution comparable to this SOI (HWHM$\sim$0$\farcs$03).  Because in our measurement, we effectively assume a constant stellar population/\ml~on the scale below our resolution, as long as this assumption is valid, our information on the stellar mass density within the SOI from the ACS/HRC imaging enables us to detect a BH despite the somewhat lower resolution of our kinematic data.  This situation is analagous to the inflated integrated dispersion measurements in UCDs; these ground based integrated dispersion measurements did not resolve the SOIs of the BHs in those objects but because the luminosity/mass models were made from higher resolution HST imaging (resolving the BH SOIs), these measurements still provided statistical evidence for their BHs that were later verified with higher resolution spectroscopy \citep[e.g.][]{Mieske13, Ahn18, Voggel18}.

\subsubsection{Possible Kinematic Errors}

The kinematic measurements of NGC~205 are challenging for two reasons: (1) the low dispersion of the nucleus, and (2) the influence of individual stars on the kinematics.  The line spread function (LSF) of our Gemini/NIFS has a median dispersion of 23~\kms; the dispersion value in the nucleus is close to this value, and thus the measured dispersion is susceptible to errors in our determined LSF.  We determine the LSF using sky line observations dithered identically to the original data and find spatial variations of up to $\sim$10\% across the FOV.  An over-estimate of 5~\kms~in the central dispersion values of NGC~205 would make our BH detection disappear.  This is comparable to our 1$\sigma$ errors on the dispersion (N18). Our kinematic measurements were made after subtracting off 32 individual stars at $r > 0\farcs35$ using the PampelMuse code  \citep{Kamann18}.  These stars were detected after subtraction of a smooth model, and significantly increase the smoothness of the velocity field (see N18).  Typically, the influence of individual stars will reduce the inferred dispersion and $V_{\rm rms}$ \citep{Lutzgendorf15}; but, this process is stochastic and individual outliers can increase the $V_{\rm rms}$ as well.  However, given the high luminosity at the center of NGC~205 ($\sim$2.1$\times10^5$\Lsun/pc$^2$, $<$0$\farcs$3), the central dispersion values should have minimal stochastic sampling issues.  We note that even using our original stellar kinematics without the stars subtracted, we obtain a similar result, with $M_{\rm BH} = 8.5\times10^3$~\Msun. 

\subsubsection{Collections of Stellar Remnants?}

The best fit BH mass in NGC 205 is just $\sim$0.5\% of the total NSC mass found in N18.  Due to the small effective radius (1.3~pc), N18 find a half-mass relaxation time of just 5.8$\times$10$^8$~years, suggesting that significant dynamical evolution can occur in the nucleus (although this evolution is undoubtedly complicated by the ongoing star formation or accretion of stars).  This dynamical evolution can enable a cluster of dark remnants, which could easily mimic the behavior of an IMBH \citep{denbrok14b, Bianchini17, Mann18}.  The complexity of the stellar populations in the NGC 205 NSC, along with the large number of stars relative to present-day simulations, and the unknown retention fraction of BHs and neutron stars make it challenging to evaluate the likelihood of the dynamical signature we are seeing as being due to remnants versus an IMBH.

\subsubsection{Accretion Evidence Based on Radio and X-ray Observations?}
 
We search for the radio and X-ray emissions in the nucleus of NGC 205 to examine if there is any evidence of an accreting BH from previous work. \citet{Markert85} observed the nucleus of NGC 205 with the {\it Einstein} Observatory and found no resolved X-ray source. However, they put an upper limit of $L_{X}<9\times10^{36}$ erg s$^{-1}$.  Even any detected source below this limit would be in a regime where identification with an AGN would be challenging due to the possibility of emission from a low-mass X-ray binary (LMXB). \citet{Lucero07} placed a radio upper limit of $<$60 $\mu$Jy beam$^{-1}$ at 1.4 GHz (20 cm), using Very Large Array C-array observations  with a beam size of $14\farcs5\times12\farcs4$.  We determine the position of NGC 205 BH's upper limits on the fundamental plane, assuming  a flat radio spectral slope of $\alpha=0$, transforming the observed 1.4 GHz emission into the standard reference frequency of 5 GHz, this flux density limit corresponds to a luminosity limit of $L_{R} < 7.58 \times 10^{31}$ erg s$^{-1}$, and following the equation from \citet{Plotkin12}:  
$\log(L_X)=(1.45\pm0.05)\log(L_R)-(0.88\pm0.06)\log(M_{\rm BH})$$-(6.07\pm1.10)$,
and plot these using our best BH mass in Figure \ref{n205_fundamental_plane} (the red dot). The position of the NGC 205 BH suggests any accretion taking place is $\lesssim$$10^{-4}$ of the Eddington limit.  About 80\% of nearby galaxy nuclei are accreting below this Eddington ratio \citep{Ho09}, including NGC205's neighboring compact elliptical satellite M32 \citep{Yang15}. 

We also turn to the newer and deeper X-ray and radio observations of the nucleus of NGC 205 existing in the archives (e.g., \emph{XMM}/\emph{Chandra} and VLA) for a better constrain on its central accretion BH. The details of these observations and their analysis are mention as follow:

 {\it Radio observations}:  we use archival observations obtained with the Karl G.~Jansky Very Large Array (project 12A-205; P.I. De Looze). These data were obtained over 2 epochs in 2012 June, with a total observing time just over 12 hr and an on-source time of 8.8 hr. The array was in B configuration and the L band (1--2 GHz) receiver was used in continuum mode, yielding a resolution of $\sim$$2.9$\arcsec\ at a central frequency of 1.5 GHz.

The data were calibrated and imaged with CASA (version 5.0.0) using a Briggs robust weighting of 0 and frequency-dependent cleaning. J0029+3456 was used as the phase calibrator and was observed before and after each 4.5-min scan of NGC 205. We used 3C48 for bandpass calibration and to set the absolute flux scale.

The rms noise in the final image was 7.6 $\mu$Jy beam$^{-1}$. While this noise value is somewhat higher than the theoretical noise of $\sim$$4~\mu$Jy beam$^{-1}$, attempts to self-calibrate the data were not successful.

We find no evidence for a radio source at the location of the putative BH (which we take as the center of the nuclear star cluster); indeed, the measured flux density at that location is negative. Hence we report an upper limit of 3 times the rms noise: $<$$22.8$ $\mu$Jy. This is about a factor of 2.6 deeper than the previous best limit  \citet{Lucero07}. Assuming a spectral index of $\alpha = 0$, at the standard reference frequency of 5 GHz, this flux density limit corresponds to a luminosity limit of $<$$8 \times 10^{31}$ erg s$^{-1}$.

{\it X-ray observations}: We also used archival \emph{Chandra} and \emph{XMM} data to get much stronger upper limits on the X-ray emission from the NGC 205 nucleus. The \emph{Chandra} analysis uses ACIS-S data from 2004 February (ObsID 4691; P.I. Terashima), which has an exposure time of 9.9 ksec. Using data reprocessed with CIAO 4.10 and CalDB 4.7.9, we extracted counts within  a $1.5\arcsec$ radius, corresponding to a 92\% fraction encircled energy at this location. We sampled the background using a larger region of $50\arcsec$ radius. For \emph{XMM}, we used archival data from 2004 January (ObsID 0204790401, P.I. Di Stefano), with an effective PN exposure time of 10.2 ksec and a MOS exposure time of 12.6 ksec; data from the different instruments were analyzed separately. For \emph{XMM} we used a source radius of $30\arcsec$ and 4 independent background regions of $60\arcsec$ radius.

No source was detected in any of the X-ray observations. To obtain upper limits, we assume  $N_H = 6.8 \times 10^{20}$ cm$^{-2}$ (consistent with only Galactic foreground absorption) and a power law emission model with $\Gamma =1.5$. All upper limits are given at the 95\% level and over the range 0.5--10 keV. We find a \emph{Chandra}/ACIS upper limit of $<$$4.9 \times 10^{-15}$ erg s$^{-1}$ cm$^{-2}$, a \emph{XMM}/PN upper limit of $<$$6.5 \times 10^{-15}$ erg s$^{-1}$ cm$^{-2}$, and a \emph{XMM}/MOS upper limit of $<$$9.9 \times 10^{-15}$ erg s$^{-1}$ cm$^{-2}$. Assuming the source has a constant flux, a conservative combination of these limits suggests $\lesssim$$4\times10^{-15}$ erg s$^{-1}$ cm$^{-2}$, suggesting an combined 0.5--10 keV X-ray luminosity of $<$$3 \times 10^{35}$ erg s$^{-1}$. This is a factor of 30 deeper than the existing X-ray upper limit \citet{Markert85}.

We plot these new upper limit estimates with our best BH mass in Figure \ref{n205_fundamental_plane} (the blue dot) in the same way we plot the estimations from \citet{Markert85} and \citet{Lucero07}. The new position of the NGC 205 BH now is more than one order-of-magnitude lower compared to its previous place in the X-ray luminosity axis suggesting an accretion taking place is $\lesssim$$10^{-5}$ of the Eddington limit. Because these new X-ray and radio observations are deeper than those from \citet{Markert85} and \citet{Lucero07}, we choose the latter accretion rate to quote for the accreting limit in the nucleus of NGC 205. 

Thus far the lack of emission in NGC 205 is not yet constraining for an accreting BH of the mass we have detected. This ambiguity is contributed by two factors: (1) the large scatter in the radio luminosities, roughly an order-of magnitude, relative to the fundamental plane usually seen in the sub-Eddington systems with known BH masses \citep{Gultekin09} and (2) the lack of simultaneous radio and X-ray observations. In the future, the deep and simultaneous detections of X-ray and radio emission in the nucleus of NGC 205 could provide important confirming evidence for the presence of its IMBH.

\subsection{JAM for the BHs in NGC 5102 and NGC 5206}\label{sec:n5102n5206} 

The BH masses in both NGC 5102 and NGC 5206 remain essentially unchanged from the results presented in N18.  We examine their new best-fit models here.
 
The best-fit Jeans model of NGC 5102 gives $M_{\rm BH} = 9.12_{-1.53}^{+1.84}\times10^5$\Msun, $\beta_z=0.06_{-0.08}^{+0.14}$, $\Gamma_{\rm F814W}=0.96_{-0.12}^{+0.14}$ ($M/L_{\rm F814W, dyn.}=0.48_{-0.06}^{+0.07}$ (\Msun/\Lsun)), and $i=70$\deg$.0_{-30.0}^{+20.0}$ (see Figure \ref{posterior_n5102}). Our new Jeans model changes the best-fit parameters (\Mbh, $\beta_z$, $\Gamma$, $i$) by (+3.5\%, $-$6.0\%, $-$16.5\%, $-$1.5\%) compared to the results of N18. The largest change is the decrease of $\Gamma$, and our new measurement brings the dynamical mass estimate into better agreement with our stellar population estimates as well as those of \citep{Mitzkus17}.  We note that this may be partly due to our improved WFC3 data; the previously available data was challenging to model due to the saturation of the center in some of the images (see appendix A of N18). 

The best-fit Jeans model of NGC 5206 give $M_{\rm BH} = 6.31_{-2.74}^{+1.06}\times10^5$\Msun, $\beta_z=-0.05_{-0.95}^{+1.05}$, $\Gamma_{\rm F814W}=0.98_{-0.96}^{+4.02}$ ($M/L_{\rm F814W, dyn.}=1.94_{-1.92}^{+9.10}$ (\Msun/\Lsun)), and $i=52.4$\deg$_{-32.5}^{+37.6}$~(see Figure \ref{posterior_n5206}). Our new Jeans model changes the best-fit parameters (\Mbh, $\beta_z$, $\Gamma$, $i$) by (+10.4\%, $-$120.0\%, $-$2.0\%, +31.0\%) compare to the results of N18.  This isotropic model reflects the improvement of our new nucleus mass map of NGC 5206, which is similar to what we found in NGC 404 and discussed in N17.

\subsection{Mass Model Uncertainties}\label{ssec:masserror}

The confidence intervals of the analysis that we have presented thus far are based on the kinematic measurement errors and do not include any systematic uncertainties in the mass model. In this section, we examine the mass model uncertainties by analyzing additional and independent mass model images.

\subsubsection{Errors in the Color--M/L Relation}\label{sssec:relationerror}

To examine the uncertainties on our IMBH mass estimates due to the color--\ml~relations, we propagate into our model the 1$\sigma$ uncertainties on our best-fit color--\ml~relations, shown as the pink regions in Figure \ref{color_m2l_relation}.  Specifically, we create mass maps and mass MGE models from the steepest and shallowest slopes of this 1$\sigma$ uncertainty region of the color--\ml~relations and we run full MCMC JAM models for both. We note that only variations in the slope matter in this case, as changes in $\Gamma$ exactly cancel out the  intercept of the relations.  The effects of variations in these mass models are quite small, with the corresponding variations in BH mass being smaller than our 3$\sigma$ confidence levels in all cases.

\subsubsection{Mass Maps from Additional Filters}\label{sssec:massmaps}

Our default models are created using F555W--F814W colors (NGC 205 and NGC 5206) and F336W--F814W (NGC 5102) color maps and their F814W images. In this section, we examine the impact of using color--\mleff~relations based on other colors and images. Specifically, for NGC 205 and NGC 5206 we have at our disposal only the F555W--F814W color map, so we can in alternative use the F555W image to create the mass map. However, with three available WFC3 filter images (F336W, F555W, and F814W) for NGC 5102, we are able to produce three color maps for this object: F336W--F814W, F547M--F814W, and F536W--F547M. Correspondingly, we can create three color--\mleff~relations and their mass models in the filters F336W, F547M, and F814W. Therefore, in total, we create nine WFC3 mass map and mass models.  These results are all presented in the Appendix (Table~\ref{sec:fulltable}).  

In the whole sample of three galaxies under our scrutiny, there is a remarkable consistency between these models. This is not the case for NGC 404, where N17 found that models using the F336W images produced inconsistent results, likely due to AGN emission in that filter.  These galaxies, with an apparent lack of any AGN component, likely do not suffer from those same uncertainties.  

\begin{figure*}  
\centering
	\includegraphics[scale=0.157]{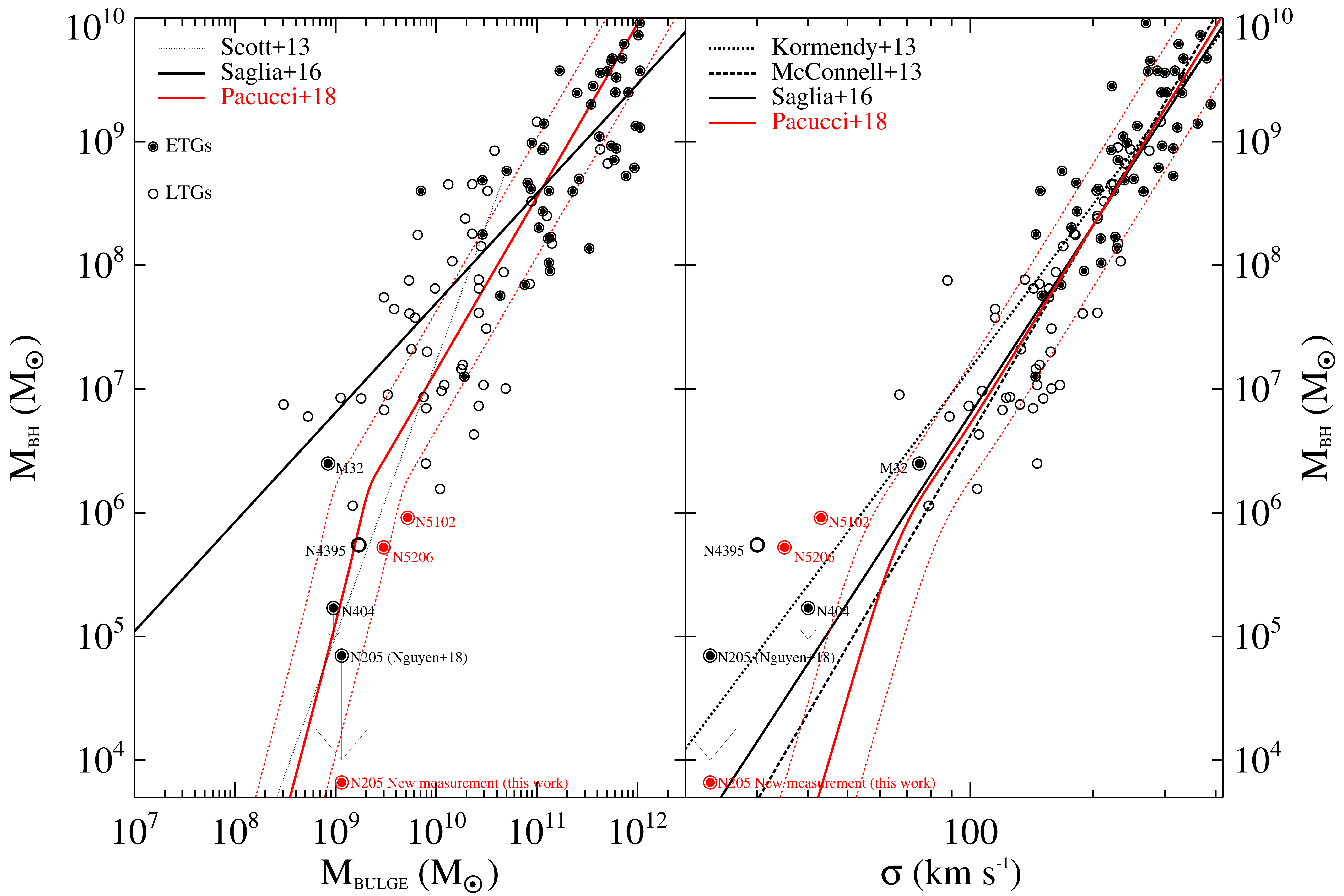}
\caption{\small{Our new BH mass constraints (red encircled data points) in the context of the $M_{\rm BH}-M_{\rm Bulge}$ (left) and $M_{\rm BH}-\sigma$ (right) scaling relations. The previous measurements of ETGs (black dots within open circles) and late-type galaxies (LTGs, black open circles) are taken from \citet{Saglia16}. The scaling relations of \citet{Scott13, Kormendy13, McConnell13, Saglia16} for ETGs and LTGs are plotted in the dotted, dashed, long-dashed lines, respectively. We also include the recent predictions for these scaling relations from \citet[][red solid lines]{Pacucci18} and its 1$\sigma$ (red dashed line) uncertainty}. The measurements of BH masses in the million/sub-million Solar mass regime are taken from N18 and are all labeled.  The downward arrows indicate upper limits. The new mass model of NGC 205 in this work obtains a detection of a BH mass of $\sim$6,760\Msun, an order of magnitude lower than the upper limit reported in N18.} 
\label{scaling}   
\end{figure*}

\begin{figure*}  
\centering
	\includegraphics[scale=0.25]{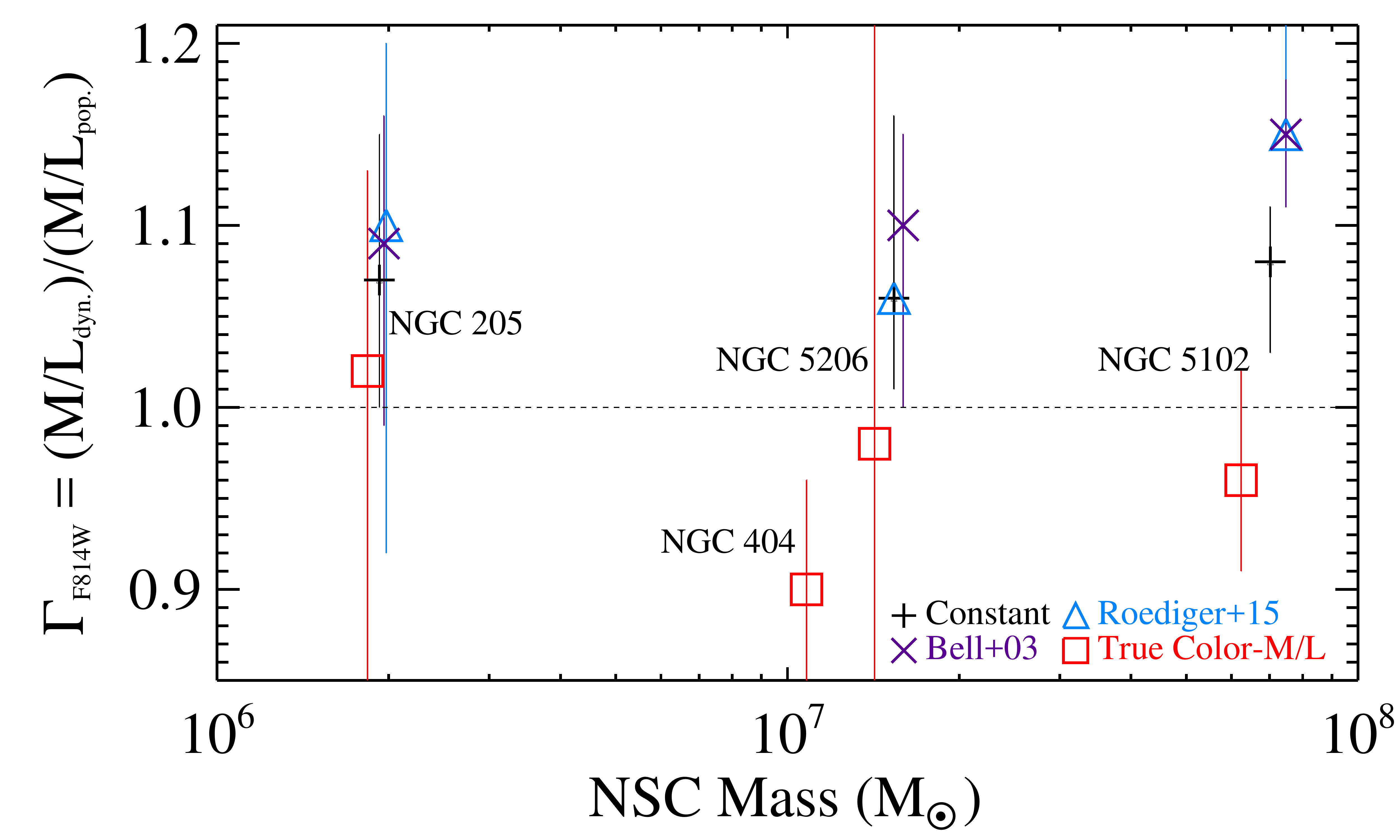}
        \caption{\small{The dynamical-to-population mass-scaling factors for each of our targets where we have measured color--\ml~relations (red points).  Additional estimates for the three galaxies in this paper based on constant color-\ml~relations as well as literature color--\ml~relations are shown as crosses \citep{Bell03} and triangles \citep{Roediger15}. } } 
\label{gamma_mnsc}   
\end{figure*}

\section{Discussion}\label{sec:dis}  

\subsection{Consequences of the detection of the BH in NGC205}

IMBHs may play the role of seeds for the growth of  SMBHs \citep[e.g.,][]{Volonteri05, Greene12}. Two common formation scenarios are the collapse of (metal-free) Population III stars \citep[e.g.,][]{Madau01} or the direct collapse of pristine gas clouds in massive halos \citep[e.g.,][]{Bromm03, Lodato06, Bonoli14, Pacucci_2015}. The accretion of gas lost in massive star clusters (SCs) is also an alternative channel for growing central IMBHs via runaway stellar mergers \citep{PortegiesZwar02, Gurkan04, PortegiesZwart04b, Vanbeveren09, Vesperini10}. Another formation channel for IMBHs is via sequential mergers of stellar mass BHs in globular clusters \citep[GCs;][]{Miller02}, or runaway tidal growth of BHs in nuclear star clusters (NSCs, \citealt{Stone17}). Although some IMBH detections have been reported in individual Galactic globular clusters (GGCs) and in the M31 GCs \citep{Camilo00, Anderson10, vanderMarel10, Gebhardt05, Ibata09, Noyola10, Lutzgendorf11, Feldmeier13, Lutzgendorf13, Bahramian17, Kiziltan17},  there is no general agreement on the presence of an IMBH in any of those objects \citep[e.g.,][]{vanderMarel10, Lutzgendorf11, Tremou18}. Accretion evidence for IMBHs in GCs at X-ray and radio wavelengths is also minimal, with very constraining upper limits. The discovery of HLX-1, located in the halo of the galaxy ESO 243-49, is the best current example of a non-nuclear BH at masses below $\sim$10$^5$\Msun. Its bright X-ray emission and X-ray variability suggests the presence of a $\sim$10$^4$\Msun~BH \citep{Servillat11, Webb12}, with observations at other wavelengths suggesting the possibility of an underlying star cluster, and perhaps a stripped galaxy nucleus \citep{Farrell14, Musaeva15, Soria17, Webb17}. 

In galactic nuclei, $<$10$^6$\Msun~BHs are found both from accretion signatures \citep[e.g.][]{Reines14, Baldassare15, Chilingarian18, Lin18} and from  dynamical measurements \citep[][N17, N18]{denbrok15}. N18 find an occupation fraction of $\sim 80\%$ in low-mass nearby ETGs; this appears to favor the presence of Population~III star seeds \citep[][]{Volonteri08, Volonteri10, Greene12, Reines16}.  In this context, the dynamical measurement for a low-mass ($\sim$10$^4$\Msun) BH in NGC 205 could suggest new evidence for a light seed formation scenario, as direct collapse models would produce higher ($\gtrsim$10$^5$\Msun) mass initial seeds \citep{Lodato06, Bonoli14, Woods_2018, Pacucci_2018}.  Furthermore, this detection would suggest that our entire sample of 5 nearby ETGs with $M_\star=$10$^{9-10}$\Msun~have central BHs and the occupation fraction would rise to 100\%.

\subsection{Comparison to Empirical Scaling Relations}

It is interesting to investigate where our new BH mass estimates fall in the parameter space of the classic scaling relations, i.e. the $M_{\rm BH}-M_{\rm Bulge}$ and $M_{\rm BH}-\sigma$ relations \citep[e.g.][]{Kormendy13, McConnell13, Saglia16} for ETGs. In Figure \ref{scaling}, we employ the bulge mass and velocity dispersion estimates already computed in N18, replicating their Figure 10, but we substitute with the new mass measurements presented in this paper.  The  best-fit BH masses of NGC 5102 and NGC 5206 are nearly identical to the results in N18, however, the new position of NGC 205 BH is one order of magnitude below the upper limit presented in N18.  With this new mass for NGC~205, the deviation of low-mass BHs from the $M_{\rm BH}-M_{\rm Bulge}$ relation seen in higher mass ETGs is even more pronounced, with the broken relation with Bulge mass by \citet{Scott13} providing a much better fit to our low mass galaxies, including NGC~205.  However, this BH mass is fully consistent with the extrapolation to lower masses for the $M_{\rm BH}-\sigma$ relation.

\subsection{Comparison of our Data and Theoretical Scaling Relations for Low-Mass Black Holes}

We compare our measurements with the predictions of the theoretical model presented in \citet{Pacucci17, Pacucci18}. This model predicts a downturn of the scaling relations ($M_{\rm BH}-M_{\rm Bulge}$ and $M_{\rm BH}-\sigma$) for BHs with mass $M_{\rm BH}\lesssim10^5$\Msun. This prediction is obtained by assuming: (i) a bimodal population of high-$z$ seeds, with the formation of light ($M_{\rm BH}\lesssim10^3$\Msun) and massive ($M_{\rm BH}\gtrsim10^4$\Msun) seeds, and (ii) that light seeds accrete inefficiently, with low accretion rates and duty cycles. Consequently, light seeds would accrete inefficiently and result in intermediate-mass black holes hosted in local galaxies, which fail to reach the black hole mass dictated by the scaling relations. On the contrary, massive seeds are assumed to grow efficiently and populate local galaxies with SMBHs \citep{Pacucci17, Pacucci18, Pacucci_2018}. The model predicts a downturn of the scaling relations $M_{\rm BH}-M_{\rm Bulge}$ and $M_{\rm BH}-\sigma$ around the critical values $\sigma_{\rm critical} = 65$ \kms and $M_{\rm Bulge,\; critical}=2\times10^9$\Msun, respectively. We show both of these theoretical scaling relations and their 1$\sigma$ uncertainty in Figure \ref{scaling}. Interestingly, the bimodal $M_{\rm BH}-M_{\rm Bulge}$ relation is consistent within 1$\sigma$ with all existing dynamical measurements at low mass. However, we observe a large departure from the prediction for the bimodal $M_{\rm BH}-\sigma$  relation. Although a fully self-consistent model for this departure does not exist yet, some previous works have provided interesting insights on this matter. For instance, \cite{Scott13, Graham_2015} suggest that a break in the $M_{bulge}-M_{BH}$ follows from the presence of a break in the $M_{sph}-\sigma$ relation, where $M_{sph}$ is the spheroid stellar mass. Additional works (e.g., \citealt{Fontanot_2015}) mention the possibility of a shift between stellar feedback and AGN feedback at a stellar spheroidal mass consistent with the break observed.

\subsection{Mass Scaling Factors and IMF Variations}\label{sssec:imf}

Our stellar population mass estimates are based on templates that assume a Chabrier IMF \citep{Vazdekis10, Vazdekis12}. As we observe a generally good agreement between our dynamical and stellar population mass estimates, this suggests that the stellar populations in these nuclei do in fact have a Chabrier IMF. In Figure \ref{gamma_mnsc}, we plot the $\Gamma$ factors in our best fitting dynamical models compared to those determined in N18 using the color--\ml~relations from \citet{Bell03} and \citet{Roediger15}.  Among the four galaxies for which we have spectroscopically determined local color--\mleff~relations, three are consistent with $\Gamma = 1$ within 1$\sigma$, while NGC 404 is somewhat lower than unity, but consistent with it at a $\sim$2$\sigma$ level.  Evidence of IMF variations at the center of massive elliptical galaxies have been claimed \citep[e.g.,][]{CappellariNa12, Conroy12, Martin15}, as also variations in the IMFs of young populations in the Milky Way nucleus \citep{Bartko10, Lu.R13}. While these observations suggest variations in the high and low-mass ends of the IMF respectively, our comparison of dynamical-to-stellar \ml s in nuclei with a range of stellar population ages generally suggests nuclear IMFs consistent with a Chabrier IMF.

\section{Conclusions} \label{sec:cons}

We presented a new analysis of three nearby early-type galactic nuclei: NGC 205, NGC 5102, and NGC 5206.  We used stellar population fits to HST/STIS spectroscopy across the nucleus in combination with \hst~imaging to create color--\ml~relations and improve the mass models and BH mass estimates for the entire sample of galaxies.  Our main results are as follows:

\begin{enumerate}

 \item Jeans modeling of the new mass model of NGC 205 suggests the detection of a central IMBH with $M_{\rm BH} = 6.8_{-6.7}^{+95.6}\times10^3$\Msun. This is the lowest central BH mass inferred for any galaxy and this measurement is significantly different from previous results for the same galaxy \citep[][and N18]{valluri05}.  However, possible systematic errors and the possibility of a cluster of dark remnants raising the central \ml~in NGC~205 suggest that we treat this detection with caution.

\item Assuming our derived mass, previous X-ray and radio upper limits suggest that the BH must be accreting at $\lesssim 10^{-5}$ the Eddington rate; this is well above the average Eddington ratio for nearby galaxies, and is thus not really informative on whether NGC~205 truly host an IMBH.

\item The new mass models in NGC~5102 and NGC~5206 suggest similar BH masses as found in N18; both have best-fit BHs below 10$^6$\Msun.  
      
\item Our derived color--\ml~relations have a wide-range of slopes, some of which are steeper than previously published relations.  In particular, this appears to be true for nuclei with prominent young populations. We construct a new spectroscopic color--\ml~relation to be applied for targets lacking detailed spectroscopic information but showing signatures of large fraction of young stars ($<$1 Gyr) in their nuclei.

\item With our new color--\ml~relations, the dynamical masses of the nuclei are fully consistent with their stellar population estimates.  This suggests the IMF in these nuclei is consistent with the Chabrier IMF assumed in our stellar population modeling.

\end{enumerate}

\vspace{2mm}
%
%
\section*{Acknowledgements}

The authors thank the anonymous referee for careful reading and useful comments which greatly improved this paper. We would also like to thank the Physics and Astronomy Department, University of Utah for supporting this work. D.D.N. and A.C.S. acknowledge financial support from NSF grant AST-1350389 and a part of the financial support from the \hst~grant GO-14742. D.D.N. also delivers his gratitude to the Willard L. and Ruth P. Eccles Foundation for their Eccles Fellowship during the 2017-2018 academic year at the University of Utah. M.C. acknowledges the support from a Royal Society University Research Fellowship. J.S. acknowledges support from NSF grant AST-1514763 and the Packard Fellowship. E.T. acknowledges financial support from the UnivEarthS Labex program of Sorbonne Paris Cit\'e (ANR-10-LABX-0023 and ANR-11-IDEX-0005-02). F.P. acknowledges support from the NASA Chandra
award No. AR8-19021A.

\vspace{2mm}

{\it Facilities:} Gemini: Gemini/NIFS/ALTAIR, ESO--VLT/SINFONI, \hst~(WFPC2, WFC3, ACS HRC), The National Radio Astronomy Observatory is a facility of the National Science Foundation operated under cooperative agreement by Associated Universities, Inc. This paper is partially based on data obtained from the Chandra Data Archive and from \emph{XMM}-Newton.

\appendix

\section{MGEs of the HST PSFs~Imaging}\label{sec:mgepsf}

The \hst~PSFs, which are created in Section~\ref{ssec:images} and use to determined the BH masses in Section~\ref{sec:steldyn} are as important as the MGE themselves. For this reason we tabulate these PSFs in Table~\ref{tab_psfmges}. The PSF was modeled by a sum of concentric elliptical Gaussians using the MGE method in a similar manner of the MGEs of the mass models. 

\begin{table}
\vspace{5mm}
\caption{MGE parameters of the \hst~PSF}
\centering
\begin{tabular}{ccccc}
\hline\hline   
$j$  &Total Counts&$\sigma$& $a/b$ & Total Counts \\
    &of Gaussian$_j$&(arcsec)&           &Normalization\\
 (1) &           (2)           &    (3)     &  (4)     & (5) \\
\hline
      &     &{\bf NGC 205}&   & \\
\hline
1 &  0.385   &   0.027  &   0.914  & 0.480 \\
2 &  0.276   &   0.075  &   0.868  & 0.343 \\
3 &  0.049   &   0.233  &   0.794  & 0.061 \\
4 &  0.093   &   0.580  &   0.957  & 0.116 \\
\hline
     &     &{\bf NGC 5102} &    &\\
\hline
1 &  0.302   &   0.026   &  0.991 & 0.306 \\
2 &  0.554   &   0.630   &  0.959 & 0.563 \\
3 &  0.063   &   0.233   &  0.852 & 0.064 \\
4 &  0.066   &   0.551   &  0.951 & 0.067 \\                              
\hline                    
     &     & {\bf NGC  5206} &   &\\
\hline       
1   &  0.281    &  0.020 & 0.264 & 0.260 \\ 
2   &  0.486    &  0.035 & 1.000 & 0.450 \\   
3   &  0.185    &  0.075 & 1.000 & 0.172 \\    
4   &  0.070    &  0.213 & 1.000 & 0.065 \\    
5   &  0.010    &  0.549 & 1.000 & 0.009 \\   
6   &  0.048    &  0.831 & 1.000 & 0.044 \\    
\hline  
\end{tabular}
\tablenotemark{}  
\tablecomments{\small{Column 1: Gaussian component number. Column 2:    the MGE models that represents for the total flux of each Gaussian.  Column 3: the Gaussian width (FWHM or dispersion) along the major axis.  Column 4: the axial ratios. Column 5: the normalization of the total counts of each Gaussian.}}
\label{tab_psfmges}
\end{table}
\section{Table of Full JAM Models}\label{sec:fulltable}

We make the complete results of JAMs testing in two color bases of F336W--F547M and F547M--F814W and mass models in three filters, F333W, F547M, and F814W for NGC 5102 in Table \ref{full_fittable} to indicate for their systemic uncertainties from various color--\mleff~relations and filters.

\begin{table*}[!ht]
\caption{Full Table of the Best-fit Model Parameters and Statistical Uncertainties for NGC 205, NGC 5102, and NGC 5206 with Different Mass Models Based on Various Colors and Filters}
\scriptsize
\hspace{-24mm}
\begin{tabular}{c|ccc|ccc|ccc} 
\hline\hline
Parameters                                 &                                                 \multicolumn{9}{c}{NGC 5102}                                                     \\
\hline           
                                                    & Best Fit & $1\sigma$ Error & $3\sigma$ Error& Best Fit & $1\sigma$ Error & $3\sigma$ Error& Best Fit & $1\sigma$ Error & $3\sigma$ Error\\
                                                    &              &    (68\% conf.)    &  (99.7\% conf.)   &              &   (68\% conf.)     &    (99.7\% conf.)&               &   (68\% conf.)    &    (99.7\% conf.) \\
                      (1)                          &    (2)     &        (3)               &            (4)           &     (5)     &                 (6)      &               (7)       &      (8)     &                (9)      &            (10)        \\
\hline                
F336W                                        &              & F336W--F547M &                            &              & F547M--F814W &                           &               & F336W--F814W &                         \\       
\hline
$\log M_{\rm BH}$ (\Msun)         &     5.94  &$-$0.03, +0.03&$-$0.08, +0.09        &     5.93  &  $-$0.03, +0.02    & $-$0.09, +0.07 &     5.94  &  $-$0.02, +0.03 & $-$0.05, +0.07   \\ 
$\beta_z$                                    &     0.07  &$-$0.02, +0.02&$-$0.06, +0.06        &     0.06  &  $-$0.03, +0.02    & $-$0.08, +0.05 &     0.05  &  $-$0.03, +0.05 & $-$0.07, +0.11   \\
$\Gamma$                                  &     0.98  &$-$0.05, +0.04&$-$0.12, +0.10        &     0.99  &  $-$0.05, +0.03    & $-$0.13, +0.09 &    1.02   &  $-$0.05, +0.07 & $-$0.12, +0.15   \\
$i$ (\deg)                                     &     71.7  &$-$14.1, +12.5&$-$30.1, +18.3       &     72.1  &  $-$15.2, +13.6    & $-$32.6, +17.9 &     70.3  &  $-$16.1, +15.7  & $-$35.3, +19.7  \\
\hline                
F547M                                         &              & F336W--F547M  &                          &               & F547M--F814W  &                         &                & F336W--F814W &                        \\       
\hline
$\log M_{\rm BH}$ (\Msun)         &     5.95  &$-$0.03, +0.03&$-$0.09, +0.07        &     5.95  &  $-$0.02, +0.02    & $-$0.06, +0.06 &   5.96    &  $-$0.03, +0.03 & $-$0.06, +0.05   \\ 
$\beta_z$                                    &     0.06  &$-$0.02, +0.04&$-$0.06, +0.11        &     0.04  &  $-$0.03, +0.04    & $-$0.08, +0.12 &   0.06    &  $-$0.07, +0.05 & $-$0.12, +0.15   \\
$\Gamma$                                  &     0.98  &$-$0.04, +0.04&$-$0.11, +0.10        &     0.98  &  $-$0.05, +0.02    & $-$0.15, +0.08 &   0.96    &  $-$0.04, +0.06 & $-$0.15, +0.17   \\
$i$ (\deg)                                     &     70.6  &$-$13.6, +11.8&$-$29.3, +19.4        &    70.5  &  $-$13.4, +12.0     & $-$26.1, +19.5 &  70.5     &  $-$15.2, +14.2 & $-$29.3, +19.5  \\
\hline    
F814W                                        &              & F336W--F547M  &                           &               & F547M--F814W  &                         &     &$\tablenotemark{a}$ $V$-band &           \\       
\hline
$\log M_{\rm BH}$ (\Msun)         &     5.95  &$-$0.03, +0.05&$-$0.07, +0.13        &     5.96  &  $-$0.04, +0.03    & $-$0.10, +0.09 &    5.95   &  $-$0.03, +0.02 & $-$0.08, +0.05 \\ 
$\beta_z$                                    &     0.07  &$-$0.03, +0.02&$-$0.07, +0.05        &     0.04  &  $-$0.02, +0.04    & $-$0.06, +0.10 &   0.07   &  $-$0.05, +0.07 & $-$0.15, +0.18 \\
$\Gamma$                                  &     0.97  &$-$0.02, +0.02&$-$0.06, +0.07        &     0.97  &  $-$0.05, +0.03    & $-$0.14, +0.10 &   0.97   &  $-$0.05, +0.05 & $-$0.14, +0.13 \\
$i$ (\deg)                                     &     70.9  &$-$14.0, +13.2&$-$28.4, +19.1        &    70.3   &  $-$13.8, +13.0    & $-$27.6, +19.7 &   71.3   &  $-$15.3, +16.3 & $-$34.6, +28.7\\
\hline \hline   
Parameters                                 &     \multicolumn{3}{c}{NGC 205}  &    \multicolumn{3}{c}{NGC 5206}    &                   \multicolumn{3}{c}{ }                                                      \\
\hline           
                                                    & Best Fit & $1\sigma$ Error & $3\sigma$ Error& Best Fit & $1\sigma$ Error & $3\sigma$ Error &              &                            &                      \\
                                                    &              &    (68\% conf.)    &  (99.7\% conf.)   &              &   (68\% conf.)     &    (99.7\% conf.) &              &                            &                       \\
\hline                
F555W                                        &              & F555W--F814W &                            &              & F555W--F814W  &                          &              &                            &                      \\       
\hline
$\log M_{\rm BH}$ (\Msun)         &   3.85    &  $-$0.59, +0.40 & $-$1.89, +1.12     &   5.70    &  $-$0.15, +0.06 & $-$0.33, +0.08   &              &                            &                       \\ 
$\beta_z$                                    & $-$0.10 &  $-$0.52, +0.51 & $-$0.92, +0.71    &  $-$0.07 &  $-$0.67, +0.61 & $-$0.93, +0.67  &              &                            &                       \\
$\Gamma$                                  &   1.03    &  $-$0.20, +0.08 & $-$0.50, +0.21    &  1.03      &  $-$0.70, +1.83 & $-$0.05, +3.97  &              &                            &                        \\
$i$ (\deg)                                    &    57.2   &  $-$18.3, +23.7 & $-$38.0, +40.1    &    53.0    &  $-$18.0, +24.2 & $-$27.0, +37.0   &              &                            &                        \\
\hline

\end{tabular}
\tablenotemark{}
\tablecomments{Column 1: The list the fitted model parameters. Columns 2--10: The best-fit value of each parameter and their uncertainties at 1$\sigma$ and 3$\sigma$ confident levels in various mass map models (F336W, F547M, and F814W), which are created from three different color maps (F336W--F814W, F336W--F547M and F547M--F814W). The parameter search range is identical in Table \ref{fittable}.\\
$^{\rm a}$Mass MGE model taken from \citet{Mitzkus17}.}
\label{full_fittable}
\end{table*}

\section{JAM Models Test for Previous Works}\label{sec:test_previous_jam}

\begin{figure*} 
\centering
	\includegraphics[scale=0.75]{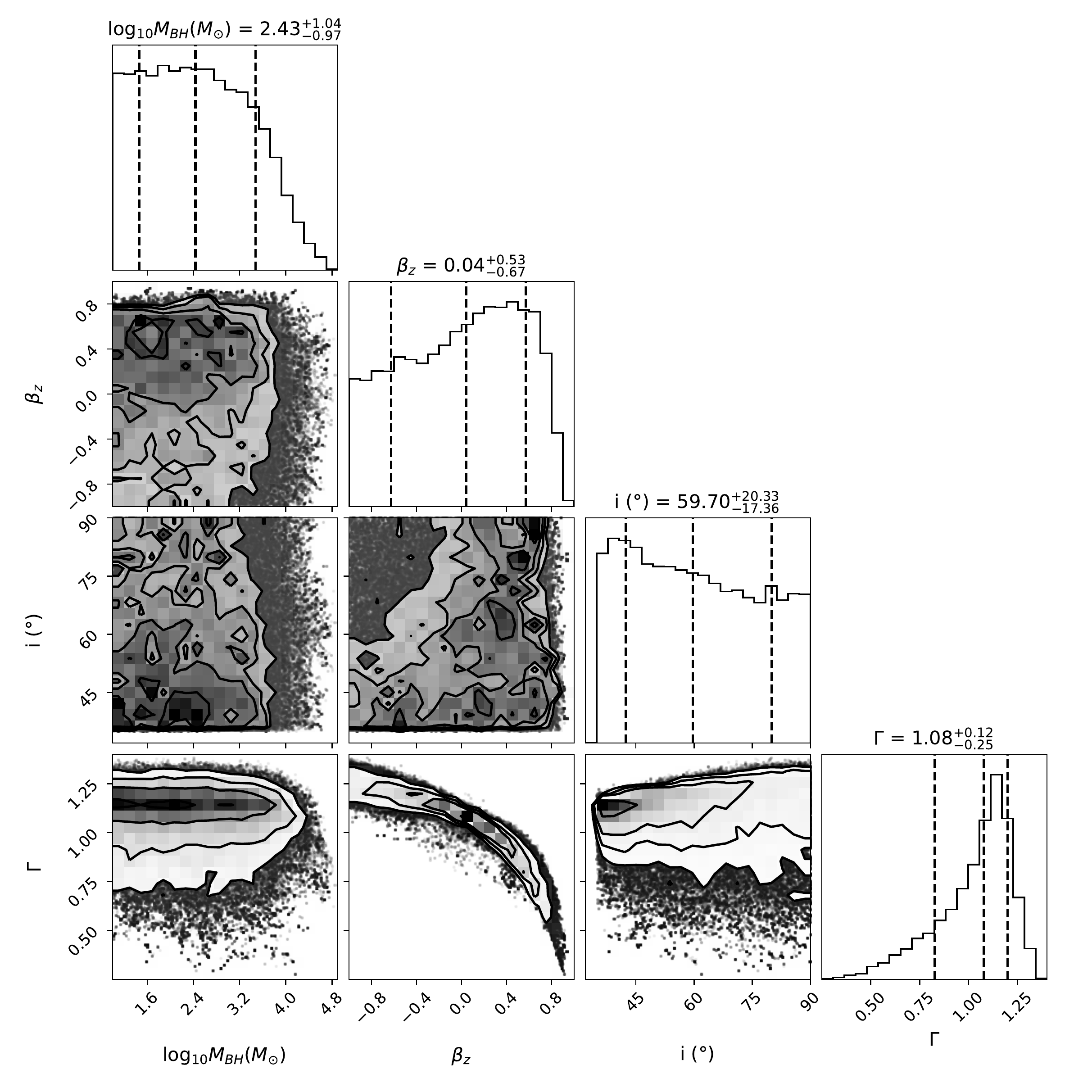}
	\vspace{-7mm}
\caption{\small{The MCMC posterior distribution of the parameter space that we explored with the JAM dynamical models for the central BH in NGC 205 using the mass and light MGEs of N18. Please refer to Fig. \ref{posterior_n205} for additional details.}}  
\label{posterior_n205_n18mges}   
\end{figure*}

We present the MCMC best-fit JAMs for NGC 205 BH in Figure \ref{posterior_n205_n18mges}. Here, we use the previous published mass/light MGE models for NGC 205 (N18) and NGC 404 (N17) to test the robustness of the dynamical model at the low-mass regime below one hundred thousand Solar masses.  

\end{document}